\documentclass[11pt,a4paper]{article}

\usepackage[utf8]{inputenc}
\usepackage[ngerman, english]{babel}
\usepackage{amsmath}
\usepackage{amsthm}
\usepackage{amssymb}
\usepackage{hyperref}
\usepackage{url}
\usepackage{mathtools}
\usepackage{tikz}
\usepackage{array}
\usepackage{pifont}
\usepackage{makecell}
\usepackage{csquotes}
\usepackage{pdflscape}
\usepackage{tabularx}
\usepackage{geometry}
\usepackage{centernot}
\usepackage[english, lined,ruled, linesnumbered]{algorithm2e}

\newcommand{\cmark}{\ding{51}}
\newcommand{\xmark}{\ding{55}}

\usetikzlibrary{arrows}
\hypersetup{
    colorlinks=true,
    linkcolor=blue,
    filecolor=magenta,      
    urlcolor=blue,
    citecolor=blue,
    pdftitle={Overleaf Example},
    }

\usetikzlibrary{%
	positioning,
	fit,
	calc,
	arrows,
	arrows.meta,
	shapes.misc
}

\usepackage{graphicx}
\graphicspath{{./images/}}

\newcounter{Figcount}
\newcounter{tempFigure}

\numberwithin{equation}{section}
\newtheorem{definition}{Definition}[section]
\newtheorem{theorem}{Theorem}[section]
\newtheorem{example}{Example}[section]

\newtheorem{lemma}[definition]{Lemma}

\newtheorem{corollary}[definition]{Corollary}
\title{Rule-based Graph Repair using Minimally Restricted Consistency-Improving Transformations}

\author{Alexander Lauer}

\date{\today}

\DeclareMathOperator{\nlvl}{nl}
\DeclareMathOperator{\inj}{\hookrightarrow}

\DeclareMathOperator{\shift}{Left}
\DeclareMathOperator{\true}{\textsf{true}}
\DeclareMathOperator{\false}{\textsf{false}}

\DeclareMathOperator{\overlay}{ol}
\DeclareMathOperator{\rep}{remv}

\DeclareMathOperator{\lay}{lay}
\DeclareMathOperator{\nvc}{nv}
\DeclareMathOperator{\ap}{ap}
\DeclareMathOperator{\nex}{exv}

\DeclareMathOperator{\track}{tr}
\DeclareMathOperator{\rem}{ned}

\DeclareMathOperator{\eol}{eol}

\DeclareMathOperator{\api}{basic}
\DeclareMathOperator{\src}{src}
\DeclareMathOperator{\tar}{tar}
\DeclareMathOperator{\id}{id}
\DeclareMathOperator{\nin}{nui}
\DeclareMathOperator{\nwo}{vio}
\DeclareMathOperator{\subcondition}{sub}
\DeclareMathOperator{\replacement}{rep}
\DeclareMathOperator{\cutted}{cut}
\DeclareMathOperator{\kmax}{k_{\max}}
\DeclareMathOperator{\interGraph}{IG}
\DeclareMathOperator{\interCondition}{IC}
\DeclareMathOperator{\ac}{ac}
\DeclareMathOperator{\maintaining}{main}
\DeclareMathOperator{\increasing}{incr}
\DeclareMathOperator{\shiftm}{Shift}

\newcommand{\scond}[2]{\subcondition_{#1}(#2)}
\newcommand{\repl}[3]{\replacement_{#1}(#2,#3)}
\newcommand{\cut}[2]{\cutted_{#1}(#2)}
\newcommand{\maxk}[2]{\kmax(#1,#2)}
\newcommand{\ig}[2]{\interGraph(#1,#2)}
\newcommand{\ic}[3]{\interCondition_{#1}(#2,#3)}
\newcommand{\nv}[2]{\nvc_{#1}(c,#2)}
\newcommand{\nvio}[3]{\nvc_{#1}(#2,#3)}

\newcommand{\apb}[2]{\api_{#1}(#2)}
\newcommand{\main}[2]{\maintaining_{#1}(#2)}
\newcommand{\incr}[3]{\increasing_{#1}(#2,#3)}
\newcommand{\remain}[2]{\rem_{#1}(#2)}
\newcommand{\ins}[2]{\nin_{#1}(#2)}
\newcommand{\wors}[2]{\nwo_{#1}(#2)}
\newcommand{\rle}[5]{#1 \overset{#2}{\longleftarrow\joinrel\rhook} #3 \overset{#4}{\lhook\joinrel\longrightarrow}#5 }
\newcommand{\vFound}[3]{\nex_{#1}(#2,#3)}
\newcommand{\vRepaired}[3]{\rep_{#1}(#2,#3)}

\begin{document}

\begin{titlepage}
    \begin{center}
    \selectlanguage{ngerman}
        \vspace*{1cm}
        \Large
        Philipps-Universität Marburg\\
        Fachbereich Mathematik und Informatik\\
        AG Softwaretechnik \\
        \vspace*{1cm}
        \Large
        \textbf{Rule-based Graph Repair using Minimally Restricted Consistency-Improving Transformations}

        \vspace{1.5cm}
            
        \textbf{Masterarbeit}\\
        zur Erlangung des akademischen Grades \\
        Master of Science
        
        \vspace{1.5cm}
        vorgelegt von\\
        \textbf{Alexander Lauer}\\
        Matrikel-Nr: 2658100

        \vspace{1.5cm}

        \today\\
        
        \selectlanguage{english}
    \end{center}
\end{titlepage}

\pagestyle{empty}
\newpage\null\newpage

\selectlanguage{ngerman}
\section*{Selbstständigkeitserklärung}

Hiermit versichere ich, Alexander Lauer, dass ich die vorliegende Arbeit mit dem Titel \emph{Rule-based Graph Repair using Minimally Restricted Consistency-Improving Transformations} selbstständig verfasst und keine anderen als die angegebenen Quellen und Hilfsmittel benutzt habe. Die Masterarbeit wurde in der jetzigen oder ähnlichen Form noch bei keiner anderen Hochschule eingereicht und hat noch keinen anderen Prüfungszwecken gedient.
\\ \\ \\ \\ 
Amöneburg, den \today 

\selectlanguage{english}
\newpage\null\newpage
\pagestyle{plain}
\section*{Abstract}
Model-driven software engineering is a suitable method for dealing with the ever-in\-crea\-sing complexity of software development processes. 
Graphs and graph transformations have proven useful for representing such models and changes to them. 
These models must satisfy certain sets of constraints. An example are the multiplicities of a class structure. 
During the development process, a change to a model may result in an inconsistent model that must at some point be repaired.
This problem is called \emph{model repair}. In particular, we will consider \emph{rule-based graph repair} which is defined as follows: Given a graph $G$, a constraint $c$ such that $G$ does not satisfy $c$, and a set of rules $\mathcal{R}$, use the rules of $\mathcal{R}$ to transform $G$ into a graph that satisfies $c$.

Known notions of consistency have either viewed consistency as a binary property, either a graph is consistent w.r.t. a constraint $c$ or not, or only viewed the number of violations of the first graph of a constraint.
In this thesis, we introduce new notions of consistency, which we call \emph{consistency-maintaining} and \emph{consistency-increasing} transformations and rules, respectively. This is based on the possibility that a constraint can be satisfied up to a certain nesting level. 
Our notion considers the graphs of the first unsatisfied nesting level and is more fine-grained than those already known. Finer in the sense that also the smallest changes, insertions or deletions of individual edges or nodes can be considered as increasing or decreasing consistency.
Furthermore, we extend these notions to \emph{direct consistency-maintaining} and \emph{direct consistency-maintaining} transformations and rules respectively, which prohibit the insertion of new violations altogether, and compare our notions with the existing ones to reveal connections and differences.

We present methods for constructing application conditions that are \emph{direct con\-sis\-ten\-cy-maintaining} or \emph{direct consistency-increasing at layer}, respectively. In the latter case, we present two types of application conditions, one for general rules and one for a particular set of rules, which we call \emph{basic increasing rules}. These application conditions for basic rules are less complex and restrictive than the ones for general rules. The notion of direct consistency-increasing at layer is a weaker version of direct consistency-increasing which allows the construction of less complex application conditions.

Finally, we present an \emph{rule-based graph repair} approach that is able to repair certain constraints, which we call \emph{circular conflict-free constraints}, and certain sets of constraints, which we call \emph{circular conflict-free set of constraints}. Intuitively, a constraint $c$ is \emph{circular conflict free}, if there is an ordering $C_0, \ldots, C_n$ of all graphs of $c$ such that there is no $j<i$ such that a repair of $C_i$ leads to the insertion of a new violation of $C_j$. Analogously, a set of constraint $\mathcal{C}$ is \emph{circular conflict free}, if there is an ordering $c_1, \ldots, c_n$ of all constraints of $\mathcal{C}$ such that there is no $j <i$ such that a repair of $c_i$ at all graphs satisfying $c_j$ leads to a graph  not satisfying $c_j$.
\section*{Zusammenfassung}
Model-driven software engineering ist eine geeignete Methode, um die ständig wachsende Komplexität von Softwareentwicklungsprozessen zu bewältigen. 
Graphen und Graphtransformationen haben sich bewährt, um solche Modelle und Änderungen der Modelle darzustellen. 
Diese Modelle müssen bestimmte Mengen von Bedingungen (constraints) erfüllen. Ein einfachen Beispiel hierfür sind die Multiplizitäten einer Klassenstruktur. 
Während des Entwicklungsprozesses kann eine Änderung eines Modelles jedoch zu einem inkonsistenten Modell führen, welches wieder in ein konsistentes Modell überführt werden muss. 
Dieses Problem heißt \emph{model repair}. Insbesondere betrachten wir das \emph{rules-based graph repair} Problem, welches folgendermaßen definiert ist: Seien ein Graph $G$, ein constraint $c$, sodass $G$ nicht $c$ erfüllt, und eine Menge von Regeln $\mathcal{R}$ gegeben, verwende die Regeln von $\mathcal{R}$, um einen Graphen zu konstruieren, der $c$ erfüllt. 

Bereits bekannte Konzepte von Konsistenz haben Konsistenz entweder als binäre Eigenschaft, entweder ein Graph ist konsistent oder nicht, oder nur den ersten Graphen eines constraints betrachtet. 
In dieser Arbeit führen wir neue Begriffe von Konsistenz ein, die wir \emph{consistency-maintaining} bzw. \emph{consistency-increasing} Transformationen und Regeln nennen. Diese basieren auf der Möglichkeit, dass constraints bis zu einem bestimmten nesting level erfüllt sein können. Unsere Begriffe betrachten daher die Graphen des ersten nicht erfüllten nesting levels und sind feiner im Vergleich zu den bereits bekannten Begriffen. Feiner in dem Sinne, dass auch kleinste Änderungen, das Einfügen oder Löschen von einzelnen Kanten oder Knoten als eine Vergrößerung oder Verschlechterung der Konsistenz angesehen werden kann. 
Des Weiteren erweitern wir diese Begriffe zu den Begriffen \emph{direct consistency-maintaining} bzw. \emph{direct consistency-increasing} Transformationen und Regeln, die das Einfügen von neuen Verletzungen gänzlich verbieten, und vergleichen unsere Begrifflichkeiten mit den bereits existierenden, um Zusammenhänge und Unterschiede aufzudecken.

Wir präsentieren Methoden, um Awendungsbedingungen zu konstruieren, die \emph{direct consistency-maintaining} bzw. \emph{direct consistency-increasing at layer} sind. Im zweiten Fall präsentieren wir zwei Arten von Anwendungsbedingungen, einmal für allgemeine Regeln und einmal für eine bestimmte Menge von Regeln, die wir \emph{basic increasing rules} nennen. 
Anwendungsbedingungen für \emph{basic increasing rules} haben den Vorteil, dass diese weniger komplex und restriktiv sind als die Anwendungsbedingungen für allgemeine Regeln. Der Begriff von \emph{direct consistency-increase rules at layer} ist eine schwächere Version von \emph{direct consistency-increase rules} und führt zu weniger komplexen Anwendungsbedingungen.

Schließlich stellen wir einen \emph{rule-based graph repair} Ansatz vor, der in der Lage ist, bestimmte constraints, die wir \emph{circular conflict free constraints} nennen, und bestimmte Mengen von constraints, die wir \emph{circular conflict free set of constraints} nennen, zu reparieren.
Intuitiv ist ein constraint $c$ \emph{circular conflict free}, wenn es eine Ordnung $C_0, \ldots, C_n$ von allen Graphen von $c$ gibt, sodass kein $j <i$ existiert, so dass eine Reparatur eines Vorkommens von $C_i$ zur Einfügung einer neuen Verletzung von $C_j$ führen kann.
Analog ist eine Menge von constraints $\mathcal{C}$  \emph{circular conflict free}, wenn es eine Ordnung $c_1, \ldots, c_n$ von allen constraints von $\mathcal{C}$ gibt, so dass kein $j <i$ existiert, so dass die Reparatur von $c_i$ an einem Graphen, der bereits $c_j$ erfüllt,  nicht zu einem Graphen führen kann, der $c_j$ nicht erfüllt.

\newpage

\tableofcontents
\newpage\null\newpage

\section{Introduction}

Model-driven software engineering is a suitable tool to deal with the increasing complexity of software development processes. 
Graphs and graph transformations have emerged as a suitable framework for model-driven software engineering, where models are represented by graphs and changes in models are represented by graph transformations. 
These models need to be consistent with respect to certain constraints, e.g. multiplicities if the model represents an object diagram. 
The concept of nested graph constraints has proven to be suitable for expressing these constraints \cite{habel2005nested}.
As models are changed during development, the new model may become inconsistent and the consistency must at some point be restored.

The problem of restoring this consistency is called \emph{graph repair}: Given a constraint $c$ and an inconsistent graph $G$, transform $G$ into a graph that satisfies $c$.
In particular, we will consider the problem of \emph{rule-based graph repair}, which is defined as follows: Given a graph $G$, a constraint $c$ and a set of rules $\mathcal{R}$, use the rules of $\mathcal{R}$ to transform $G$ into a graph that satisfies $c$.
Because of the versatility of graphs and graph transformations, the concept of graph repair can be used to resolve inconsistencies for all kinds of graph-like structures.  

There are several rule-based graph repair approaches and we will now briefly discuss some of them.
A rule-based graph repair approach for so-called proper constraints has been introduced by Sandmann and Habel \cite{sandmann2019rule}.
Nassar et al. have presented a repair approach to repair multiplicities for EMF models \cite{nassar2017rule,nassar2017rule1}. 
Habel and Pennemann have introduced the notions of \emph{consistency-preserving} and \emph{consistency-guaranteeing} transformations \cite{habel2009correctness}. As the name suggests, these binary notions allow to decide whether a transformation guarantees or preserves the consistency of the derived graph. In addition, they presented application conditions which guarantee that a transformation via rules equipped with them is consistency-preserving or consistency-guaranteeing respectively. 
Kosiol et al. have presented the graduated notions of \emph{consistency-sustaining} and \emph{consistency-improving} transformations and rules \cite{kosiol2022sustaining}. In contrast to the notions of consistency-preserving and consistency-guaranteeing, these notions allow an evaluation  of the magnitude of the inconsistency of a graph. They have also presented consistency-sustaining application conditions.

None of these approaches is able to repair multiple constraints with a nesting level higher than $2$, and the application conditions constructed are very complex. 

In this thesis, we will introduce a rule-based graph repair approach for certain constraints in alternating normal-form (ANF), so-called \emph{circular conflict-free constraints}, and for specific sets of these constraints, called \emph{circular conflict-free sets of constraints}.

The main idea of our approach for one constraint is to increase the consistency of a graph level by level. That is, given a constraint $c = \forall(C_1, \exists(C_2, \forall(C_3, \exists(C_4, \ldots)))$ and an inconsistent graph $G$, in the first step we repair the graph such that the derived graph satisfies $\forall(C_1, \exists(C_2, \true))$. 

In the next step, we repair the graph until it satisfies $\forall(C_1, \exists(C_2, \forall(C_3, \exists(C_4, \true)))$ and so on until finally, the derived graph satisfies $c$.
Note that we always repair two nesting levels at a time, since $\forall(C_1, \true)$ is always satisfied, and the satisfaction of $\forall(C_1, \false)$ immediately implies the satisfaction of $c$. Therefore, a process that repairs only one nesting level would return a consistent graph after at most two iterations, with the drawback that many occurrences of graphs are simply deleted. 

To do this, we will introduce new notions of consistency, called \emph{consistency-main\-tain\-ing} and \emph{con\-sis\-ten\-cy-increasing} transformations and rules. As the name suggests, \emph{consistency-maintaining} transformations do not decrease consistency, and  \emph{con\-sis\-ten\-cy-increasing} transformations increase the consistency. 
These notions are based on the first two levels of a constraint that are not satisfied. That is, given a graph $G$ that satisfies $\forall(C_1, \exists(C_2, \true))$ but not $\forall(C_1, \exists(C_2, \forall(C_3, \exists(C_4, \true)))$  only occurrences of $C_3$ are considered. This is because only occurrences of $C_3$ need to be repaired in order to derive a graph from $G$ that satisfies  $\forall(C_1, \exists(C_2, \forall(C_3, \exists(C_4, \true)))$. In particular, our notions are also able to detect the smallest changes in consistency, namely, the insertion or deletion of single elements, which will lead to a more consistent graph. 
Therefore, our notions are more fine-grained than the notions of consistency-preserving, consistency-guaranteeing, consistency-sustaining and consistency-improving transformations and rules. 

In addition, we will refine these notions into the notions of \emph{direct consistency-main\-tain\-ing} and \emph{direct consistency-in\-crea\-sing} transformations and rules, which can be expressed via second-order logic formulas. These notions completely prohibit the insertion of new violations. We will formally compare our newly introduced notions with those described above to ensure that they are indeed new notions of consistency, and to point out similarities and differences. 

We will introduce weaker notions of (direct) consistency-maintaining and (direct) consistency-increasing rules, called (direct) consistency-maintaining rules at layer and (direct) consistency-increasing rules at layer. 
Intuitively, a rule is (direct) consistency-maintaining at layer or (direct) consistency-increasing at layer if all its applications to graphs satisfying the constraint up to that layer are (direct) consistency-maintaining or (direct) consistency-increasing with respect to that constraint. With these notions, we will be able to construct less complex application conditions.
We will present direct consistency-maintaining and two types of direct consistency-increasing application conditions at layer. One for general rules and one for a specific set of rules called \emph{basic increasing rules}. For basic increasing rules, we are able to construct direct consistency-increasing application conditions at layer that are less restrictive and less complex compared to the general ones. 
We will show that each of the constructed application conditions produces consistency-maintaining rules or consistency-increasing rules at layer.

Finally, we present two rule-based graph repair approaches. The first  for one \emph{circular conflict-free} constraint and the second for a \emph{circular conflict-free set of constraints}, which uses the repair approach for one constraint. 
Intuitively, a constraint $c$ is \emph{circular conflict free}, if there is an ordering $C_0, \ldots, C_n$ of all graphs of $c$ such that there is no $j<i$ such that a repair of $C_i$ leads to the insertion of a new violation of $C_j$. Analogously, a set of constraint $\mathcal{C}$ is \emph{circular conflict free}, if there is an ordering $c_1, \ldots, c_n$ of all constraints of $\mathcal{C}$ such that there is no $j <i$ such that a repair of $c_i$ at all graphs satisfying $c_j$ leads to a graph  not satisfying $c_j$. 

Both processes make use of a given set of rules $\mathcal{R}$ and we present a characterisation of when such a set of rules is able to repair the constraint or the set of constraints, respectively. We will use the notions of \emph{consistency-maintainment} and \emph{consistency-increasement} to characterise sequences of transformations, that are able to repair an occurrence of a graph $C_k$ without introducing new violations of certain other graphs of the constraint, and in particular of $C_k$ itself.
Of course, this is only a sufficient criterion, since it depends strongly on the input graph whether a consistent graph can be derived by applying the rules of $\mathcal{R}$. In addition, we show the correctness and termination of our approach.

This thesis is structured as follows:
Formal prerequisites are introduced in Section \ref{preliminaries}. The notions of (direct) consistency-maintaining and (direct) consistency-increasing transformations and rules are given in Section \ref{increase_maintainment} and the construction of application conditions and characterisation of basic rules is given in Section \ref{appl_conds}. 
The repair process is presented in Section \ref{repair}. We summarise related graph repair approaches in Section \ref{rel_work}, before concluding the paper with Section \ref{conclusion}.

\section{Preliminaries}\label{preliminaries}
Our graph repair process is based on the concept of the double-pushout approach \cite{hartmut2006fundamentals}. 
In this chapter, we introduce some formal prerequisites such as graphs, graph morphisms, nested graph conditions and constraints, and graph transformations. 
  
\subsection{Graphs and Graph morphisms}
We start by introducing graphs and graph morphisms according to \cite{hartmut2006fundamentals}.

\begin{definition}[\textbf{graph} \cite{hartmut2006fundamentals}]
	A \emph{graph} $G = (V,E,\src, \tar)$ consists of a set of
	 vertices (or nodes)
	$V$, a set of edges $E$ and two mappings $\src, \tar: E \to 
	V$ that assigns the source and target vertices to an edge.
	The edge $e \in E$ connects the vertices $\tar(e)$ and $\src(e)$. 
	
	If no tuple as above is given, $V_G$, $E_G$, $\tar_G$ and
	$\src_G$ denote the sets of vertices, edges and target and source mappings, 
	respectively.

\end{definition}

\begin{definition}[\textbf{graph morphism} \cite{hartmut2006fundamentals}]
	Let graphs $G$ and $H$ be given. A \emph{graph morphism}
	$f : G \to H$ consists of two mappings $f_V: V_G \to V_H$
	and $f_E: E_G \to E_H$ such that the source and target 
	functions are preserved. This means
	\begin{equation*}
		\begin{split}
			&f_V \circ \src_G = \src_H \circ f_E \text{ and} \\
			&f_V \circ \tar_G = \tar_H \circ f_E
		\end{split}
	\end{equation*}
	holds.
	A graph morphism $f$ is called injective (surjective) if 
	$f_E$ and $f_V$ are injective (surjective) mappings. An injective morphism $f: G \to H$ is called \emph{inclusion} if 
	$f_E(e) = e$ and $f_V(v) = v$ for all edges $e \in E_G$ and all nodes $v \in V_G$.
	If $f$ is injective, it is denoted with  $f: G \inj H$.
	Two morphisms $f_1 :G_1 \to H$ and $f_2: G_2 \to H$ 
	are called \emph{jointly surjective} if for each element
	$e$ of $H$ either an element $e' \in G_1$ with $f_1(e') =e$
	or an element $e' \in G_2$ with $f_2(e') = e$ exists. 

\end{definition}

\begin{definition}[\textbf{typed graph and typed graph morphism \cite{hartmut2006fundamentals}}]
	Given a graph $TG$, called the \emph{type graph}. A \emph{typed graph} over $TG$ is a tuple $(G, type)$ which consists of a graph $G$ and a graph morphism $type\colon G \to TG$. Given two typed graphs $G= (G', type_1)$ and $H = (H', type_2)$, a \emph{typed graph morphism} $f \colon G \to H$ is a graph morphism $f\colon G' \to H'$ such that 
	$$type_2 \circ f = type_1.$$
\end{definition}

In the following, we assume that all graphs are typed over a common type graph, and will simply refer to them as graphs. 
For our newly introduced notions of consistency increase and 
maintainment, we also need to consider \emph{subgraphs}, \emph{overlaps} of graphs, and so-called \emph{intermediate graphs}. Intuitively, intermediate graphs are graphs $G'$ which lie between two given graphs $G$ and $H$. That is, $G$ is a subgraph of $G'$ and $G'$ is a subgraph of $H$.

\begin{definition}[\textbf{subgraph}]
	Let the graphs $G$ and $H$ be given. Then $G$ is called a 
	\emph{subgraph} of $H$ if an inclusion $p: G \inj H$
	exists. $G$ is called a \emph{proper subgraph of $H$} if the morphism $p$ is not bijective. 
\end{definition}
Note that since the inclusion can also be surjective, by this definition every graph $G$ is a subgraph of itself.

\begin{definition}[\textbf{intermediate graph}]
	Let $G$ and $H$ be graphs such that $G$ is a subgraph of $H$. 
	A graph $C$ is called an \emph{intermediate graph} of $G$ 
	and $H$, if $G$ is a proper subgraph of $C$ and $C$ is a subgraph of $H$. 
	The set of intermediate graphs of $G$ and 
	$H$ is denoted by $\ig{G}{H}$.
\end{definition}

\begin{definition}[\textbf{overlap}]
	Let the graphs $G_1$ and $G_2$ be given. An \emph{overlap} $P = (H, 
	i_{G_1}, i_{G_2})$ consists of a graph $H$ and a jointly surjective pair of  
	injective morphisms $i_{G_1}: G_1 \inj H$ and $i_{G_2}: G_2 \inj H$ with 
	$i_{G_1}(G_1) \cap i_{G_2}(G_2) \neq \emptyset$. 
	The set of all overlaps of $G_1$ and $G_2$ is denoted by 
	$\overlay(G_1,G_2)$.
	If a tuple as above is not given, then $G_P$, $i_{G_1}^P$ and  $i_{G_2}^P$ 
	denote the graph and morphisms of a given overlap $P \in \overlay(G_1,G_2)$. 
\end{definition}

Note that $(H, i_{G_1}, i_{G_2})$ where $i_{G_1}$ and $i_{G_2}$ are jointly surjective and  $i_{G_1}(G_1) \cap i_{G_2}(G_2)= \emptyset$ could also be considered as an overlap of $G_1$ and $G_2$.
In this thesis we only need to consider overlaps with $i_{G_1}(G_1) \cap i_{G_2}(G_2)\neq \emptyset$.
So we have embedded this property directly into the definition.
 
As mentioned above, our approach also considers intermediate graphs. Therefore a notion of restricted graph morphisms is needed. For this, we introduce the notion of \emph{restrictions of morphisms}, which intuitively is the restriction of the domain and co-domain of a morphism $p: G \inj H$ with subgraphs of $G$ and $H$ respectively.

\begin{definition}[\textbf{restriction of a morphism}]
	Let the graphs $G$, $H$ and a morphism $f : G \to H$ be given. 
	Then, a morphism $f' : G' \to H'$ is called a 
	\emph{restriction} of $p$ if inclusions $i: G' \inj G$
	and $i': H' \inj H$ exist , i.e. $G'$ is a subgraph of $G$ and $H'$ is a subgraph of $H$, such that 
	\begin{equation*}
		\begin{split}
			&i'_E \circ f'_E = f_E \circ i_E \text{ and} \\
			&i'_V \circ f'_V = f_V \circ i_V.		
		\end{split}
	\end{equation*}
	A restriction of $p$ is denoted by $p^r$.
\end{definition}
Note that given a morphism $p: G \to H$ a restriction 
$p^r: G' \to H'$ of $p$ is uniquely determined by $G'$ and $H'$.
Assume two restrictions $p^r: G' \to H'$ and $q^r: G' \to H'$ of $p$ are given. 
It holds that $i' \circ p^r = p \circ i =  i' \circ q^r$ and $p^r = q^r$ follows with the injectivity of $i'$.  

\subsection{Nested Graph Conditions and Constraints}
\emph{Nested graph constraints} are useful for specifying graph 
properties. The more general notion of \emph{nested graph conditions} allows the specification of properties for graph morphisms and the definition of graph conditions and constraints in a recursive manner.  Within these conditions, only quantifiers and Boolean operators are used \cite{habel2009correctness}.

\begin{definition}[\textbf{nested graph condition} \cite{habel2009correctness}]
A \emph{nested graph condition} over a graph $C_0$ is defined recursively as

\begin{enumerate}
	\item \textsf{true} is a  graph condition over every graph.
	\item $\exists(a_0:C_0 \inj C_1,d)$ is a graph condition over $C_0$ if $a_0$ is an inclusion and $d$ is a graph condition over $C_1$. 
	\item $\neg d$, $d_1 \wedge d_2$ and $d_1 \vee d_2$ are 
		  graph conditions over $C_0$ if $d$, $d_1$ and $d_2$
		  are graph conditions over $C_0$.
	
\end{enumerate}
	Conditions over the empty graph $\emptyset$ are called \emph{constraints}.
We use the abbreviations $\forall(a_0:C_0 \xhookrightarrow{} C_1,d) := \neg \exists(a_0:C_0 \xhookrightarrow{} C_1,\neg d)$ and $\false = \neg \true$.

Conditions of the form $\exists(a_0:C_0 \inj C_1, d)$ are called 
\emph{existential} and the graph $C_1$ is called
\emph{existentially bound}. 
Conditions of the form $\forall(a_0:C_0 \inj C_1, d)$ are called 
\emph{universal} and the graph $C_1$ is called
\emph{universally bound}. 

\end{definition}

Since these are the only types of conditions that will be used in this paper, we will refer to them only as \emph{conditions} and \emph{constraints}. We will use the more compact notations $\exists(C_1,d)$ for $\exists(a_0: C_0 \inj C_1, d)$ and $\forall(C_1,d)$ for $\forall(a_0: C_0 \inj C_1, d)$ if $C_0$ and $a_0$ are clear from the context.

\begin{example}
	Given a condition $c = \forall(a_0 :\emptyset \inj C_1, \exists(a_2: C_1 \inj C_2, \true))$.
	The compact notation of this condition is given by $\forall(C_1, \exists(C_2, \true))$.
\end{example}

\begin{definition}[\textbf{semantic of graph conditions} \cite{habel2009correctness}]
	Given a graph $G$, a condition $c$ over $C_0$ and a graph 
	morphism $p: C_0 \inj G$. Then $p$ satisfies $c$,
	denoted by $p \models c$, if
	\begin{enumerate}
		\item	
			$c = \true$.
		\item 
			$c = \exists(a_0:C_0 \inj C_1,d)$ and there exists an injective 
			morphism $q: C_1 \inj G$ with $p = q \circ a_0$ and $q \models d$.
		\item 
			$c = \neg d$ and  $p \not \models d$. 
		\item 
			$c = d_1 \wedge d_2$ and  
			$p \models d_1$ and $p \models d_2$.
		\item 
			$c = d_1 \vee d_2$ and $p \models d_1$ or $p \models d_2$.			
	\end{enumerate}
	A graph $G$ satisfies a constraint $c$, denoted by $G \models 
	c$, if the empty morphism $p: \emptyset \inj G$ satisfies $c$.
\end{definition}

Our approach is designed to repair a specific type of constraint, constraints  without any boolean operators. Each of these conditions can be transformed into an equivalent condition in so-called \emph{alternating quantifier normal form} \cite{sandmann2019rule}. As the name suggests, these are conditions with alternating quantifiers and without any Boolean operators.

\begin{definition}[\textbf{alternating quantifier normal form (ANF)} \cite{sandmann2019rule}]
	Conditions in \emph{alternating quantifier normal form} (ANF) 
	are defined recursively as 
	
	\begin{enumerate}
		\item $\true$ and $\false$ are conditions in ANF.
		\item \label{c_2} 
			$\exists(a_0: C_0 \inj C_1,d)$ is a condition in ANF
			if either $d$ is a universal condition over $C_1$ in ANF 
			or $d =\true$.
		\item\label{c_3}
			$\forall(a_0: C_0 \inj C_1,d)$ is a condition in ANF
			if either $d$ is an existential condition over $C_1$ in ANF  
			or $d = \false$.
	\end{enumerate}
	Every condition is a \emph{subcondition} of itself.
	In cases \ref{c_2} and \ref{c_3}, $d$ is called a \emph{subcondition} of $\exists(a: C_0 \inj 
	C_1,d)$ or $\forall(a: C_0 \inj C_1,d)$ 
	respectively. All subcondition of $d$ are also subconditions of $\exists(a: 
	C_0 \inj C_1,d)$ or $\forall(a: C_0 \inj C_1,d)$ respectively. 
	The \emph{nesting level} $\nlvl(c)$ of a condition $c$ is 
	recursively defined as $\nlvl(\true)= \nlvl(\false) = 0$ and  $\nlvl
	(\exists(a: P \inj Q, d)) = \nlvl(\forall(a:P \inj Q,d)) := \nlvl(d) +1$.
\end{definition}

In the literature, conditions in ANF also allow conditions that end with conditions of the form $\exists( C_1, \false)$ or $\forall(C_1, \true)$. We exclude these cases so that conditions in ANF can only end with conditions of the form $\exists(C_1, \true)$ or $\forall( C_1, \false)$, since it is easily seen that every morphism $p: C_0 \inj G$ satisfies $\forall( C_1, \true)$ and does not satisfy $\exists(C_1, \false)$. Therefore, these conditions can be replaced by $\true$ and $\false$ respectively.

In the following, we assume that all graphs and the nesting level of a condition are finite. 

Using the \emph{shift over morphism} construction, we are able to transform a 
nested condition over $C$ into a nested condition over $C'$ via an injective 
morphism $i : C \inj C'$ \cite{habel2009correctness}.

\begin{figure}
\center
	\begin{tikzpicture}
		\node (L) at (0,2) {$C_0$};
		\node (K) at (2,2) {$C_0'$};
		
		\node (G) at (0,0) {$C_1$};
		
		\node (H) at (2,0) {$C_1'$};
		
		\draw[right hook-stealth] (L) edge node [above] {$i$}  (K);
		\draw[left hook-stealth] (L) edge node [left] {$a_0$}  (G);
		\draw[right hook-stealth] (G) edge node [above] {$i'$}  (H);
		\draw[right hook-stealth] (K) edge node [right] {$a_0'$}  (H);
		
	\end{tikzpicture}
	\caption{Diagram for the $\shiftm$ operator.}\label{fig_shift}
\end{figure}
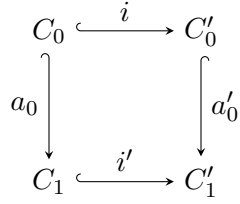

\begin{definition}[\textbf{shift over morphism} \cite{habel2009correctness}]
	Let a condition $c$ over $C_0$ and a morphism $i:C_0 \inj C_0'$ be given.
	The \emph{shift of $c$ over $i$}, denoted by $\shiftm(c,i)$, is given by 
	\begin{enumerate}
		\item 
			If $c = \true$, $\shiftm(c,i) = \true$.
		\item 
			If $c = \exists(a_1:C_0 \inj C_1, d)$, $\shiftm(c,i) = 
			\bigvee_{(a',i') \in \mathcal{F}} \exists(a', \shiftm(d,i'))$
			with $\mathcal{F}$ being the set of all pairs $(a',i')$ of injective 
			morphisms that are jointly surjective and $i' \circ a = a' \circ i$, 
			i.e., the diagram shown in Figure \ref{fig_shift} commutates.
		\item 
			If $c = \neg d$, $\shiftm(c, i) = \neg \shiftm(d, i)$
		\item
			If $c = d_1 \wedge d_2$, $\shiftm(c,i) = \shiftm(d_1,i) \wedge 
			\shiftm(d_2,i)$
		\item
			If $c = d_1 \vee d_2$, $\shiftm(c,i) = \shiftm(d_1,i) \vee 
			\shiftm(d_2,i)$ 
	\end{enumerate}
	
\end{definition} 

\begin{lemma}[\cite{habel2009correctness}]
	Let a condition $c$ over $C_0$ and a morphism $i: C_0 \inj C_0'$ be given.
	Then, for each morphism $m: C_0' \inj G$, 
	$$m \models \shiftm(c,i) \iff m \circ i \models c$$
\end{lemma}
\subsection{Rules and Graph Transformations}
Via \emph{rules} and \emph{graph transformation} graphs can be modified by inserting or deleting nodes and edges.
We will use the concept of the double-pushout approach for rules and transformations, which is based on category theory \cite{hartmut2006fundamentals}. A rule consists of the three graphs $L$, called the \emph{left-hand side}, $K$, called \emph{context},  and $R$, called \emph{right-hand side}, where $K$ is a subgraph of $L$ and $R$. During a transformation, denoted by $G 
\Longrightarrow H$, elements of $L \setminus K$
are removed and elements of $R \setminus K$ are inserted 
so that a new morphism $p: R \inj H$ is created.
In addition, the so-called \emph{dangling edge condition} must be satisfied. Intuitively, for every edge $e \in E_H$ there are vertices $u,v \in V_H$ such that $\tar(e) = u$ and 
$\src(e) = v$ or vice versa.
We also define application conditions. These are nested conditions over $L$ and $R$ that  prevent the transformation if they are not satisfied. 
Later, we will use application conditions to ensure that transformations cannot reduce consistency. 
For example, application conditions that prevent a transformation 
if $G \models c$ and $H\not \models c$.

\begin{definition}[\textbf{rules and application conditions} \cite{hartmut2006fundamentals}]
	A \emph{plain rule} $\rho = \rle{L}{l}{K}{r}{R}$ consists of 
	graphs $L,K,R$ and inclusions $l: K \inj L$ 
	and $r:  K \inj R$. The rule $\rho^{-1} = \rle{R}{r}{K}{l}{L}$
	is called the \emph{inverse rule of $\rho'$}. 
	
	An \emph{application condition} is a nested condition over 
	$L$ or $R$ respectively. A \emph{rule} $(\ap_L, \rho, \ap_R)$ 
	consists of a plain rule $\rho$ and application conditions 
	$\ap_L$ over $L$, called \emph{left application condition}, 
	and $\ap_R$	over $R$, called \emph{right application 
	condition} respectively. 
\end{definition}

\begin{figure}
\center
	\begin{tikzpicture}
		\node (L) at (0,2) {L};
		\node (K) at (2,2) {K};
		\node (R) at (4,2) {R};
		\node (G) at (0,0) {G};
		\node (D) at (2,0) {D};
		\node (H) at (4,0) {H};
		\node (1) at (1,1) {(1)};
		\node (2) at (3,1) {(2)};
		
		\draw[left hook-stealth] (K) edge node [above] {$l$}  (L); 
		\draw[right hook-stealth] (K) edge node [above] {$r$}  (R); 
		\draw[left hook-stealth] (D)   edge node[above]{$g$}(G); 
		\draw[right hook-stealth] (D)  edge node [above]{$h$}(H);
		\draw[left hook-stealth] (K) edge node[fill = white] {$k$} (D);  
		\draw[left hook-stealth] (L) -- node[left]{$m$} (G);
		\draw[right hook-stealth] (R) edge node [right] {$n$} (H);    
	\end{tikzpicture}
	\caption{Diagram of a transformation in the double-pushout approach.}\label{fig_dpo}
\end{figure}
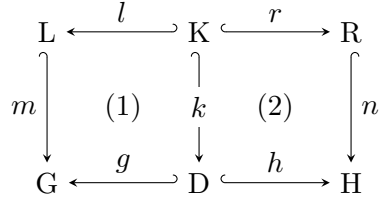

\begin{definition}[\textbf{graph transformation} \cite{hartmut2006fundamentals}]
	Let a rule $\rho = (\ap_L,\rho', \ap_R)$, a graph $G$ 
	and a morphism $m: L \inj G$,
	called the \emph{match}, be given. Then, a \emph{graph 
	transformation}, denoted by $t: G \Longrightarrow_{\rho,m} H$, is given
	in Figure \ref{fig_dpo} if the squares $(1)$ and $(2)$ are
	pushouts in the sense of category theory, $m\models \ap_L$
	and the morphism $n:L \inj H$, called the co-match of $t$, 
	satisfies $\ap_R$.
	The morphisms $g: D \inj G $ and $h: D \inj H$ are called the \emph{transformations morphisms} of $t$. 
\end{definition}

The presence of right application conditions leads to unpleasant side effects.
The satisfaction of a right application condition can only be checked after 
the transformation. The transformation must therefore be reversed if the co-match does not satisfy this condition. To avoid this, we introduce the \emph{shift over rule} operation, which is capable of transforming a right into an equivalent
left application condition \cite{habel2009correctness}. 
\begin{definition}[\textbf{shift over rule} \cite{habel2009correctness}]\label{def:shift}
	Let a rule $\rho = \rle{L}{l}{K}{r}{R}$ with the right
	application condition $\ap$  be given. The \emph{shift of $\ap$ over 
	$\rho$}, denoted with $\shift(\ap, \rho)$, is defined as
	\begin{enumerate}
		\item If $\ap = \true$, $\shift(\ap, \rho) := \true$.
		\item 
			If $\ap = \neg d$, $\shift(\ap, \rho) := \neg \shift(d, \rho)$.
		\item 
			If $\ap = d_1 \wedge d_2$, $\shift(\ap, \rho) := \shift(d_1, \rho) 
			\wedge \shift(d_2, \rho)$.
		\item 
			If $\ap = d_1 \vee d_2$, $\shift(\ap, \rho) := \shift(d_1, \rho) 
			\vee \shift(d_2, \rho)$. 
		\item 
			If $\ap = \exists(a_0: R \inj C_1,d)$, $\shift(\ap, \rho) := 
			\exists(a_0': L \inj C_0', \shift(d, \rho')) $ where 
			$\rho' = \rle{C_0}{g}{D}{h}{C_0'} $ is the rule shown in Figure \ref{fig_left} which is derived by applying $\rho^{-1}$ at match $a_0$. 
			If this transformation does not exist, we set 
			$\shift(\ap, \rho) := \false$.
	
	\end{enumerate}
	 
\end{definition}

Shift over rule produces an equivalent left application condition, 
meaning that, given a right application condition $\ap$ and a plain rule $\rho$ , a match of a transformation satisfies $\shift(\ap,
\rho)$ if and only if the co-match satisfies $\ap$ \cite{habel2009correctness}. 

\begin{lemma}[\cite{habel2009correctness}]
	Let a plain rule $\rho = \rle{L}{l}{K}{r}{R}$, a right application condition 
	$\ap$ for $\rho$ and a transformation $t: G \Longrightarrow_{\rho,m}H$ be 
	given. Then, 
	$$m \models \shift(\ap,\rho) \iff n \models \ap.$$
\end{lemma}
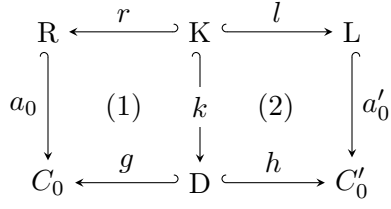
\begin{figure}
\center
	\begin{tikzpicture}
		\node (L) at (0,2) {R};
		\node (K) at (2,2) {K};
		\node (R) at (4,2) {L};
		\node (G) at (0,0) {$C_0$};
		\node (D) at (2,0) {D};
		\node (H) at (4,0) {$C_0'$};
		\node (1) at (1,1) {(1)};
		\node (2) at (3,1) {(2)};
		
		\draw[left hook-stealth] (K) edge node [above] {$r$}  (L); 
		\draw[right hook-stealth] (K) edge node [above] {$l$}  (R); 
		\draw[left hook-stealth] (D)   edge node[above]{$g$}(G); 
		\draw[right hook-stealth] (D)  edge node [above]{$h$}(H);
		\draw[left hook-stealth] (K) edge node[fill = white] {$k$} (D);  
		\draw[left hook-stealth] (L) -- node[left]{$a_0$} (G);
		\draw[right hook-stealth] (R) edge node [right] {$a_0'$} (H);    
	\end{tikzpicture}
	\caption{Transformation for the shift over rule operator.}\label{fig_left}
\end{figure}

Since every right application condition can be transformed into an equivalent left application condition, we will assume from now on that each rule contains only left application conditions. These rules are denoted by $(\ap, \rho)$.
The following Theorem shows, under which conditions a rule $\rho$ is applicable at a match $m$, i.e. there is transformation via $\rho$ at match $m$.

\begin{figure}
\center
	\begin{tikzpicture}
		\node (L) at (0,2) {L};
		\node (K) at (2,2) {K};
		\node (R) at (4,2) {R};
		\node (G) at (0,0) {G};
		\node (D) at (2,0) {D};

		\node (1) at (1,1) {(1)};

		\draw[left hook-stealth] (K) edge node [above] {$l$}  (L); 
		\draw[right hook-stealth] (K) edge node [above] {$r$}  (R); 
		\draw[left hook-stealth] (D)   edge node[above]{$g$}(G); 
		
		\draw[left hook-stealth] (K) edge node[fill = white] {$k$} (D);  
		\draw[left hook-stealth] (L) -- node[left]{$m$} (G);
		    
	\end{tikzpicture}
	\caption{Applicability of a rule $\rle{L}{l}{K}{r}{R}$ at match $m$.}\label{fig_appl}
\end{figure}
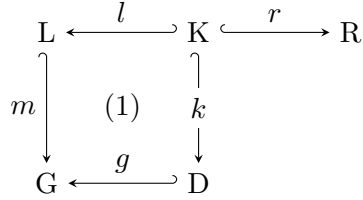
\begin{theorem}[\cite{hartmut2006fundamentals}]
Given a rule $\rho = (\ap, \rle{L}{l}{K}{r}{R})$, a graph $G$ and a match $m: L \inj G$ such that $m \models \ap$. The rule $\rho$ is applicable at match $m$, i.e. there is a transformation $t: G \Longrightarrow_{\rho, m} H$ if and only if there is a graph $D$ such that the square $(1)$ in Figure \ref{fig_appl} is a pushout. 
This is equivalent to $\rho$ and $m$ satisfying the \emph{dangling-edge condition}, i.e. $m \models \ap$ and
$$IP_{\rho,m} \cup DP_{\rho,m} \subseteq GP_{\rho,m}$$ 
where $GP$ is the set of all nodes and edges in $L$ that are not deleted by $\rho$, i.e.
$$GP_{\rho,m} = l(K),$$ 
$IP$ is the set of nodes and edges in $L$ that are identified by $m$, i.e. 
\begin{equation*}
	\begin{split}
		IP_{\rho,m} = &\{v \in V_L \mid \exists v' \in V_L (v' \neq v \text{ and } m_v(v) = m_V(v'))\} \cup \\
		&\{e \in E_L \mid \exists e' \in E_L(e \neq e' \text{ and } m_v(e) = m_V(e'))\},
	\end{split}
\end{equation*}
and DP is the set of nodes in $L$ whose images under $m$ are the source or target of an edge in $G$ that does not belong to $m(L)$, i.e. 
\begin{equation*}
	DP_{\rho,m} = \{v\in V_L \mid \exists e \in E_G \setminus m_E(E_L)(\src(e) = m_V(v) \text{ or } \tar(e) = m_V(v))\}.
\end{equation*}

\end{theorem}
Since we will only consider injective matches, the set $IP_{\rho,m}$ is always empty, and therefore $DP_{\rho,m} \subseteq GP_{\rho,m}$ is sufficient to state that the \emph{dangling-edge condition} is satisfied.

Via the \emph{track morphism} it is possible to track elements across a transformation \cite{plump2005confluence}.

\begin{definition}[\textbf{track morphism} \cite{plump2005confluence}]
	Consider the transformation $t$ shown in figure \ref{fig_dpo}.
	The \emph{track morphism} of $t$, denoted by  $\track_t: G \dashrightarrow H$, is 
	a partial morphism defined as
	$$\track_t = \begin{cases}
					h (g^{-1}(e)) & \text{if $ e \in g(D)$} \\
					\text{undefined} & \text{otherwise.}
				 \end{cases}$$

\end{definition}

For example, given a transformation $t: G \Longrightarrow H$, the track morphism can be used to check whether a morphism $p:C \inj G$ extends to the derived graph $H$ by checking whether $\track_t \circ p$ is total, or whether a new morphism $q:C \inj H$ has been inserted by checking that no morphism $p : C \inj H$ with $q = \track_t \circ p$ exists \cite{kosiol2022sustaining}. We will use these results later on.

\begin{lemma}[\cite{kosiol2022sustaining}]
	Let a transformation $t: G \Longrightarrow H$ with transformation morphisms $g: D \inj G$, $h: D \inj H$ and an occurrence $p: C \inj G$ of a graph $C$ be given. Then,
	\begin{enumerate}
		\item The track morphism $\track_t$ of $t$ is total when restricted to $p(C)$, i.e. $\track_t \circ p$ is total, if and only if there is an injective morphism $p': C \inj D$ such that $p = g \circ p'$. 
		\item 
			Given an injective morphism $p: C \inj H$, $p(C)$ is contained in $\track_t(G)$ if and only if there is an injective morphism $p': C \inj D$ such that $p = h \circ p'$.
	\end{enumerate}
\end{lemma}

\begin{lemma}[\cite{kosiol2022sustaining}]
	Given a transformation $t:G \Longrightarrow H$ with the transformation morphisms $g: D \inj G$, $h:D \inj H$ and a constraint $c$ in ANF that contains the morphism $a_i: C_{i-1} \inj C_i$. Given injective morphisms $p_{i-1}: C_{i-1} \inj G$ and $p_i:C_i \inj G$ such that $p_{i-1} = p_i \circ a_i$ and the track morphism $\track_t : G \dashrightarrow H$ is total when restricted to $p_i(C_i)$. 
	Then, $p'_{i-1} = p'_i \circ a_i$ where $p'_{i-1} = \track_t \circ p_{i-1}$  and $ p'_{i} = \track_t \circ p_{i}$.
	Also, given injective morphisms $p'_{i-1}:C_{i-1} \inj H$ and $p'_{i}:C_{i} \inj H$ such that $p'_{i-1} = p'_i \circ a_i$ and 
	$p_i(C_i)$ is contained in $\track_t(G)$, then $p_{i-1} = p_i \circ a_i$ where $p_{i-1} = \track_t^{-1} \circ p'_{i-1}$ and $p_i = \track_t^{-1} \circ p'_{i}$. 
\end{lemma}

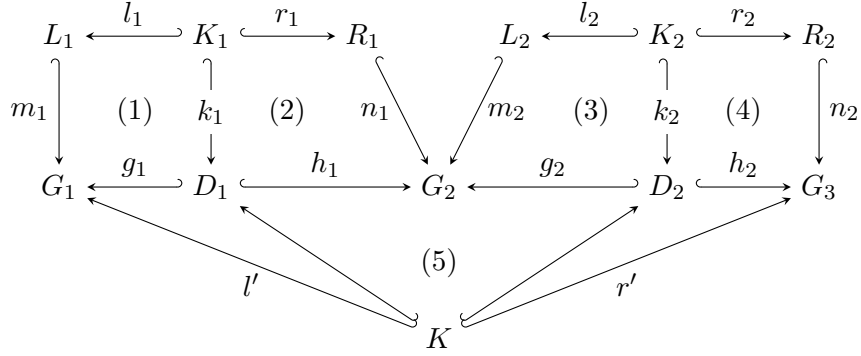
\begin{figure}
\center
	\begin{tikzpicture}
		\node (L1) at (0,2) {$L_1$};
		\node (K1) at (2,2) {$K_1$};
		\node (R1) at (4,2) {$R_1$};
		\node (G1) at (0,0) {$G_1$};
		\node (D1) at (2,0) {$D_1$};
		\node (G2) at (5,0) {$G_2$};
		
		\node(L2) at (6,2) {$L_2$};
		\node(K2) at (8,2) {$K_2$};
		\node(R2) at (10,2) {$R_2$};
		\node(D2) at (8,0) {$D_2$};
		\node(G3) at (10,0) {$G_3$};
		\node(K) at (5, -2) {$K$};
		\node (1) at (1,1) {(1)};
		\node (2) at (3,1) {(2)};
		\node (3) at (7,1) {(3)};
		\node (4) at (9,1) {(4)};
		\node (5) at (5,-1) {(5)};
		
		\draw[left hook-stealth] (K1) edge node [above] {$l_1$}  (L1); 
		\draw[right hook-stealth] (K1) edge node [above] {$r_1$}  (R1); 
		\draw[left hook-stealth] (D1)   edge node[above]{$g_1$}(G1); 
		\draw[right hook-stealth] (D1)  edge node [above]{$h_1$}(G2);
		\draw[left hook-stealth] (K1) edge node[fill = white] {$k_1$} (D1);  
		\draw[left hook-stealth] (L1) -- node[left]{$m_1$} (G1);
		\draw[right hook-stealth] (R1) edge node [left] {$n_1$} (G2); 
		\draw[left hook-stealth] (L2) edge node[right]{$m_2$}(G2);
		\draw[left hook-stealth] (K2) edge node[above]{$l_2$}(L2);
		\draw[right hook-stealth] (K2) edge node[above]{$r_2$}(R2);
		\draw[left hook-stealth] (D2) edge node[above]{$g_2$}(G2);
		\draw[right hook-stealth] (D2) edge node[above]{$h_2$}(G3); 
		\draw[left hook-stealth] (K2) edge node[fill = white]{$k_2$}(D2);
		\draw[right hook-stealth] (R2) edge node[right]{$n_2$}(G3); 
		\draw[left hook-stealth] (K) edge node[below]{$l'$}(G1);
		\draw[right hook-stealth] (K) edge node[below]{$r'$}(G3);
		\draw[left hook-stealth] (K) edge node[below]{}(D1);
		\draw[right hook-stealth] (K) edge node[below]{}(D2);  
	\end{tikzpicture}
	\caption{Pushout diagram of the transformation sequence 
	$G_1 \Longrightarrow_{\rho_1,m_1} G_2 \Longrightarrow_{\rho_2, m_2}G_3$ using 
	the rules $\rho_1  = \rle{L_1}{l_1}{K_1}{r_1}{R_1}$ and $\rho_2  = \rle{L_2}{l_2}{K_2}{r_2}{R_2}$.}\label{fig_concurrent_rule}
\end{figure}

Given a sequence of transformations, the notion of \emph{concurrent rules}
can be used to describe this sequence by a rule. In other words, any sequence of transformations can be replaced by a transformation via its concurrent rule \cite{ehrig2010parallelism}.
\begin{definition}[\textbf{concurrent rule} \cite{ehrig2010parallelism}]
	Let the rules $\rho_1 = \rle{L_1}{l_1}{K_1}{r_1}{R_1}$, $\rho_2 = \rle{L_2}
	{l_2}{K_2}{r_2}{R_2}$ and a sequence of transformations 
	$$G_1 \Longrightarrow_{\rho_1,m_1} G_2 \Longrightarrow_{\rho_2,m_2}G_3$$ 
	be given. Then, $\rho' = \rle{G_1}{l'}{K}{r'}{G_3}$ is called the 
	\emph{concurrent rule of the transformation sequence} if the 
	square $(5)$ in Figure \ref{fig_concurrent_rule} is a pullback.
	
\end{definition}
A transformation sequence $G_1 \Longrightarrow_{\rho_1,m_1} G_2 \Longrightarrow_{\rho_2,m_2}G_3$ can be replaced by a transformation
$G_1 \Longrightarrow_{\rho', \id} G_3$ via its concurrent rule. By inductive application, a concurrent rule for a transformation sequence $G_1 \Longrightarrow_{\rho_0} \ldots \Longrightarrow_{\rho_n} G_n$ of arbitrary finite length can be derived.

\subsection{Concepts of Consistency}\label{sec_consistency}

Now, we will introduce familiar consistency concepts. 
Namely, the notions of consistency preserving and guaranteeing transformations 
\cite{habel2009correctness} and the notions of (direct) consistency sustaining and improving transformations \cite{kosiol2022sustaining}.
Later, we will examine how these concepts differ from and correlate to our newly introduced concept of consistency.

\begin{definition}[\textbf{consistency preserving and guaranteeing transformations} \cite{habel2009correctness}]
	Let a constraint $c$ and a transformation $t: G \Longrightarrow H$ be given. 
	Then, $t$ is called \emph{$c$-preserving} if 
	$$ G \models c \implies H \models c.$$
	The transformation $t$ is called \emph{$c$-guaranteeing} if $H \models c$.

\end{definition}

While consistency preserving and guaranteeing transformations are defined for nested conditions, the finer-grained notions of (direct) consistency sustaining and improving transformations are defined only for conditions in ANF.

\begin{definition}[\textbf{consistency sustaining and improving transformations} \cite{kosiol2022sustaining}]
	Let a constraint $c$ in ANF and a transformation $t: G 
	\Longrightarrow_{\rho} H$ be given.
	If $c$ is existentially bound, $t$ is called \emph{consistency sustaining 
	w.r.t. $c$} if it is $c$-preserving and $t$ is called \emph{consistency 
	improving w.r.t. $c$} if it is $c$-guaranteeing.
	If $c = \forall(a_0: \emptyset \inj C_1,d)$ is universal, $t$ is 
	called \emph{consistency sustaining w.r.t. $c$} if
	$$ |\{p:C_1 \inj G \mid p \not \models d\}| \geq |\{p:C_1 \inj H \mid p \not 
	\models d\}|$$
	and $t$ is called \emph{consistency improving w.r.t. $c$} if 
	$$ |\{p:C_1 \inj G \mid p \not \models d\}| > |\{p:C_1 \inj H \mid p \not 
	\models d\}|.$$
	The number of elements of these sets is called the \emph{number of 
	violations in $G$} and \emph{number of violations in $H$} respectively.
\end{definition}

The even stricter notion of direct sustaining and improving transformations prohibits the insertion of new violations altogether.
\begin{definition}[\textbf{direct sustaining and improving transformations} \cite{kosiol2022sustaining}]
	Let a constraint $c$ in ANF and a transformation $t: G \Longrightarrow_{\rho} 
	H$ be given. If $c$ is existential, $t$ is called \emph{direct 
	consistency sustaining w.r.t. $c$} if $t$ is $c$-preserving and 
	$t$ is called \emph{direct consistency improving w.r.t. $c$} if $t$ is 
	$c$-guaranteeing. 
	
	If $c = \forall(a_0:\emptyset \inj C_1,d)$, $t$ is called 
	\emph{consistency sustaining w.r.t. $c$} if 
	\begin{equation*}
		\begin{split}
			&\forall p:C_0 \inj G((p \models d \wedge \track_t \circ p \text{ is total}) 
	\implies \track_t \circ p \models d) \text{ and} \\& \forall p':C_0 \inj H(\neg 
	\exists q:C_0 \inj G(p' = \track_t \circ q) \implies p' \models d))
		\end{split}
	\end{equation*} 
	and $t$ is called \emph{consistency improving w.r.t. $c$} if additionally 
	\begin{equation*}
		\begin{split}
			&\exists p:C_0 \in G(p \not \models d \text{ and} \track_t \circ p \text{ 
			is total} \wedge \track_t \circ p \models d) \vee \\
			&\exists p:C \inj G (p \not \models d \wedge \track_t \circ p \text{ 
			is not total}).
		\end{split}
	\end{equation*}
\end{definition}

\section{Consistency Increase and Maintainment}\label{increase_maintainment}
In the following we will introduce the notions of \emph{satisfaction up to layer}, \emph{(direct) con\-sis\-ten\-cy-maintaining} and \emph{(direct) consistency-increasing} transformations and rules and compare them with the notions of consistency introduced in the previous section.
\begin{definition}[\textbf{layer of a subcondition}]
	Let a condition $c$ in ANF and a subcondition $d$ of $c$ be given.
	The \emph{layer of d} is defined as $\lay(d) := \nlvl(c) - \nlvl(d)$.
\end{definition}

Our approach is based on the idea that the consistency of a constraint increases layer by layer, and that even small improvements, such as inserting single elements of existentially bound graphs, should be detectable as increasing.
To formalise this, we introduce the notions of \emph{consistency increasing} and \emph{consistency maintaining} transformations and rules, where consistency increasing indicates that the consistency has actually increased and consistency maintaining indicates that the consistency has not decreased.
\subsection{Universally quantified ANF}
The definition of consistency increase and maintainment requires that each condition begins with a universal quantifier. Otherwise, case discrimination is required.  Therefore, we will only consider a subset of the set of conditions in ANF, namely the set of universally quantified conditions in ANF, called \emph{universally quantified ANF} (UANF). Furthermore, we will show that these sets are expressively equivalent by showing that every condition in ANF can be transformed into an equivalent condition in UANF.

\begin{definition}[\textbf{universally quantified alternating quantifier normal form}]
A condition $c$ in ANF is in \emph{universally quantified ANF} (UANF) if it is universally bound. 
\end{definition}

Note that in our notation, given a condition $c$ in UANF, any subcondition of $c$ at layer $0 \leq k \leq \nlvl(c)$ is universal if $k$ is an even number and existential if $k$ is an odd number. Furthermore, a graph $C_k$ of $c$ is  universally bound if $k$ is an odd number and existentially bound if 
$k$ is an even number.  
It is already known that an existentially bound condition $c$ 
can be extended to the equivalent condition $\exists(\id_{C_0}: C_0 \inj C_0, d)$ \cite{habel2005nested}.
Analogously, we show that every condition in ANF has an 
equivalent condition in UANF.

\begin{lemma}
Any condition in ANF can be transformed into an equivalent condition in UANF. 
\end{lemma}

\begin{proof}
Let a graph $G$ and a constraint $c$ in ANF be given.
If $c$ is universal, then $c$ is already in UANF. 
If $c = \exists(a_0: C_0 \inj C_1,d)$, we show that $c$ is equivalent to $c' := \forall(\id_{C_0}: C_0 \inj C_0, c)$.
\begin{enumerate}
\item 
Let $p :C_0 \inj G$ be a morphism such that $p \models c$. 
Then $p \models c'$, because $p$ is the only morphism from $C_0$ to $G$ with $p = p \circ \id_{C_0}$ and $p \models c$.

\item 
Let $p:C_0 \inj G$ be a morphism with $p \models c'$, then all morphisms $q:C_0 \inj G$ with $p = q \circ \id_{C_0}$ satisfy $c$. Since $p = p \circ \id_{C_0}$, it immediately follows that $p \models c$. \qedhere
\end{enumerate}
\end{proof}

For the rest of this thesis, given a condition $c = \forall(a_0 : C_0 \inj C_1, d)$ in UANF, we assume that no morphism in $c$, except $a_0$, is bijective, since it can be shown that every condition in ANF can be transformed into an equivalent condition in ANF that satisfies this property, by showing that
$\exists(\id_{C_0} : C_0 \inj C_0, \forall(a_1: C_0 \inj C_2,d))$ is equivalent to
$\forall(a_1: C_0 \inj C_2,d)$ and that $\forall(\id_{C_0} : C_0 \inj C_0, \exists(a_1: C_0 \inj C_2,d))$ is equivalent to
$\exists(a_1 \circ a_0 : C_0 \inj C_2,d)$ \cite{habel2005nested}.
In addition, given a condition $c$ in UANF, we will denote the first graph of $c$ with $C_0$, the first morphism with $a_0$, the second graph with $C_1$, the second morphism with $a_1$, and so on. This means that we always write constraints as $\forall(a_0:C_0 \inj C_1, \exists(a_1:C_1 \inj C_2, \ldots))$.

\subsection{Satisfaction up to Layer}

\begin{figure}
\centering
\includegraphics[scale=1]{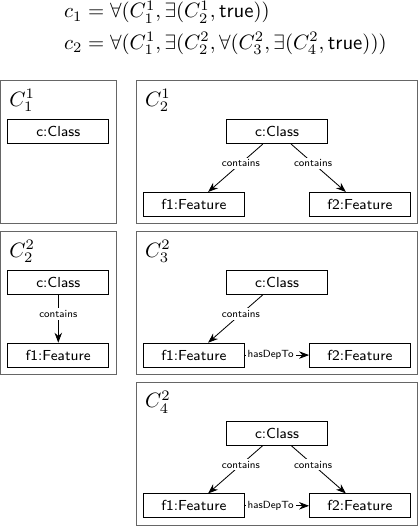}

\caption{Constraints used for examples throughout the thesis.}\label{fig:constraints}
\end{figure}
\begin{figure}
\centering
\includegraphics[scale=1.0]{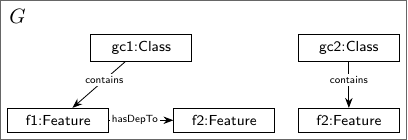}

\caption{The graph used in example \ref{ex_graph}.}\label{fig:graph}
\end{figure}

The goal of our approach is to increase the consistency of a constraint layer by layer, as we have already mentioned. To do this, we introduce a notion of partial consistency, called \emph{satisfaction up to layer}, which allows us to check whether a constraint is satisfied at a particular layer by checking whether the so-called \emph{truncated condition after layer} is satisfied at that layer.

We first define the \emph{subcondition at layer} $-1 \leq k \leq \nlvl(c)$ of a condition $c$. As the name suggests, the subcondition at layer $0 \leq k \leq \nlvl(c)$ denotes the subcondition of $c$ with layer $k$. 
We also define the subcondition at layer $-1$, which is $\true$. This will be useful for evaluating the satisfaction up to layer when a graph does not satisfy any layer of a constraint.
 
\begin{definition}[\textbf{subcondition at layer}]
Let a condition $c$ in ANF be given. The \emph{subcondition at layer $-1 \leq k \leq \nlvl(c)$}, denoted by $\scond{k}{c}$, is the subcondition $d$ of $c$ with $\lay(d) = k$ if $0 \leq k \leq \nlvl(c)$ and $\true$ if $k = -1$. 
\end{definition}

Note that by definition the subcondition at layer $k$ is always a condition over the graph $C_k$ and the morphism is denoted by $a_k$.
\begin{example}
	Consider the condition $ c =\forall(a_0:C_0 \inj C_1, \exists(a_1: C_1
	\inj C_2, \forall(a_2:C_2 \inj C_3, \false)))$.
	Then, 
	\begin{enumerate}
		\item $\scond{-1}{c} = \true$.
		\item $\scond{0}{c} =  \forall(a_0:C_0 \inj C_1, \exists(a_1: C_1
	\inj C_2, \forall(a_2:C_2 \inj C_3, \false))) = c$.
		\item $\scond{1}{c} =  \exists(a_1: C_1
	\inj C_2, \forall(a_2:C_2 \inj C_3, \false))$.
	\item $\scond{2}{c} =  \forall(a_2:C_2 \inj C_3, \false)$.
	\item $\scond{3}{c} =   \false$.
	\end{enumerate}
\end{example}

Let us first introduce an operator which allows to replace a subcondition $\scond{k}{c}$ by an arbitrary condition over $C_k$, called \emph{replacement starting from layer}.

\begin{definition}[\textbf{replacement starting from layer}]
Given a condition $c = Q(a_0: C_0 \inj C_1, d)$ in ANF  with $Q \in \{\forall, \exists\}$ and a condition $e$ over $C_k$ in ANF.  
The \emph{replacement starting from layer $k$ in $c$ by $e$}, denoted by $\repl{k}{c}{e}$, is defined recursively as
\begin{equation*}
	\repl{k}{c}{e} := 
		\begin{cases}
			e & \text{if $k = 0$} \\
			Q(a_0:C_0 \inj C_1, \repl{k-1}{d}{e}) &\text{otherwise.}	
		\end{cases}
\end{equation*}
\end{definition}

\begin{example}
Consider the conditions $c := \forall(a_0: C_0 \inj C_1, \exists(a_1: C_1 \inj C_2, \true))$ and $e = \exists(a'_1: C_1 \inj C_3, d)$. 
The replacement starting from layer $1$ in $c$ by $e$ is given by
$$\repl{1}{c}{e} = \forall(a_0: C_0 \inj C_1, \exists(a'_1: C_1 \inj C_3, d)).$$ 
\end{example}

We now define \emph{truncated conditions after layer} using the concept of replacement starting from layer.
Intuitively, a condition is truncated after a particular layer by replacing the subcondition at the next layer with $\true$ or $\false$, depending on which quantifier the replaced subcondition is bound by.

\begin{definition}[\textbf{truncated condition after layer}]
	Let a condition $c$ in UANF be given. 
	The \emph{truncated condition of $c$ after layer $-1 \leq k < \nlvl(c)$}, 
	denoted by $\cut{k}{c}$, is defined as

$$ \cut{k}{c} := \begin{cases}
					\true &\text{if $k = -1$} \\
					\repl{k +1}{c}{\true}&\text{if $\scond{k}{c}$ is a existential condition,
					 i.e. $k$ is odd} \\
					\repl{k+1}{c}{\false}&\text{if $\scond{k}{c}$ is a universal condition, 
					i.e. $k$ is even.} \\
				\end{cases}$$

\end{definition}

\begin{example}
Consider constraint $c_2$ given in Figure \ref{fig:constraints}.
The truncated conditions of $c_2$ after layer $-1 \leq k < 3$ are given by 
\begin{enumerate}
	\item $\cut{-1}{c_2} = \true$.
	\item $\cut{0}{c_2} = \forall(C_1^1, \false)$.
	\item $\cut{1}{c_2} = \forall (C_1^1, \exists (C_2^2,\true))$.
	\item $\cut{2}{c_2} = \forall (C_1^1, \exists (C_2^2,(\forall C_3^2, \false)))$.
	\item $\cut{3}{c_2} = \forall (C_1^1, \exists (C_2^2,(\forall C_3^2, \exists(C_4^2, \true)))) = c_2$.
\end{enumerate}
\end{example}

Note that the truncated condition of a condition $c$ at layer $\nlvl(c) -1$ is $c$ itself.
With these prerequisites, we can now introduce \emph{satisfaction up to layer}, which allows us to check whether a condition is satisfied up to a given layer. A morphism or graph satisfies a condition or constraint up to a layer if it satisfies the truncated condition after that layer.

\begin{definition}[\textbf{satisfaction up to layer}]
Let a graph $G$ and a condition $c$ in UANF be given.
A morphism $p: C_0 \inj G$ \emph{satisfies $c$ up to layer  $-1 \leq k < \nlvl(c)$}, denoted by $p \models_k c$, if $$p\models \cut{k}{c}.$$

A graph $G$ satisfies a constraint $c$ up to layer $-1 \leq k < \nlvl(c)$, denoted by $G \models_k c$, if $q: \emptyset \inj G$ satisfies $\cut{k}{c}$.
The largest $-1 \leq k < \nlvl(c)$ such that $G \models_k c$ and there is no $k < j < \nlvl(c)$ with $G \models_j c$, called the \emph{largest satisfied layer}, is denoted by $\maxk{c}{G}$. 
When $c$ and $G$ are clear from the context, we use the abbreviation $\kmax$.
\end{definition}

Note that given a graph $G$ and a constraint $c$, $\maxk{c}{G}$ always exists, since every graph satisfies $\true$ and $\cut{-1}{c} = \true$.
Moreover, if $p \models_{\nlvl(c)-1} c$, i.e. $\kmax = \nlvl(c)-1$, it immediately follows that $p \models c$. 
\begin{example}\label{ex_graph}
Consider the graph $G$ given in Figure \ref{fig:graph} and the constraint $c_2$ given in Figure \ref{fig:constraints}. 
This graph does not satisfy $c_2$ because the occurrence of \emph{\texttt{Class}} denoted with \emph{\texttt{gc2}}  does not satisfy $\scond{1}{c_2} = \exists(C_2^2,(\forall C_3^2,(\exists C_4^2,\true)))$, but it satisfies $\cut{1}{c_2} = \forall(C_1^1, \exists(C_2^2, \true))$ and therefore $$G \models_1 c_2 \text{ and } \kmax = 1.$$
\end{example}

\begin{table}
\centering
\begin{tabular}{c|c|c|c|c|c}

$p \models_k c$ & \multicolumn{2}{c|}{$p \models_{j < k} c$} & \multicolumn{2}{c|}{$p \models_{j > k} c$} &$ p \models c $ \\
 & $j$ even & $j$ odd & $j$ even & $j$ odd & \\

\hline
$k$ even & ? & \cmark &\cmark & \cmark & \cmark \\
$k$ odd & ? & \cmark & ? & ? & ? \\

\end{tabular}

\caption{Overview of the inferences made about satisfaction up to layer, where \enquote{\cmark} indicates that $p \models_j c$ or $p \models c$ if $p \models_k c$ and \enquote{?} indicates that it cannot be inferred from $p \models_k c$ whether $p \models_j c$ or $p \not \models_j c$.}\label{table_satisfaction_at_layer}

\end{table}

Given a graph $G$, a condition $c$ and a morphism $p : C_0 \inj G$.
Suppose that $p \models_k c$ with $0 \leq k < \nlvl(c)$. 
Then we can infer results for the satisfaction until up to other layers.
If $k$ is even, i.e. $\scond{k}{c}$ is a universal condition,  we can conclude that $p \models_j c$ for all $k < j < \nlvl(c)$ and especially $p \models c$.
It also follows that $p \models_j c$ for all odd $0 \leq j < k$, i.e. $\scond{j}{c}$ is an existential condition. 
We present these results in the following lemmas,  an overview is given in Table \ref{table_satisfaction_at_layer}.

We start by examining the consequences for the satisfaction up to layer $\nlvl(c) > j > k$. Our first lemma shows that replacing the subcondition $\scond{k+1}{c}$ by any condition over $C_{k+1}$ leads to a condition that is satisfied by $p$ if $k$ is even.

\begin{lemma}\label{lemma_help_lay_sat}
Given a graph $G$, a condition $c$ in UANF and a morphism $p : C_0 \inj G$ with $p \models_k c$ where $-1 \leq k < \nlvl(c)$ is even.
Then, for any condition $f$ over $C_{k+1}$ it holds that
$$p \models \repl{k + 1}{c}{f}.$$
\end{lemma}

\begin{proof}
We start by showing the statement for the smallest $-1 \leq j < \nlvl(c)$ such that $\scond{j}{c}$ is universally bound and $p \models_j c$. 
After this, we can conclude that this statement holds for all $-1 \leq i < \nlvl(c)$ such that $\scond{i}{c}$ is universally bound and $p \models_i c$.

Let $q: C_{j} \inj G$ be a morphism such that $q \models \forall(a_j:C_{j} \inj C_{j+1}, \false)$. 
This morphism must exist, since $j$ is the smallest even number with $p \models_j c$.
Therefore, there is no morphism $q': C_{j+1} \inj G$ with $q = q' \circ a_j$.
Hence, for every condition $f$ over $C_{j+1}$ a morphism $q': C_{j+1} \inj G$ with $q \not \models f$ and $q = q' \circ a_j$ cannot exist. 
It follows immediately that $q \models \forall(a_j:C_{j} \inj C_{j+1}, f)$ and with that $p \models \repl{j + 1}{c}{f}$.

We can now conclude that for every even $j < k \leq \nlvl(c)$, such that $p \models_k c$, and every condition $d$ over $C_{k + 1}$ it holds that $p \models \repl{k + 1}{c}{d}$ because $\repl{k + 1}{c}{d} = \repl{j+1}{c}{\scond{j+1}{\repl{k+1}{c}{d}}}$.
\end{proof}

As a direct consequence of the previous lemma, a morphism which satisfies a condition up to layer $k$, where $k$ is even, also satisfies the condition at layer
$j$ for all $j > k$.

\begin{lemma}\label{lemma_lay_sat}

Given a graph $G$, a morphism $p:C_0 \inj G$ and a condition $c$ in UANF.
If $0 \leq k < \nlvl(c)$ is even, i.e. $\scond{k}{c}$ is a universal condition, then for all $k < j < \nlvl(c)$ it holds that
	$$p \models_k c \implies p \models_j c.$$

\end{lemma}
\begin{proof}
Follows immediately by using Lemma \ref{lemma_help_lay_sat} and setting $f$ equal to $\scond{k+1}{\cut{j}{c}}$.
\end{proof}

Since a morphism $p$ satisfies a condition $c$ in UANF if and only if $p$ satisfies $c$ up to layer $\nlvl(c)-1$, we can conclude the following.

\begin{corollary}\label{corol:satisfaction}
	Given a graph $G$, a morphism $p:C_0 \inj G$ and a condition $c$ in UANF. If 
	$0 \leq k < \nlvl(c)$ is even, it holds that 
	$$p \models_k c \implies p \models c.$$
\end{corollary}
In the following, we will mostly assume that $\kmax$ is odd. This is because an even $\kmax$ implies that the condition is already satisfied.
Furthermore, this allows us to make statements about the satisfaction of other conditions. Given a graph $G$, a morphism $p : C_0 \inj G$ and a condition $c$ such that $p \models_k c$ for an even $-1 \leq k < \nlvl(c)$. It follows that $p \models c$ and in particular $p \models c'$ for each condition $c'$ with  $\cut{k}{c} = \cut{k}{c'}$.

Let us now examine the satisfaction up to layer $j$ with $-1 < j < k$. If $j$ is odd, i.e. $\scond{j}{c}$ is an existential condition, we can conclude that $p \models_j c$ as shown in the next lemma.
If $j$ is even, i.e. $\scond{j}{c}$ is universally bound, we can only make statements that depend on $\kmax$. If $\kmax < \nlvl(c) -1$, then $p \not \models_j c$. Otherwise, Corollary \ref{corol:satisfaction} implies  that $p \models c$ and therefore $\kmax = \nlvl(c)-1$.
If $\kmax = \nlvl(c)-1$, we can say that there is at least one even $j \leq \kmax$ with $p \models_j c$ if $c$ ends with $\forall(C_{\nlvl(c)}, \false)$.
An overview of these relations is given in Table \ref{table_abh_kmax}.

\begin{table}
\centering
\begin{tabular}{c|c|c|c}

$p \models_k c$ & $\kmax < \nlvl(c)-1$  &$ \kmax = \nlvl(c) -1$ \\
 & $k \leq \kmax$  & $k < \kmax$  \\

\hline
$k$ even & \xmark & ? \\
$k$ odd & \cmark &  \cmark \\

\end{tabular}

\caption{Overview of the satisfaction up to layer $k$ with respect to $\kmax$, where \enquote{\cmark} indicates that $p \models_k c$, \enquote{\xmark} indicates that $p \not \models_k c$ and \enquote{?} indicates that it cannot be concluded from $p \models_k c$ whether $p \models_j c$ or $p \not \models_j c$.  }\label{table_abh_kmax}

\end{table}

\begin{lemma}\label{lem_ex_lower}
Given a graph $G$, a morphism $p: C_0 \inj G$ and a constraint $c$ in UANF.
Then for all odd $-1 \leq k \leq \kmax$, i.e. $\scond{k}{c}$ is an existential condition, we have
$$p \models_k c.$$

\end{lemma}
\begin{proof}
If there is an even $0 \leq j < \kmax$, i.e. $\scond{j}{c}$ is universal, with $p \models_j c$, let $j'$ be the smallest of these.
Lemma \ref{lemma_help_lay_sat} implies that $p \models_{\ell} c$ for all $j' \leq \ell < \nlvl(c)$. Otherwise we set $j' = \kmax$.

Let $\ell < j'$, such that $\scond{\ell}{c}$ is a existential condition and let $d = \scond{\ell}{\cut{j'}{c}} = \exists(a_{\ell}: C_{\ell} \inj C_{\ell+1}, e)$ be the subcondition at layer $\ell $ of the truncated condition after layer $j'$ of $c$. 
	Since $\ell < j'$, there must be a morphism $q: C_{\ell} \inj G$ with $q \models d$ and therefore there must be a morphism $q': C_{\ell +1}\inj G$ with $q = q' \circ a_{\ell}$ and $q' \models e$.
	It follows that $q \models \exists(a_{\ell}: C_{\ell} \inj C_{\ell+1}, \true)$ and thus $p \models_{\ell} c$.
\end{proof}

\begin{example}
	We will show counterexamples for all \enquote{?} in Table 
	\ref{table_satisfaction_at_layer} and Table \ref{table_abh_kmax}.
	Consider constraint $c_2 = \forall(C_1^1, \exists(C_2^2, \forall (C_3^2, 
	\exists(C_4^2, \true))))$ given in Figure \ref{fig:constraints}.
	We begin with Table \ref{table_satisfaction_at_layer}.
	\begin{enumerate}
		\item $j < k$, $j$ and $k$ are even. We set $k = 2$ and $j = 0$. 
			 It follows that $\cut{k}{c} = \forall(C_1^1, \exists(C_2^2, \forall (C_3^2, \false)))$ and $\cut{j}{c} = \forall(C_1^1,\false)$.  Then,
			 \begin{enumerate}
			 \item $C_2^2 \models_2 c_2$ and $C_2^2 \not \models_0 c_2$.
			  
			  \item $\emptyset \models_2 c_2$ and $\emptyset \models_0 c_2$.
			  
\end{enumerate}
		\item $j <k$, $k$ is odd and $j$ is even. We set $k = 3$ and $j = 0$. 
		It follows that $\cut{k}{c_2} = c_2$ and $\cut{j}{c} = \forall(C_1^1,\false)$.
		Then,
		\begin{enumerate}
			\item $C_4^2 \models_3 c_2$ and $C_4^2 \not \models_0 c_2$.
			\item $\emptyset \models_3 c_2$ and $\emptyset \models_0 c_2$.
\end{enumerate}

		\item $j > k$, $k$ is odd and $j$ is even. We set $k = 1$ and $j = 2$. It follows that 
		$\cut{k}{c} = \forall(C_1^1, \exists(C_2^2,\true))$ and $\cut{j}{c} = \forall(C_1^1, \exists(C_2^2, \forall (C_3^2, \false)))$. Then,
		\begin{enumerate}
			\item $C_2^2 \models_1 c_2$ and $C_2^2 \models_2 c_2$.
			\item $C_3^2 \models_1 c_2$ and $C_3^2 \not \models_2 c_2$.	
\end{enumerate}				
		  
		\item $j >k$, $k$ and $j$ are odd. We set $k = 1$ and $j = 3$. It follows that $\cut{k}{c} = \forall(C_1^1, \exists(C_2^2,\true))$ and $\cut{j}{c} = c_2$. Then,
		\begin{enumerate}
		 \item $C_2^2 \models_1 c_2$ and $C_2^2 \models_3 c_2$.
		 \item $C_3^2 \models_1 c_2$ and $C_3^2 \not \models_3 c_2$.
\end{enumerate}	 Since  $\cut{j}{c} = c_2$ this is also a counterexample for the \enquote{?} in the column denoted by \enquote{$p \models c$}.
		
	\end{enumerate}
	For the \enquote{?} in Table \ref{table_abh_kmax}, the graph $C_2^2$ satisfies $c_2$ and therefore $\kmax = 3 = \nlvl(c_2)-1$. 
	But, $C_2^2 \models_2 c_2$ and $C_2^2 \not \models_0 c_2$.
\end{example}

\subsection{Consistency Increasing and Maintaining Transformations and Rules}
Using satisfaction up to layer, an increase of consistency can be detected in the following way: Let $t: G \Longrightarrow H$ be a transformation. 
If the largest satisfied layer in $H$ is greater than the largest satisfied layer in $G$, i.e. $\maxk{c}{G} < \maxk{c}{H}$, we consider the transformation as consistency-increasing. 
However, the notion of consistency-increasing should also be able to detect the smallest changes made by a transformation that leads to an increase of consistency, namely the insertion of a single edge or node of an existentially bound graph.
To do this, we introduce \emph{intermediate conditions}, which are used to detect this type of increase by checking whether an intermediate condition not satisfied by $G$ is satisfied by $H$.
A decrease of consistency can be detected in a similar way, by checking whether an intermediate condition satisfied by $G$ is not satisfied by $H$.  
Intuitively, the last graph of a truncated condition $c$ is replaced by an intermediate graph of the penultimate graph and the last graph of that truncated condition.

If $c$ ends with an existential condition, the constructed intermediate condition is weaker than $c$, in the sense that the satisfaction of $c$ implies the satisfaction of the intermediate condition, as shown by Lemma \ref{lemma_intermediateCondition}.

Conversely, if $c$ ends with a universal condition, the opposite holds: The satisfaction of an  intermediate condition implies the satisfaction of $c$. 
For this reason, we have designed intermediate conditions so that they only replace graphs on existential layers.

\begin{definition}[\textbf{intermediate condition}]
	Given a condition $c$ in UANF  and let $0 \leq k < \nlvl(c)$ be odd, i.e. $\scond{k}{c}$ is an existential condition. The \emph{intermediate condition}, denoted by $\ic{k}{c} {C'}$, of $c$ at layer $k$ with $C' \in \ig{C_k} {C_{k+1}}$ is defined as
	$$\ic{k}{c}{C'} := \repl{k}{c}{\exists(a_k^r:C_k \inj C', \true)}.$$
\end{definition}

\begin{lemma}\label{lemma_intermediateCondition}
	Given a condition $c$ in UANF, a graph $G$, $0 \leq k < \nlvl(c)$ odd, i.e. $\scond{k}{c}$ is a existential condition, and $C' \in \ig{C_k}{C_{k+1}}$. Then, 
	$$G \models \cut{k}{c} \implies G \models \ic{k}{c}{C'}.$$  
\end{lemma}
\begin{proof}
	Assume that $G \models \cut{k}{c}$, i.e. $G \models_k c$. If there is an even $-1 \leq j < k$ such that $G \models_j c$, $G \models \ic{k}{c}{C'}$ follows with Lemma \ref{lemma_help_lay_sat}. 
	Otherwise, if there is no such $j$, for all morphism $p: C_k \inj G$ such that there is a morphism $p': C_{k+1} \inj G$ with $p = p' \circ a_k$, there is also a morphism $q: C' \inj G$ with $p = q \circ a_k^r$ where $a_k^r: C_k \inj C$ is the restriction of $a_k$ and $q$ is a restriction of $p$ to the domain $C$.  It follows that $G \models \ic{k}{c}{C'}$.
\end{proof}

%

\begin{example}
Consider constraint $c_1$ given in Figure \ref{fig:constraints}. 
Since $C_2^2 \in \ig{C_1^1}{C_2^1}$, we can construct an intermediate condition of $c_1$ at layer $1$ with $C_2^2$ as $\ic{1}{c_1}{C_2^2} = \forall (C_1^1 \exists (C_2^2, \true))$. 
While $c_1$ checks whether each node of type \emph{\texttt{Class}} is connected to at least two nodes of  type \emph{\texttt{Feature}}, the intermediate condition checks whether each node of type \emph{\texttt{Class}} is connected to at least one node of type \emph{\texttt{Feature}} which is trivially satisfied if $c_1$ is satisfied.
\end{example}

With the results above, we are now ready to define the notions of \emph{consistency-increasement} and \emph{consistency-maintainment}, where increasement is a special case of maintainment. A transformation $t$ is  consistency-maintaining if it does not decrease consistency in the finer-grained sense as described above, while $t$ is 
consistency-increasing if it increases the consistency.

These notions are designed to detect only transformations that maintain (or increase) the consistency of the first two unsatisfied layers of a constraint $c$. 
That means, given a graph $G$ and a constraint $c$,  a transformation $ t:G \Longrightarrow H$ is  consistency-maintaining if the largest satisfied layer has not decreased, i.e. if $\maxk{c}{G} \leq \maxk{c}{H}$, and at least as many increasing insertions or deletions have been made as decreasing ones.
An increasing deletion is the deletion of an occurrence of $C_{\maxk{c}{G} +2}$ that does not satisfy $\exists(C_{\maxk{c}{G}+3},\true)$, an increasing insertion is the insertion of elements, such that for at least one occurrence $p$ of $C_{\maxk{c}{G}+2}$ it holds that $p \not \models \exists(C',\true)$ and $\track_t \circ p \models \exists(C',\true)$ for an intermediate graph $C' \in \ig{C_{\maxk{c}{G}+2}}{ C_{\maxk{c}{G}+3}}$.
Decreasing insertions and deletions are the opposite of increasing ones. 
A decreasing insertion is the insertion of an occurrence of $C_{\maxk{c}{G}+2}$ that does not satisfy $\exists(C_{\maxk{c}{G}+3}, \true)$ and a decreasing deletion is the deletion of elements such that for one occurrence $p$ of 
$C_{\maxk{c}{G}+2}$ with $p \models \exists(C', \true)$ it holds that 
$\track_t \circ p \not \models \exists(C', \true)$ for an intermediate graph
$C'\in \ig{C_{\maxk{c}{G}+2}}{ C_{\maxk{c}{G}+3}}$.
If $\maxk{c}{G} < \maxk{c}{H}$ or the number of increasing insertions and deletions is greater than the number of decreasing ones, $t$ is consistency-increasing.

To evaluate this, we define the \emph{number of violations}.
Intuitively, for all occurrences $p$ of $C_{\kmax+2}$ the number of graphs $C' \in \ig{C_{\kmax+2}}{C_{\kmax+3}}$ with $p \not \models \exists (C', \true)$ is added up, and by comparing these numbers for $G$ and $H$ it can be determined whether there have been more increasing insertions and deletions than decreasing ones.

The number of violations is defined for each layer of the constraint, but only for the first unsatisfied layer the sum is calculated as described above.
For all layers $k$ with $k \leq \kmax$ it is set to $0$ and for all layers $k$ with $k > \kmax + 1$ it is set to $\infty$. 
In this way, a transformation $t: G \Longrightarrow H$ that increases the largest satisfied layer can be easily detected, since the number of violations in $H$ at layer $\kmax + 1$ will be set to $0$.

\begin{definition}[\textbf{number of violations}]\label{def:num_violations}
Given a graph $G$, a constraint $c$ in UANF and let $e = \scond{\kmax + 2}{c}$.
The \emph{number of violations $\nv{j}{G}$ at layer $-1 \leq j < \nlvl(c)$ in $G$} is defined as:

\begin{equation*}
	\nv{j}{G} := \begin{cases}
					0 & \text{if $j < \kmax +1$} \\
					\sum_{C' \in \ig{C_{j+1}}{C_{j+2}}} |\{q \mid q:C_{j+1} \inj 
					G \wedge q \not \models \ic{0}{e}{C'}\}| &  \text{if $e \neq 
					\false$ and $j = \kmax+1$} \\
					|\{q \mid q:C_{j+1} \inj G\}| & \text{if $e = \false$ and 
					$j = \kmax + 1$} \\
					\infty &\text{if $j > \kmax + 1$}
				 \end{cases}
\end{equation*}
\end{definition}
Note that the second and third cases of Definition \ref{def:num_violations} only occur if $G \not \models c$ and $\scond{\kmax}{c}$ is an existential condition. So $e$ is also an existential condition or equal to $\false$ if $c$ ends with $\forall(C_{\nlvl(c)}, \false)$ and $\kmax = \nlvl(c) -2$. 
Also note that the sets described above do contain occurrences of $C_{\kmax +2}$ whose removal (or repair so that they satisfy $\exists(C_{\kmax +3}, \true)$) will never lead to an increase of the largest satisfied layer. 
In particular, only the occurrences of $p: C_{\kmax +2} \inj G$, which are so-called \emph{potentially increasing occurrences at layer $\kmax$ w.r.t. $c$}, need to be considered. 

\begin{definition}[\textbf{potentially increasing occurrences at layer}]
	Given a graph $G$, a constraint $c$ in UANF and an occurrence $p: C_{k+2} \inj G$ of a universally bound graph $C_k$. 
	Then $p$ is called a \emph{potentially increasing occurrence at layer $k$ w.r.t. $c$} if 
	\begin{enumerate}
		\item $p \not \models \cut{0}{\scond{k+2}{c}}$.
		\item $p = a_{k+1} \circ \ldots \circ a_0 \circ q$ where $a_i \circ \ldots \circ q \models \scond{i+1}{\cut{k}{c}}$ for all $0 \leq i \leq k$ and $q: \emptyset \inj G$ is the empty morphism. 
	\end{enumerate}
\end{definition}

In the following lemma, we will show that only these occurrences need to be considered.
\begin{lemma}\label{num_violations}
	Given a graph $G$, a constraint $c$ in UANF and an odd $-1 \leq k < \nlvl(c)-2$ such that $G \models_k c$. 
	Then, $$G \models_{k+2} c$$ if for all occurrences $p: C_{k+2} \inj G$ of $C_{k+2}$ where  $p = a_{k+1} \circ \ldots \circ a_0 \circ q$, $a_i \circ \ldots \circ q \models \scond{i+1}{\cut{k}{c}}$ for all $0 \leq i \leq k$ and $q: \emptyset \inj G$ is the empty morphism, it holds that $p \models \cut{0}{\scond{\kmax+2}{c}}$. 
\end{lemma}
\begin{proof}
	Assume that $G \not \models_{k+2} c$ and for all occurrence $p: C_{k+2} \inj G$ where $p = a_{k+1} \circ \ldots \circ a_0 \circ q$ and $a_i \circ \ldots \circ q \models \scond{i+1}{\cut{k}{c}}$ for all $0 \leq i \leq k$ it holds that $p \models \cut{0}{\scond{\kmax+2}{c}}$.
	Since $G \models_{k} c$ and  $G \not \models_{k+2} c$ there must be a morphism $p :C_{k+2} \inj G$ such that $p \not \models \cut{0}{\scond{\kmax+2}{c}}$, $p = a_{k+1} \circ \ldots \circ q$ and $a_i \circ \ldots \circ q \models \scond{i+1}{\cut{k}{c}}$ for all $0 \leq i \leq k$. This is a contradiction.
\end{proof}
Only considering the occurrence of $C_{\kmax+2}$ as described above will lead to a more precise definition of the number of violations, and therefore to a more precise definition of consistency maintaining and increasing transformations and rules, with the drawback that the application conditions designed for this more precise version will be much more complex, since it will be necessary to check that repaired occurrences satisfy the additional condition.
We have therefore decided to use this less precise definition of the number of violations.

Using the number of violations, we are now ready to define \emph{consistency-maintaining} and \emph{consistency-increasing} transformations and rules by checking that the number of violations has not increased or, in the case of consistency-increasing, has decreased.
In addition, we will also introduce weaker notions, called \emph{consistency-maintaining rules at layer} and \emph{consistency-increasing rules at layer}.
Intuitively, a rule is consistency-maintaining or consistency-increasing w.r.t. $c$ at layer $k$ if all of its applications at graphs $G$ with $G \models_k c$ are consistency-maintaining or consistency-increasing w.r.t. $c$. 
This weaker notion will be important for our consistency-increasing application condition, since the graphs at the first unsatisfied layer must be considered, and constructing a consistency-increasing application condition will lead to huge application conditions.

\begin{definition}[\textbf{consistency maintaining and increasing transformations and rules}]
	Given a graph $G$, a constraint $c$ in UANF and a rule $\rho$.
	A transformation $t: G \Longrightarrow_{\rho,m} H$ is called \emph{consistency-maintaining w.r.t. $c$}, if  $$\nv{k}{ H} \leq \nv{k}{ G} $$
	for all $-1 \leq k < \nlvl(c)$.
	The transformation is called \emph{consistency-increasing w.r.t. $c$} 
	if it is consistency-maintaining w.r.t. $c$ and 
	$$\nv{\maxk{c}{G}+1}{ H} < \nv{\maxk{c}{G}+1}{ G}.$$
	A rule $\rho$ is called \emph{consistency maintaining or increasing  w.r.t. $c$}, 
	if all of its transformations are. 
	
	A rule $\rho$ is called \emph{consistency maintaining w.r.t. $c$ at layer $-1 \leq k < \nlvl(c)$} if all transformations $t: G \Longrightarrow_{\rho,m} H$ with $\maxk{c}{G} = k$ are consistency maintaining w.r.t. $c$. 
	Analogously, a rule $\rho$ is called \emph{consistency increasing w.r.t. $c$ at layer $-1 \leq k < \nlvl(c)$} if all transformations $t: G \Longrightarrow_{\rho,m} H$ with $\maxk{c}{G} = k$ are consistency-increasing w.r.t. $c$.
\end{definition}

Note that if $G \models c$, there is no consistency-increasing transformation $G \Longrightarrow H$ w.r.t. $c$, since $\nv{j}{G} = 0$ for all $0 \leq j < \nlvl(c)$. 
No plain rule $\rho$ is consistency-increasing w.r.t $c$, since a graph $G$ satisfying $c$ such that a transformation $t: G \Longrightarrow_{\rho,m} H$ exists can always be constructed. 
Therefore, every consistency-increasing rule must have at least one application condition.

As mentioned above, a transformation is consistency-increasing if the largest satisfied layer is increased. This property is already indirectly embedded in the definition of consistency-increasing transformations. 

\begin{theorem}
	Given a rule $\rho$, a constraint $c$ in UANF and a graph $G$ with $G \not 
	\models c$.
	A transformation $t: G \Longrightarrow_{\rho,m} H$ is 
	consistency-increasing w.r.t. $c$ if $$\maxk{c}{G} < \maxk{c}{H}.$$.	
\end{theorem}

\begin{proof}
There is no $\ell >\maxk{c}{G}$ with $G\models_{\ell} c$. So, $\nv{\maxk{c}{G} + 1}{G} > 0$. 
Since $\maxk{c}{H} > \maxk{c}{G}$, it follows that $\nv{\maxk{c}{G}+1}{H} = 0$, which immediately implies that $t$ is consistency-increasing w.r.t. $c$.
\end{proof}

Since there are no consistency-increasing transformations starting from consistent graphs, there are no infinitely long sequences of consistency-increasing transformations.

\begin{theorem}
Let $c$ be a constraint in UANF. 
Every sequence of consistency-increasing transformations w.r.t. $c$ is finite.

\end{theorem}

\begin{proof}
Let 
$$G_0 \Longrightarrow_{\rho_0,m_0} G_1 \Longrightarrow_{\rho_1,m_1} G_2 \Longrightarrow_{\rho_2,m_2} \ldots$$
be a sequence of consistency-increasing transformations w.r.t. $c$.
We assume that $\maxk{c}{G_0} < \nlvl(c)-1$, otherwise $\nv{j}{G_0} = 0$ for all $0 \leq j < \nlvl(c)$ and there is no consistency-increasing transformation $G_0 \Longrightarrow H$ with respect to $c$.

We show that $G_x \models_{\maxk{c}{G_0} + 2} c$ holds after a maximum of $x := \nv{\maxk{c}{G_0}+1}{G_0}$ transformations. 
Note that $x$ must be finite, since $G_0$ contains only a finite number of occurrences of $C_{\maxk{c}{G_0}+2}$. Since every transformation is consistency-increasing 
w.r.t. $C$, it follows that $\nv{\maxk{c}{G_{i}}+1}{G_{i+1}} \leq \nv{\maxk{c}{G_i}+1}{G_{i}}-1$ after each transformation.
Therefore, after at most $x$ transformations, $\nv{\maxk{c}{G_0}+1}{G_{j}} \leq \nv{\maxk{c}{G_0}+1}{G_{0}}-x = 0$ and thus $G_x \models_{\maxk{c}{G_0}+ 2} c$.
If this is applied iteratively, it follows that after a finite number of transformations, there must exist a graph $G_k$ with $G_k \models c$. 
Since there is no consistency increasing transformation $G_k \Longrightarrow_{\rho_k, m_k} G_{k+1}$, the sequence must be finite.
\end{proof}

\subsection{Direct Consistency Maintaining and Increasing Transformations}
We will now introduce stricter versions of consistency-increasing and consistency-main\-taining transformations, called \emph{direct consistency-maintaining} and \emph{direct consistency-increasing} transformations and rules.
These are consistency-maintaining and consistency-increasing  transformations which do not perform any unnecessary insertions and deletions.
For example, given a constraint $c$ in UANF and graphs $G$ with $G \not \models c$ and $H$ with $H \models c$, the transformation $t:G \Longrightarrow_{\rho, \id_G} H$ via the rule $\rho = \rle{G}{l}{\emptyset}{r}{H}$ is a consistency-increasing transformation. 
Therefore, the notions of consistency-increasing and consistency-maintaining transformations  allow insertions or deletions that are unnecessary in order to increase or maintain consistency. 
That is, deleting occurrences of existentially bound graphs, deleting occurrences $p: C_k \inj G$ of universally bound graphs $C_k$ satisfying $\exists(C_{k+1}, \true)$ or inserting occurrences of universally bound graphs and inserting occurrences $p$ of intermediate graphs $C'\in \ig{C_{k-1}}{C_{k}}$ such that each occurrence $q$ of $C_{k-1}$ with $q = p \circ a^r_{k-1}$ already satisfies $\exists(C', \true)$.

\emph{Direct consistency-increasing} and \emph{direct consistency-maintaining} transformations are more restricted, in the sense that these unnecessary deletions and insertions cause a transformation not to be direct consistency-increasing or direct consistency-maintaining, respectively. 
In addition, we can use second-order logic formulas to characterise these transformations. 
Furthermore, these formulas ensure that no new violations are inserted.
Thus, the removal of one violation is sufficient to state that the transformation is (direct) consistency-increasing, which can also be expressed using a second-order logic formula.
We start by introducing \emph{direct consistency-maintaining} transformations, rules and the weaker notion of \emph{direct consistency-maintaining rules at layer}. The definition of direct consistency maintaining transformations consists of the following formulas:
\begin{enumerate}
	\item 
		\emph{No new violation by deletion:} 
			This condition ensures that the consistency is not reduced 
by deleting intermediate graphs $C' \in \ig{C_{\kmax +2}}{C_{\kmax +3}}$. This leads to the insertion of new violations only if an occurrence of $C_{\kmax +2}$ which satisfies $\exists(C',\true)$ in the originating graph does not satisfy $\exists(C',\true)$ in the derived graph of the transformation. 
Therefore, this condition checks that this case does not occur.
			 
	\item	
		\emph{No new violation by insertion:}
		This condition ensures that the consistency is not decreased by inserting an occurrence of $C_{\kmax +2}$. Again, this will only cause a new violation if the new occurrence does not satisfy $\exists(C_{\kmax +3}, \true)$. The condition checks that this is not the case.
	\item
		\emph{No satisfied layer reduction by insertion:}
			This condition ensures that the largest satisfied layer is not reduced by inserting a universally bound graph $C_j$. This can only happen if $j \leq \kmax$, and the condition checks that no occurrences of such universally bound graphs are inserted.
			 
	\item
		\emph{No satisfied layer reduction by deletion:}
			This condition ensures that the largest satisfied layer is not reduced by deleting an existentially bound graph $C_j$. Again, this can only happen if $j \leq \kmax$. The condition checks that no occurrences of such existentially bound graphs are deleted.
\end{enumerate}

The \emph{no new violation by deletion} and \emph{no new violation by insertion} formulas ensure that the number of violations is not increased, and the \emph{no satisfied layer reduction by insertion} and \emph{no satisfied layer reduction deletion} formulas ensure that the largest satisfied layer is not reduced.
Of course, the insertion of universal and deletion of existential graphs  does not necessarily lead to a decrease of the largest satisfied layer, but it can also be considered as an unnecessary insertion or deletion. 

Since a condition, $c$ in UANF is also allowed to end with $\forall(C_{\nlvl(c)},\false)$, the \emph{no new violation by deletion} and \emph{no new violation by insertion} formulas contain case discrimination.  
If the constraint $c$ ends with $\forall(C_{\nlvl(c)},\false)$ and $\kmax = \nlvl(c) -2$, there is no graph $C_{\kmax +3}$ and thus no intermediate graphs. Therefore, there is no new violation by deletion, this formula is set equal to $\true$ and the \emph{no new violation by insertion} formula will check that no new occurrence of $C_{\kmax +2}$ are introduced at all. 

For the rest of this thesis, we will assume that the empty conjunction is always equal to $\true$.

\begin{definition}[\textbf{direct consistency maintaining transformations and rules}]
	Given a graph $G$, a rule $\rho$ and a constraint $c$ in UANF.
	If $G \models c$, a transformation $t: G\Longrightarrow_{\rho,m}H$ is
	called \emph{direct consistency-maintaining w.r.t. $c$} if $H \models c$.
	Otherwise, if $G \not \models c$, let $\kmax = \maxk{c}{G}$ and $e = \scond{\kmax +2}{c}$ .
	A transformation $t: G\Longrightarrow_{\rho,m}H$ is called \emph{direct
	consistency maintaining w.r.t. $c$} if the following formulas are 
	satisfied.
	
	\begin{enumerate}
		\item
			\emph{No new violation by deletion:}
			If $e \neq \false$, then each occurrence of $C_{\kmax+2}$ in $G$ 
			which satisfies $\ic{0}{e}{C'}$ for any 	$C' \in \ig{C_{\kmax+2}}
			{C_{\kmax+3}}$ still satisfies $\ic{0}{e}{C'}$ in $H$:
			\begin{equation}\label{direct_improving_1}
				\begin{split}
					\forall p: C_{\kmax +2} \inj G\Big( \bigwedge_{C' \in \ig{C_{\kmax+2}}
					{C_{\kmax+3}}}
					\big( &p \models \ic{0}{e}{C'} \wedge \track_t \circ p \text{ 	
					is total}\big)\\&\implies  \track_t \circ p \models \ic{0}{e}
					{C'} \Big) 
				\end{split}
			\end{equation}
			Otherwise, if $e = \false$, this formula is equal to $\true$.
		\item
			\emph{No satisfied layer reduction by insertion:}
			Let $d = \ic{0}{e}{C_{\kmax+3}}$ if $e \neq \false$ and $d = \false$ 		
			otherwise. Each newly inserted occurrence of $C_{\kmax+2}$ satisfies 
			$d$.
			\begin{equation}\label{direct_improving_1_2}
				\begin{split}
					\forall p': C_{\kmax+2} \inj H \big(\neg \exists p : C_{\kmax+2} \inj G(p' 
					= \track_t \circ p) \implies \ p' \models d\big)
				\end{split}
			\end{equation}
		\item
		\emph{No satisfied layer reduction by insertion:}
			No occurrence of a universally bound graph $C_j$ with $j \leq \kmax$ 
			is inserted. 
			\begin{equation}\label{direct_improving_3}
				\bigwedge_{\substack{i < \kmax \\ C_i \textit{ universal}}} 
				\forall p: 
				C_i \inj H ( \exists p': C_i \inj G (p = \track_t \circ p'))
			\end{equation}
		\item
			\emph{No satisfied layer reduction by deletion:}
			No occurrence of an existentially bound graph $C_j$ with $j \leq 
			\kmax+1$ is deleted. 
			\begin{equation}\label{direct_improving_4}
				\bigwedge_{\substack{i \leq \kmax \\ C_i \textit{ 
				existential}}} \forall 
				p: C_i \inj G( \track_t \circ p \textit{ is total})
			\end{equation}
	\end{enumerate}
	A rule $\rho$ is called \emph{direct consistency-maintaining w.r.t. $c$} if all of its transformations are. 
	A rule $\rho$ is called \emph{direct consistency maintaining w.r.t. $c$ at layer $-1 \leq k < \nlvl(c)$} if all 
	transformations $t: G \Longrightarrow_{\rho,m}$ with $\maxk{c}{G} = k$ are direct consistency-maintaining w.r.t. $c$.
\end{definition}
 
Before continuing with the definition of direct consistency-increasing transformations and rules, let us first show that every direct consistency-maintaining transformation is indeed consistency-maintaining.  To do this, we first show that satisfying the \emph{no satisfied layer reduction by insertion} and \emph{no satisfied layer reduction by insertion} formulas guarantees that the largest satisfied layer is not decreased.

\begin{lemma}\label{lemma_consistent}
	Given a transformation $t: G \Longrightarrow H$ and a constraint $c$ in UANF 
	such that the \emph{no satisfied layer reduction by insertion} and \emph{no satisfied layer reduction by insertion} formulas are satisfied. Then
	$$H \models_{\maxk{c}{G}} c.$$
\end{lemma}
\begin{proof}
Let us assume that $H \not \models_{\maxk{c}{G}}c$. 
Then either a new occurrence of a universally bound graph $C_i$ with $i < \maxk{c}{G}$ has been inserted, or an occurrence of an existentially bound graph $C_j$ with $j \leq \maxk{c}{G}$ has been destroyed.
Therefore, the following applies:
$$ \exists p: C_i \inj H(\neg \exists p': C_i \inj G(p = \track_t \circ p')) \vee \exists p:C_j \inj G(\track_t \circ p \textit{ is not total})$$
where $i, j \leq \maxk{c}{G}$, $i$ is even and $j$ is odd, i.e. $C_i$ is universally and $C_j$ is existentially bound.
It follows immediately that either the \emph{no satisfied layer reduction by insertion} and \emph{no satisfied layer reduction by insertion} formula is not satisfied. 
This is a contradiction. 
\end{proof}

With this, we are now going to show that a direct consistency-maintaining transformation is also a consistency-maintaining transformation.

\begin{theorem}\label{thm:direct_maintaining_to_maintaining}
	Given a graph $G$, a constraint $c$ in UANF, a rule $\rho$ and a direct consistency-maintaining transformation $t: G \Longrightarrow_{\rho,m} H$ w.r.t. $c$. 
Then, $t$ is also a consistency-maintaining transformation.
\end{theorem}
\begin{proof}
	Lemma \ref{lemma_consistent} implies  that $\maxk{c}{G} \leq \maxk{c}{H}$ 
	and it immediately follows that $\nv{\maxk{c}{G}+1}{H} \neq \infty$.
	It remains to show that $\nv{k}{H} \leq \nv{k}{G}$ for all $0 \leq k < 
	\nlvl(c)$.
	In particular, we only need to show that $\nv{\maxk{c}{G}+1}{H} \leq 
	\nv{\maxk{c}{G}+1}{G}$ since for all $-1 \leq j < \maxk{c}{G}+1$ it holds 
	that $\nv{j}{H} = \nv{j}{G} = 0$. And since $\nv{j}{G} = \infty$ for all $\maxk{c}{G} + 1 < j < \nlvl(c)$, it follows that $\nv{j}{H} \leq \nv{j}{G}$ for all $\maxk{c}{G} + 1 < j < \nlvl(c)$.
	
	Let  $\kmax = \maxk{c}{G}$ and $d = \scond{\kmax +2}{c}$.
	We show that the satisfaction of the \emph{no new violation by deletion} and \emph{no new violation by insertion} formulas imply 
	that $\nv{\kmax +1}{H} \leq \nv{\kmax +1}{G}$. 
	
	Let us assume that $\nv{\kmax +1}{H} > \nv{\kmax +1}{G}$. 
	Therefore, there is a morphism $p: C_{\kmax +2} \inj H$ 
	with $p \not \models \ic{0}{d}{C'}$ for some $C' \in \ig{C_{\kmax +2}}
	{C_{\kmax +3}}$ 
	such that either \ref{proof_direct_minimal_1}. or 
	\ref{proof_direct_minimal_2}. below is satisfied. Note that this is only the case if 
	$d \neq \false$. Otherwise, there must be a morphism $p$ which satisfies 
	\ref{proof_direct_minimal_2}.
	\begin{enumerate}
		\item \label{proof_direct_minimal_1}
			There is a morphism $q': C_{\kmax +2} \inj G$ with $q' \models \ic{0}
			{d}{C'}$ and $p = \track_t \circ q'$. 

		\item \label{proof_direct_minimal_2}
			There is no morphism $q : C_{\kmax +2} \inj G$ with $p = \track_t 
			\circ q$. 		
	\end{enumerate}
	This is a contradiction if \ref{proof_direct_minimal_1}. is satisfied, $q'$ 
	does not satisfy the \emph{no new violation by deletion} formula. If 
	\ref{proof_direct_minimal_2}. is satisfied, $p$ does not satisfy the
	\emph{no new violation by insertion} formula since $p$ only satisfies  $\ic{0}{d}{C_{k+2}}$ 
	if $p$ satisfies $\ic{0}{d}{C'}$ for all $C' \in \ig{C_{k+1}}{C_{k+2}}$.
	It follows that 
	$$\nv{k}{H} \leq \nv{k}{G}$$ holds and $t$ is a consistency-maintaining transformation.
	
\end{proof}
The following corollary arises as a direct consequence of Theorem \ref{thm:direct_maintaining_to_maintaining}.
\begin{corollary}
	Given a constraint $c$ in UANF and a rule $\rho$. 
	If $\rho$ is a direct consistency-maintaining rule w.r.t. $c$, the $\rho$ is also a consistency-maintaining rule w.r.t. $c$.
	If $\rho$ is a direct consistency-maintaining rule w.r.t. $c$ at layer $-1 \leq k \leq \nlvl(c)$, then $\rho$ is also a consistency-maintaining rule w.r.t. $c$ at layer $k$.
\end{corollary}
Let us now introduce the notions of \emph{direct consistency-increasing} transformations, rules and  \emph{direct consistency-increasing rules at layer}. Similar to the definition of con\-sis\-ten\-cy-maintaining and consistency-increasing transformations, the notion of \emph{direct consistency-increasing transformations} is based on the notion of direct consistency-maintaining transformations, in the sense that a direct consistency-increasing transformation is also a direct consistency-maintaining one. Since a direct consistency-maintaining transformation $t$ does not introduce any new violations, it is sufficient that $t$ removes at least one violation to say that $t$ is direct consistency-increasing.

Again, we need case discrimination if the constraint ends with $\forall(C_{\nlvl(c)}, \false)$ and $\kmax = \nlvl(c) -2$. 
So we will use two second-order logic formulas, one for the general case and one for this special case.

\begin{enumerate}
	\item	
		\emph{General increasing formula}:
			This formula is satisfied if either an occurrence of $C_{\kmax +2}$ that does not satisfy $\exists(C_{\kmax+3}, \true)$ is deleted, or an occurrence of $C_{\kmax +2}$ which does not satisfy $\exists(C', \true)$ in the first graph of the transformation satisfies $\exists(C', \true)$ in the second graph of the transformation where
$C' \in \ig{C_{\kmax +2}}{C_{\kmax +3}}$. Both cases result in the removal of a violation.
	\item 
		\emph{Special increasing formula}:
		This formula is satisfied if an occurrence of $C_{\kmax+2}$ is 
		removed. 
		In the special case, this is the only way to remove a violation.
\end{enumerate}

\begin{definition}[\textbf{direct consistency-increasing transformations and rules}]\label{def_direct_improving}
	
	Given a constraint $c$ in UANF, a rule $\rho$, a graph $G$ with 
	$G \not \models c$ and let $e = \scond{\kmax +2}{c}$.  
	
	A transformation $t: G \Longrightarrow_{\rho,m} H$ is called \emph{direct consistency-increasing w.r.t. $c$} if it is direct consistency-maintaining w.r.t. $c$ and either the special increasing condition is satisfied if  $\scond{\nlvl(c)-1}{c} = \forall(C_{\nlvl(c)},\false)$ and $\kmax = \nlvl(c) -2$ or the general increasing condition is satisfied otherwise.
	
	\begin{enumerate}
		\item
			General increasing formula:
			\begin{equation}\label{direct_improving_2}
			\begin{split}
				\exists p: C_{\kmax+2} \inj G&\Big(\bigvee_{C' \in \ig{\kmax+2}{\kmax+3}}\big( p \not \models \ic{0}{e}{C'} \wedge 
		\\&(\track_t \circ p \text{ is not total } \vee \track_t \circ p  \models 
		\ic{0}{e}{C'})\big)\Big)
			\end{split}
			\end{equation}
		\item 
			Special increasing formula:
			\begin{equation}
				\exists p: C_{\kmax+2} \inj G(\track_t \circ p \text{ is not total})
			\end{equation}
	\end{enumerate}
	A rule $\rho$ is called \emph{direct consistency-increasing w.r.t. $c$} if all of its transformations are. 
	A rule $\rho$ is called \emph{direct consistency-increasing w.r.t. $c$ at layer $-1 \leq k < \nlvl(c)$} if all transformations 
	$t: G \Longrightarrow_{\rho,m} H$ with $\maxk{c}{G} = k$ are direct consistency-increasing w.r.t. $c$.
\end{definition}

Note that the satisfaction of the \emph{no satisfied layer reduction by insertion} and \emph{no satisfied layer reduction by deletion} formulas  not only ensure that the largest satisfied layer does not decrease, as shown in Lemma \ref{lemma_consistent}, but also prevent further unnecessary insertions and deletions, since inserting a universally bound graph and deleting an existentially bound graph will never lead to an increase in consistency.

Now, we will show the already indicated relation between direct consistency-increasing and consistency-increasing transformations, namely that a direct consistency-increasing transformation is also a consistency-increasing transformation.
Counterexamples in which the inversion of the implication does not hold can be easily constructed to show that these notions are not identical but related. 

\begin{theorem}\label{lemma:direct_implies_normal}
	Given a constraint $c$ in UANF, a rule $\rho$, a graph $G$ with 
	$G \not \models c$ and a direct consistency-increasing transformation
	$t: G \Longrightarrow_{\rho,m} H$ w.r.t. $c$. 
	Then, $t$ is also a consistency-increasing transformation. 
\end{theorem}

\begin{proof}
	Theorem \ref{thm:direct_maintaining_to_maintaining} implies that 
	$t$ is a consistency-maintaining transformation. Therefore, it is 
	sufficient to show that $\nv{\maxk{c}{G} + 1}{H} < \nv{\maxk{c}{G} + 1}{G}$.
	Let $\kmax = \maxk{c}{G}$ and $d = \scond{\kmax+2}{c}$ with $d \neq \false$.

	Then, the general increasing formula is satisfied, so there exists an intermediate graph $C' \in \ig{C_{\kmax+2}}{C_{\kmax+3}}$ and a morphism $p:C_{\kmax+2} \inj G$ with $p \not \models \ic{0}{d}{C'}$, such that either $\track \circ p$ is total and $\track_t \circ p  \models \ic{0}{d}{C'}$ or $\track \circ p$ is not total.
	In both cases, the following applies: 
	\begin{equation*}
		\begin{split}
			p &\in \{q \mid q:C_{\kmax+2} \inj G \text{ and } q \not \models \ic{0}{d}{C'}\} 
			\text{ and} \\
			\track \circ p &\notin \{q \mid q:C_{\kmax+2} \inj H \text{ and } q \not \models 
			\ic{0}{d}{C'}\}
		\end{split}
	\end{equation*}
	Since $t$ is direct consistency maintaining, it follows that 
	$$|\{q \mid q:C_{\kmax+2} \inj G \text{ and } q \not \models \ic{0}{d}{C}\}| \leq |\{q 
	\mid q:C_{\kmax+2} \inj H \text{ and } q \not \models \ic{0}{d}{C}\}|.$$
	for all $C \in \ig{C_{\kmax+2}}{C_{\kmax+3}}$. Furthermore, this inequality 
	is strictly satisfied if $C = C'$.
	It immediately follows that $\nv{k}{H} < \nv{k}{G}$ and $t$ is a consistency-
	increasing transformation.

	If $d = \false$, i.e. $\scond{\nlvl(c)-1}{c} = \forall(C_{\nlvl(c)},\true)$ and $\kmax = \nlvl(c)-2$, the special increasing formula is satisfied.
	It holds that $$|\{q \mid q:C_k \inj G\}| \leq |\{q \mid q:C_k \inj H\}|,$$ and since $t$ is a direct consistency-maintaining transformation, it can be shown in a similar way as above that satisfying the special increasing formula implies that  
	$$|\{q \mid q:C_k \inj G\}| < |\{q \mid q:C_k \inj H\}|.$$
	It follows that $t$ is a consistency-increasing transformation.
\end{proof}

Again, the following corollary is a direct consequence of Theorem \ref{lemma:direct_implies_normal}. 
\begin{corollary}
	Given a constraint $c$ in UANF and a rule $\rho$. 
	If $\rho$ is a direct consistency-increasing rule w.r.t. $c$, then $\rho$ is also a consistency-increasing rule w.r.t. $c$.
	If $\rho$ is a direct consistency-increasing rule w.r.t. $c$ at layer $-1 \leq k \leq \nlvl(c)$, then $\rho$ is also a consistency-increasing rule w.r.t. $c$ at layer $k$.
\end{corollary}
\begin{figure}
\centering

\includegraphics[scale=1.0]{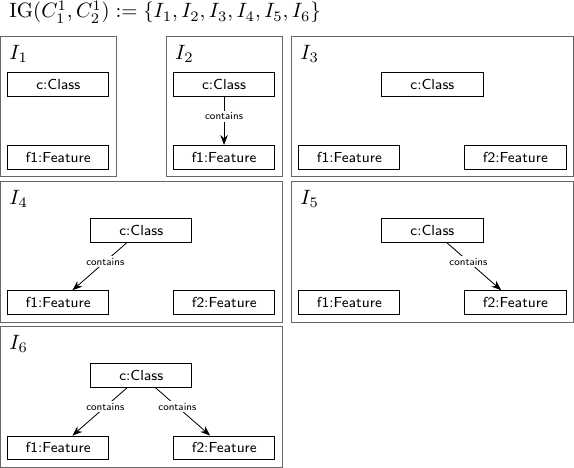}

\vspace{0.3cm}
\includegraphics[scale=0.8]{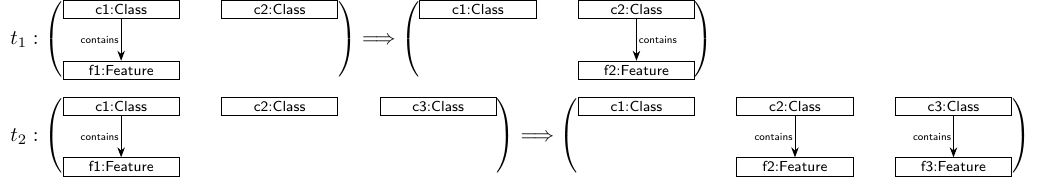}

\caption{Example of the set of intermediate graph. Transformation $t_1$ is consistency-maintaining and not direct consistency-maintaining. Transformation $t_2$ is consistency-increasing and not direct consistency-increasing.}\label{fig:example_direct}
\end{figure}

\begin{example} \label{ex_direct}
	Consider constraint $c_1$ given in Figure \ref{fig:constraints}, 
	the transformations $t_1$, $t_2$ and the set $\ig{C_1^1}{C_2^1}$ given 
	in Figure \ref{fig:example_direct}.  
 	Then, $t_1$ is a consistency-maintaining transformation w.r.t. $c_1$.
	The number of violations in both graphs is  $9$. 
	In the first graph, the occurrence \emph{\texttt{c1}} does not satisfy $\exists(I_3, \true)$, $\exists(I_4, \true)$, $\exists(I_5, \true)$ and $\exists(I_6, \true)$, the occurrence \emph{\texttt{c2}} does not satisfy $\exists(I_2, \true)$, $\exists(I_3, \true)$, 
	$\exists(I_4, \true)$, $\exists(I_5, \true)$ and $\exists(I_6, \true)$.
	In the second graph, these roles are swapped, i.e. \emph{\texttt{c1}} satisfies exactly the intermediate conditions that \emph{\texttt{c2}} satisfied in the first graph, and vice versa.
	But, $t_1$ is not a direct consistency-maintaining transformation, since the occurrence \emph{\texttt{c1}} satisfies $\exists(I_2,\true)$ in the first but not in the second graph. Therefore, the \emph{no new violation by deletion} formula is not satisfied.
	
	The transformation $t_2$ is consistency increasing w.r.t. $c_1$. The number of violations in the first graph is equal to $14$. The occurrence \emph{\texttt{c1}} does not satisfy $\exists(I_3, \true)$, $\exists(I_4, \true)$, $\exists(I_5, \true)$ and $\exists(I_6, \true)$. Both occurrences \emph{\texttt{c2}} and \emph{\texttt{c3}} do not satisfy $\exists(I_2, \true)$, $\exists(I_3, \true)$, 
	$\exists(I_4, \true)$, $\exists(I_5, \true)$ and $\exists(I_6, \true)$.
	In the second graph, \emph{\texttt{c1}} does not satisfy $\exists(I_2, \true)$, $\exists(I_4, \true)$, $\exists(I_5, \true)$ and $\exists(I_6, \true)$ and both \emph{\texttt{c2}} and \emph{\texttt{c3}} do not satisfy $\exists(I_6,\true)$. Therefore, the number of violations in the second graph is $6$.
But $t_2$ is not a direct consistency increasing transformation, since \emph{\texttt{c1}} satisfies $\exists(I_3,\true)$ in the first but not in the second graph, and the \emph{no new violation by deletion} formula is not satisfied.
\end{example}
\subsection{Comparison with other concepts of Consistency}\label{comp_general}
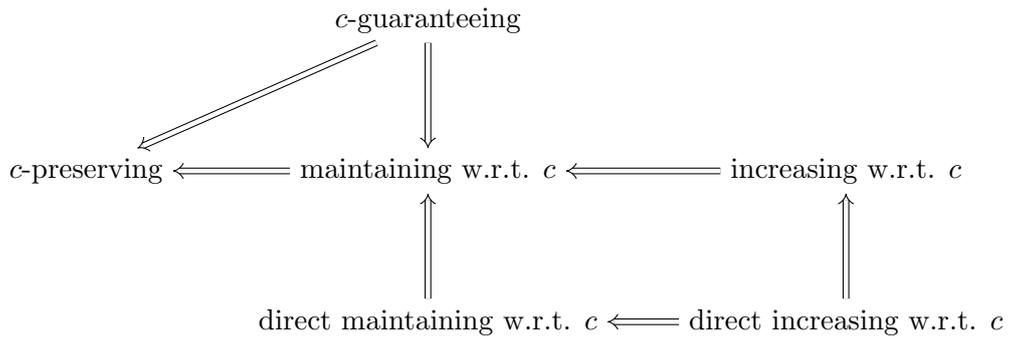
\begin{figure}
\center
	\begin{tikzpicture}

		\node(guaran) at (4.5,2) {$c$-guaranteeing};
		\node(pres) at (0,0) {$c$-preserving};
		\node (incr) at (10,0) {increasing w.r.t. $c$};
		\node (direct) at (10,-2) {direct increasing w.r.t. $c$};
		\node (maintaining) at (4.5,0) {maintaining w.r.t. $c$};
		\node (dir_maintaining) at (4.5,-2) {direct maintaining w.r.t. $c$};

		\draw[-implies,double equal sign distance] (guaran) --  (maintaining); 
		\draw[-implies,double equal sign distance] (guaran) --  (pres);
		\draw[-implies,double equal sign distance] (incr) --  (maintaining);
		\draw[-implies,double equal sign distance] (maintaining) --  (pres);
		\draw[-implies,double equal sign distance] (direct) --  (incr);   
		\draw[-implies,double equal sign distance] (dir_maintaining) --  (maintaining);
		\draw[-implies,double equal sign distance] (direct) --  (dir_maintaining); 
		    
	\end{tikzpicture}
	\caption{Relations of consistency notions.}\label{fig:consistency_relations}
\end{figure}

In this chapter, the notions of (direct) consistency increase and maintainment are compared to the already known notions of consistency-guaranteeing, consistency-preserving \cite{habel2009correctness}, (direct) consistency-increasing and sustaining \cite{kosiol2022sustaining}, in order to reveal relations between them and to ensure that (direct) consistency-increase and 
maintainment are indeed new notions of consistency. 
These relationships are summarised in Figure \ref{fig:consistency_relations}.

First, we compare (direct) consistency increase and maintainment with the notions of consistency-guaranteeing, preserving, sustaining and improving in the general case and later on, for some  special cases.
We begin by examining the implications that can be drawn about a consistency-maintaining or consistency-increasing transformation.

\begin{theorem}[\textbf{Implications of a consistency-maintaining or consistency-increasing transformation}]
	Given a condition $c$ in UANF and a transformation $t: G \Longrightarrow H$.
	Then,
	\begin{alignat*}{3}
			&t \text{ is consistency-maintaining w.r.t. $c$} && \implies \text{$t$ is 
			$c$-preserving} & \text{ and}\\
			&t \text{ is consistency-maintaining w.r.t. $c$} &&\centernot \implies t \text{ 
			is $c$-guaranteeing} &\text{ and} \\
		&t \text{ is direct consistency-increasing w.r.t. $c$ } &&\centernot \implies t \text{ is consistency sustaining w.r.t. $c$ }
	\end{alignat*}
\end{theorem}

\begin{figure}
\centering
\includegraphics[scale=1]{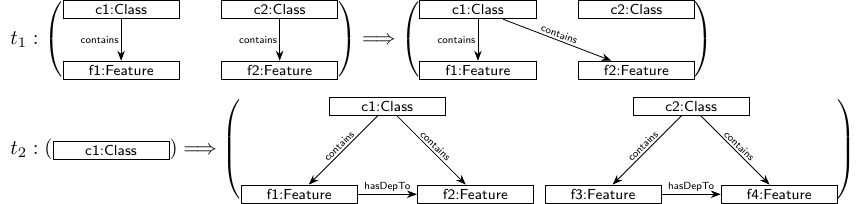}
\caption{Transformations for the comparison of consistency concepts.}\label{fig:counterexamples_increase_improving}
\end{figure}

\begin{proof}
	\begin{enumerate}
		\item 
		$t$ is consistency-maintaining w.r.t. $c$ $\implies$ $t$ is $c$-preserving: Let $t$ be a consistency-maintaining transformation w.r.t. $c$. 
			If $G \not \models c$, then $t$ is a $c$-preserving transformation.
			If $G \models c$, then $\nv{j}{G} = 0$ for all 
			$0 \leq j < \nlvl(c)$. Since $t$ is consistency maintaining
			it follows that $\nv{j}{H} = 0$ for all 
			$0 \leq j < \nlvl(c)$ and hence $H \models c$. 
			It follows that $t$ is a $c$-preserving transformation.
		
		\item
		$t$ is consistency maintaining w.r.t. $c$ $\centernot \implies$ $t$ is $c$-guaranteeing:
		Consider the transformation $t_2: G \Longrightarrow H$ shown in Figure 
			\ref{fig:example_direct} and constraint
			$c_1$ shown in Figure \ref{fig:constraints}.
			As discussed in Example \ref{ex_direct}, $t_2$ is consistency-increasing and thus consistency-maintaining w.r.t. $c_1$. 
			But $t_2$ is not a $c$-guaranteeing transformation, since all 
			occurrences of nodes of type \texttt{Class} do not satisfy 
			$\exists(C_2^1, \true)$.
		\item 
		$t$ is direct consistency maintaining w.r.t. $c$ $\centernot \implies$ $t$ is consistency sustaining w.r.t. $c$:
		Consider the constraint $c = \forall (C_1^1, \exists (C_2^1, \forall (C_4^2,d)))$, where $d$ is an existentially bound constraint in ANF with $d \neq  \false$ 	composed of the graphs given in Figure \ref{fig:constraints}. And consider  transformation $t_2$ given in Figure \ref{fig:counterexamples_increase_improving}. Then, $t$ is direct consistency increasing; the \emph{no new violation by deletion}, \emph{no new violation by insertion}, \emph{no satisfied layer reduction by insertion} and \emph{no satisfied layer reduction by deletion}  formulas are satisfied and the  \emph{general increasing} formula is  satisfied because an occurrence of $C_1^1$ that did not satisfy $\exists (C_2^1,\true)$ in $G$ satisfies $\exists (C_2^1, \true)$ in $H$.  But this transformation is not consistency-sustaining since the number of occurrences of $C_1^1$ not satisfying $\exists (C_2^1, \forall (C_4^2,d))$ in $H$ is greater than the number of occurrences of $C_1^1$ in $G$ not satisfying $\exists (C_2^1, \forall (C_4^2,d))$.
	\end{enumerate} \qedhere
\end{proof}

These results are not surprising, since consistency-maintaining and con\-sis\-ten\-cy-in\-crea\-sing are much stricter notions than guaranteeing and sustaining, in the sense that the notion of violation is more fine-grained.
For example, for guaranteeing transformations, an arbitrary number of violations can be introduced as long as the derived graph satisfies the constraint, and thus guaranteeing does not imply direct increasing, since a direct increasing transformation is not allowed to introduce new violations.
Let us now examine whether a concept of consistency implies the notions of 
consistency-maintaining and increasing.

\begin{theorem}[\textbf{Implications of preserving, guaranteeing, sustaining and improving transformations}]
 	Given a condition $c$ in UANF and a transformation $t:G \Longrightarrow H$.
 	Then,
 	\begin{alignat*}{3}
		&t \text{ is $c$-guaranteeing} &&\implies t \text{ is consistency-maintaining w.r.t. 
		$c$} &\text{ and} \\
		&t \text{ is $c$-guaranteeing} &&\centernot \implies t \text{ is consistency-increasing w.r.t. 
		$c$} &\text{ and} \\
		&t \text{ is $c$-preserving} &&\centernot \implies t \text{ is 
		consistency-maintaining w.r.t. $c$} &\text{ and} \\
		&t \text{ is direct consistency improving w.r.t $c$ } &&\centernot \implies t \text{ is consistency-maintaining w.r.t. $c$ } 
	\end{alignat*}
\end{theorem}
\begin{proof}
	\begin{enumerate}
		\item
		$t$ is $c$-guaranteeing $\implies$ $t$ is consistency-maintaining w.r.t. $c$:
		Let $t$ be a $c$-guaranteeing transformation. Then, $t$ is also a 
		consistency-maintaining transformation w.r.t. $c$ since 
		$H \models c$ and therefore $\nv{j}{H} = 0$ for all $-1 \leq j < \nlvl(c)$.
		\item
		$t$ is $c$-guaranteeing $\centernot \implies$ $t$ is consistency-increasing w.r.t. $c$:
		Assume that $G \models c$ and $H \models c$. Then, $t$ is a $c$-guaranteeing transformation, but not a consistency-increasing one.
		\item
		$t$ is $c$-preserving $\centernot \implies$ $t$ is consistency-maintaining w.r.t. $c$:
		Consider graphs $C_1^1$, $C_2^2$ and constraint $c_1$ given in 
			Figure \ref{fig:constraints}. The transformation
			$t: C_2^2 \Longrightarrow C_1^1$ is $c_1$-preserving, since 
			$C_2^2 \not \models c_1$, but not consistency maintaining
			w.r.t. $c_1$ since $\nvio{0}{c_1}{C_2^2} = 4$ and $\nvio{0}{c_1}{C_1^1} = 6$.
		\item
		$t$ is direct consistency-improving w.r.t. $c$ $\centernot \implies$ $t$ is consistency-maintaining w.r.t. $c$:
		Consider transformation $t_1$ given in Figure 
			\ref{fig:counterexamples_increase_improving} and constraint
			$c_1$ given in Figure \ref{fig:constraints}. The transformation
			$t_1$ is direct consistency-improving since no occurrence of $C_1^1$ is inserted, no occurrence of $C_1^1$ satisfying $\exists(C_2^1, \true)$ is deleted, and one occurrence of $C_1$ satisfies $\exists(C_2^1, \true)$. But, this 
			transformation is not consistency-maintaining w.r.t. $c$ since the 
			number of violations in the first graph is $2$ and the number of 
			violations in the second graph is $3$.
	\end{enumerate}
\end{proof}
This shows, that in general the notions of (direct) consistency increase and maintainment are not related to (direct) consistency improvement and sustainment.
We have shown only some of these relationships. Since (direct) consistency improvement (direct) consistency sustainment, consistency-preserving and consistency-guaranteeing are related, we can conclude results for all pairs of consistency types \cite{kosiol2022sustaining}.
An overview of these is given in Table \ref{table_cons_relations}.

For some special cases, we can infer other types of relationships.

\begin{theorem}[\textbf{Relations of consistency concepts in special cases}]
	Given a constraint $c$ in UANF and a transformation $t: G \Longrightarrow H$.
	\begin{enumerate}
	\item 
		If $G \not \models c$, then 
		$$ \text{$t$ is $c$-guaranteeing $\implies$ $t$ is consistency-increasing w.r.t. $c$.}$$
	\item If $\nlvl(c) =1$, then
	$$t \text{ is consistency improving w.r.t $c$ } \iff t \text{ is consistency increasing w.r.t $c$}$$ 
		
	\end{enumerate}

\end{theorem}

\begin{proof}
	\begin{enumerate}
		\item Let $t$ be a $c$-guaranteeing transformation with $G \not \models c$. Then, $t$ is a con\-sis\-ten\-cy-increasing transformation w.r.t. $c$ since 
		$0 < \nv{\maxk{c}{G}+1}{G} < \infty$ and $\nv{j}{H} = 0$ for all 
		$-1 \leq j \leq \nlvl(c)$.
		\item Let $\nlvl(c) = 1$. 
		Since $c$ is in UANF, $\scond{1}{c} = \false$ and $\nv{0}{G}$ is the number of occurrences of $C$ in $G$. 
This is exactly the definition of the number of violations for consistency-improving transformations, and the statement immediately follows.
	\end{enumerate}
\end{proof}
\begin{table}
\centering
\begin{tabular}{c|c|c|c|c|c|c|c|c|c|c}

$\implies$ &(1) & (2) & (3) &(4) & (5) & (6) & (7) & (8) & (9) & (10) \\

\hline
maintaining (1) & \cmark & \xmark & \xmark & \xmark & \xmark &\xmark & \xmark &\xmark & \xmark &\cmark \\
increasing (2)& \cmark & \cmark & \xmark & \xmark & \xmark &\xmark & \xmark &\xmark & \xmark &\cmark\\
direct maintaining(3)& \cmark & \xmark & \cmark & \xmark & \xmark &\xmark & \xmark &\xmark & \xmark &\cmark \\
direct increasing (4) & \cmark & \cmark & \cmark & \cmark & \xmark &\xmark & \xmark &\xmark & \xmark &\cmark \\
improving (5) & \xmark & \xmark & \xmark & \xmark & \cmark * &\cmark * & \xmark * &\xmark * & \xmark *&\cmark * \\
sustaining(6)  & \xmark & \xmark & \xmark & \xmark & \xmark * &\cmark * & \xmark * &\xmark * & \xmark *&\cmark *\\
direct improving (7) & \xmark & \xmark & \xmark & \xmark & \cmark * &\cmark * & \cmark * &\cmark * & \xmark *&\cmark *\\
direct sustaining (8) & \xmark & \xmark & \xmark & \xmark & \xmark * &\cmark * & \xmark * &\cmark * & \xmark *&\cmark *\\
guaranteeing(9)& \cmark & \xmark & \xmark & \xmark & \cmark * &\cmark * & \cmark * &\cmark * & \cmark **&\cmark ** \\
preserving (10) & \xmark & \xmark & \xmark & \xmark & \xmark * &\xmark * & \xmark * &\xmark * & \xmark **&\cmark **\\

\end{tabular}

\caption{Overview of the relationships between consistency concepts, \enquote{\cmark}
indicates that the notion in this row implies the notion in the column, and 
\enquote{\xmark} indicates that this implication does not hold. 
All results marked with \enquote{*} are from \cite{kosiol2022sustaining} and those marked with \enquote{**} are from \cite{habel2009correctness}.}\label{table_cons_relations}

\end{table}

\section{Consistency-Maintaining and Increasing Application Conditions}\label{appl_conds}

\begin{figure}
	\centering
	\includegraphics[scale=0.8]{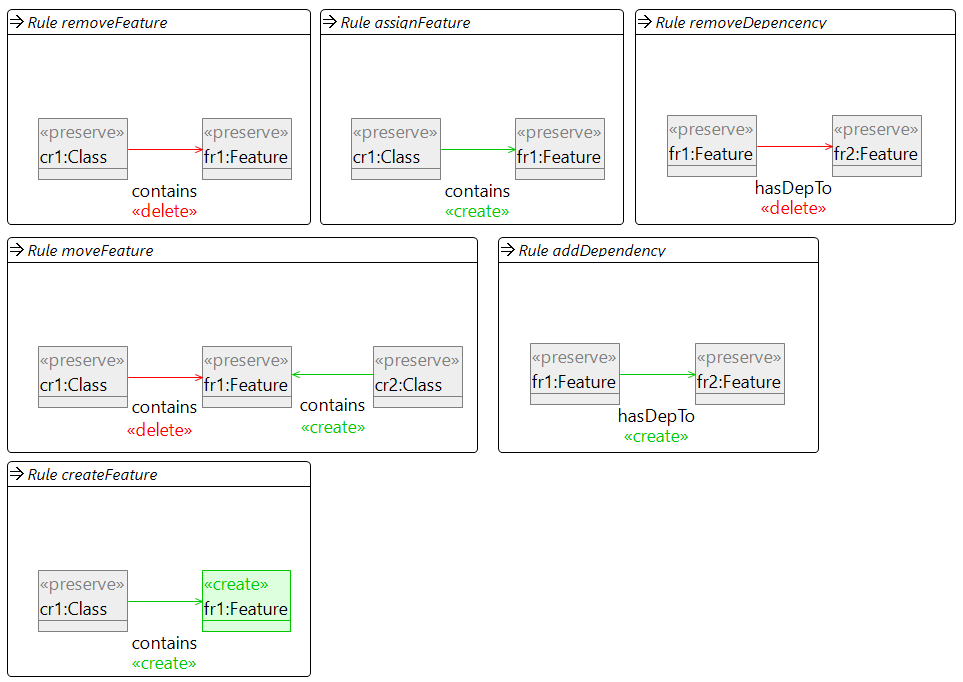}

	\caption{The rules used throughout the examples.}\label{fig:rules}
\end{figure}
\begin{figure}
\centering
\includegraphics[scale=0.6]{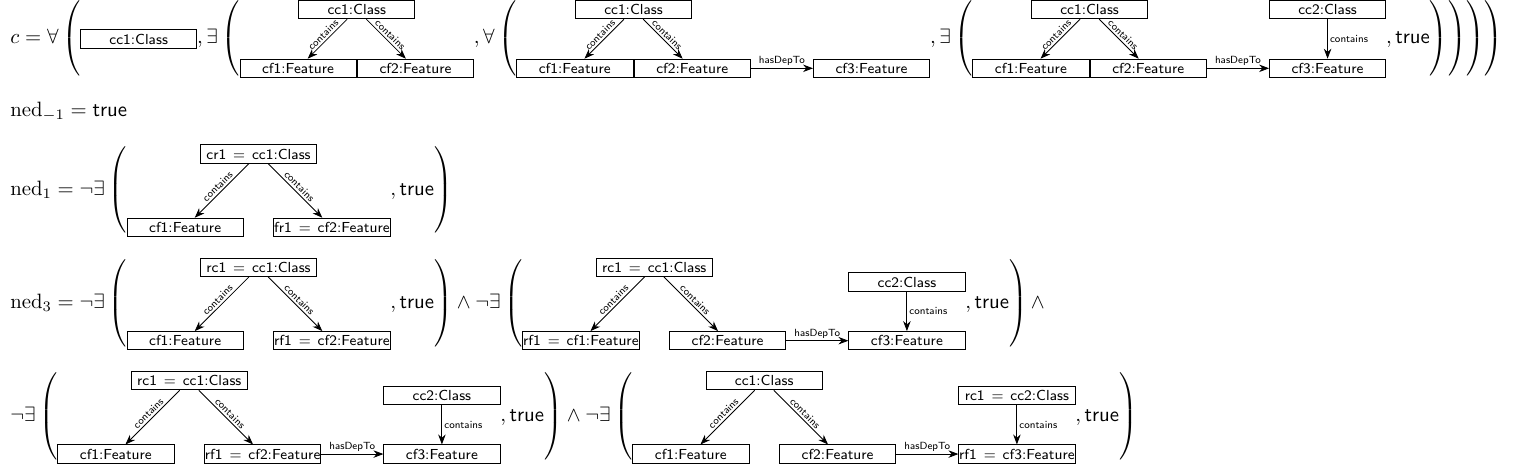}

\caption{no existentially deleted conditions for the constraint and rule \texttt{removeFeature.}}\label{fig:appl_cond}
\end{figure}

In the following, we present application conditions which ensure that every rule equipped with this application condition is (direct) consistency-increasing at layer, (direct) con\-sis\-ten\-cy-maintaining at layer or (direct) consistency-maintaining.
In particular, we present application conditions in the general case and for specific rules, called  \emph{basic increasing} rules.
For basic rules, less complex application conditions can be constructed.
Similar to the notions of consistency-maintaining and consistency-increasing, these application conditions will only consider graphs of the constraint up to a certain layer  and we will show that rules equipped with these application conditions are direct consistency-maintaining and direct consistency-increasing rules at layer, respectively.  


\subsection{Application Conditions for general Rules}
We will now introduce consistency-maintaining and consistency-increasing application conditions at layer for general rules. 
This means that a rule equipped with these application conditions is consistency-maintaining at layer or consistency-increasing at layer. We will also introduce an application condition such that any rule equipped with it is indeed a consistency-maintaining rule.

Let us start with the consistency-maintaining application condition at layer. 
This  application condition has an odd parameter $-1 \leq k < \nlvl(c)$ which specifies which graphs of a constraint are to be considered. 
In particular, only the graphs $C_j$ with $0 \leq j \leq k+3$ are considered.
Note that there is no graph with $\kmax$ even and $\kmax \neq \nlvl(c)-1$. In this case, it can be shown that the application condition with $k = \nlvl(c)-2$ produces a consistency-preserving application condition.  
So it is not a restriction that $k$ must be odd. 
The maintaining application condition consists of the following three parts, which also use the $k$ parameter:
\begin{enumerate}

	\item \emph{No violation inserted ($\wors{k}{} $):} 
		This application condition checks that no new violations are introduced by removing occurrences of intermediate graphs from the set $\ig{C_{k+2}}{C_{k+3}}$.  
		It corresponds to the \emph{no new violation by deletion} formula in the sense that a rule that satisfies this application condition also satisfies the \emph{no new violation by deletion} formula when applied to a graph with $\kmax = k$.
		There are several cases for this application condition. If $k = \nlvl(c)-2$, the constraint ends with $\forall(C_{\nlvl(c)}, \false)$. Therefore no violations can be introduced by removing occurrences of intermediate graphs. In particular, there is no graph $C_{k+3}$.
If $k = \nlvl(c) -1$, every transformation $t: G \Longrightarrow H$ with $G \models c$ is direct consistency maintaining w.r.t. $c$ if $H \models c$. This is ensured by $\remain{k}{}$ and $\ins{k}{}$.
  Therefore, if $k \geq \nlvl(c)-2$, we set the application condition to $\true$.
  
  \item \emph{No universally inserted ($\ins{k}{}$):}
		This application condition checks that no occurrences of universally bound graphs $C_j$ with $1 \leq j \leq k+2$ are inserted. It corresponds to the \emph{no satisfied layer reduction by insertion} and \emph{no new violations by insertion} formulas, in the sense that a rule that satisfies this application condition also satisfies the \emph{no satisfied layer reduction by insertion} and \emph{no new violations by insertion} formulas when applied to a graph with $\kmax = k$.

	\item \emph{No existentially destroyed ($\remain{k}{}$):}
		 This application condition checks that no occurrences of existentially bound graphs $C_j$ with $2 \leq j \leq k+1$ are removed. It corresponds to the \emph{no satisfied layer reduction by deletion} formula since a rule  that satisfies this application condition also satisfies the \emph{no satisfied layer reduction by deletion} formula when applied to a graph with $\kmax = k$.

		\end{enumerate}

Recall that given a constraint $c$ in UANF, each subcondition $\scond{k}{c}$
is a condition over the graph $C_k$ and the morphism with domain $C_k$  is denoted by $a_k$.

\begin{definition}[\textbf{maintaining application condition at layer}]\label{def_main}
	Given a  rule $\rho = (\ac, \rho')$ with $\rho' = \rle{L}{}{K}{}{R}$, 
	a constraint $c$ in UANF and an odd $-1 \leq k < \nlvl(c)$.
	The \emph{maintaining application condition} of $c$ for $\rho$ at layer 
	$k$  is defined  as $\ac \wedge \main{k}{\rho'}$ with
	$$\main{k}{\rho'} :=  \remain{k}{\rho'} \wedge \ins{k}{\rho'} \wedge \wors{k}
	{\rho'} $$ 
	where $\remain{k}{\rho'}$, $ \ins{k}{\rho'}$ and $\wors{k}$ are defined as

	\begin{enumerate}
	\item No violation inserted:
			 Let $\mathbf{P}_{C'}$ be the set of all overlaps $P$ of $L$ and $C'$ 
			 with 
			 $i_L^{P}(L \setminus K) \cap i_{C'}^{P}(C'\setminus C_{k+2})\neq \emptyset$ for 
			 $C' \in \ig{C_{k+2}}{C_{k+3}}$: 
			 \begin{equation*}
				\wors{k}{\rho'} :=			 	
			 	\begin{cases}
			 	 	\true &\text{if $k \geq \nlvl(c)-2$} \\
			 	 	\bigwedge_{C' \in \ig{C_{k+2}}{C_{k+3}}}\bigwedge_{P \in \mathbf{P}_{C'}} \neg \exists(i_L^{P}: L 
			\inj P', \true) &\text{otherwise}
			 	\end{cases}
			 \end{equation*}
			 
		\item No universally inserted:
		
			Let $\mathbf{U}$ be the set of all universally bound graphs $C_j$ with $1 \leq j \leq
			k+2$, and $\mathbf{P}_{C_j} $ be the set of all overlaps $P'$ of $R$ 
			and $C_j$ with $i_R^{P'}(R \setminus K) \cap i_{C_j}^{P'}(C_j\setminus C_{j-1}) \neq 
			\emptyset$:
			$$\ins{k}{\rho'} := \bigwedge_{C \in \mathbf{U}} \bigwedge_{P' \in \mathbf{P}
			_{C}} \shift(\neg \exists(i_R^{P'}: R \inj P', \true), \rho')$$

		\item No existentially destroyed:
			If $k = -1$, we set $\remain{k}{\rho'} := \true$.
			Otherwise, let $\mathbf{E}$ be the set of all existentially bound graphs $C_j$ 
			with $2 \leq j \leq k+1$ and  $\mathbf{P}_{C_j}$ be the set all overlaps 
			$P'$ 
			of $L$ and $C_j$ with $i_L^{P'}(L \setminus K) \cap i_{C_j}^{P'}(C_j\setminus C_{j-1})
			\neq \emptyset$: 
			$$\remain{k}{\rho'} := \bigwedge_{C \in \mathbf{E}} \bigwedge_{P' \in 
			\mathbf{P}_{C}} \neg \exists(i_L^{P'}: L \inj P', \true)$$

	\end{enumerate}
\end{definition}

\begin{figure}
\centering
\includegraphics[scale=0.7]{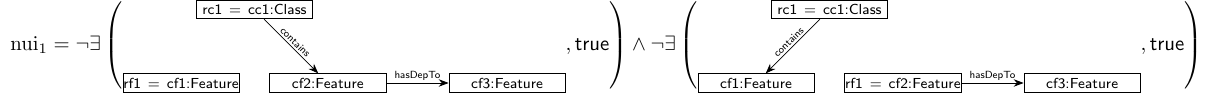}

\caption{No universally inserted conditions for the constraint given in Figure \ref{fig:appl_cond} and rule \texttt{assignFeature.}}\label{fig:nui}
\end{figure}

\begin{figure}
\centering
\includegraphics[scale=0.9]{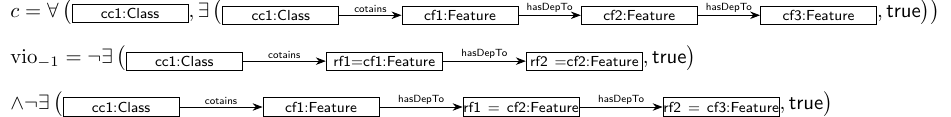}

\caption{No violation inserted conditions for the constraint $c$ and rule \texttt{removeDependency.}}\label{fig_vio}
\end{figure}

We are aware that there are optimisations for $\ins{k}{} $ and $\remain{k}{\rho'}$, since according to \cite{behr2020efficient} not all overlaps need to be considered. But for the simplicity of our definition, we have not implemented these optimisations.
\begin{example}
	\begin{enumerate}
		\item
			Consider the constraint $c$ given in Figure \ref{fig:appl_cond} and the rule \emph{\texttt{re\-move\-Feature}}. The conditions $\remain{-1}{\emph{\texttt{removeFeature}}}$, $\remain{1}{\emph{\texttt{removeFeature}}}$ and $\remain{3}{\emph{\texttt{removeFeature}}}$ are also given in Figure  \ref{fig:appl_cond}; $\remain{-1}{\emph{\texttt{removeFeature}}}$ is equal to $\true$. $\remain{1}{\emph{\texttt{removeFeature}}}$ checks that no occurrences of $C_2$ are inserted, while $\remain{3}{\emph{\texttt{removeFeature}}}$ checks that no occurrences of $C_2$ and $C_4$ are inserted. Obviously, $\remain{1}{\emph{\texttt{removeFeature}}}$ is contained in $\remain{3}{\emph{\texttt{removeFeature}}}$. Note that there are conditions in $\remain{3}{\emph{\texttt{removeFeature}}}$ that imply each other. For example, the first condition implies the second and third. 
		Using the optimisations according to \cite{behr2020efficient}, these can be removed.
		\item
		Again consider constraint $c$ given in Figure \ref{fig:appl_cond} and 
		the rule \emph{\texttt{assignFeature}}. The condition $\ins{-1}{\emph{\texttt{assignFeature}}}$ is equal to $\true$ since \emph{\texttt{assignFeature}} cannot create any occurrences of the first universally bound graph of the constraint. The condition $\ins{1}{\emph{\texttt{assignFeature}}}$ is given in Figure \ref{fig:nui}.
		
		The condition  $\ins{3}{\emph{\texttt{as\-sign\-Fea\-ture}}}$ is equal to $\ins{1}{\emph{\texttt{assignFeature}}}$ since there are only two universally bound graphs and $\ins{1}{\emph{\texttt{assignFeature}}}$ already considers both.
		\item
		Consider the constraint $c$ given in Figure \ref{fig_vio} and the rule \emph{\texttt{addDependency}}.
		Then, $\wors{-1}{\emph{\texttt{addDependency}}}$ is given in Figure \ref{fig_vio}. The condition $\wors{1}{\emph{\texttt{addDependency}}}$ is 
		equal to $\true$ since $1 > \nlvl(c)-2 = 2-2$.
		
	\end{enumerate}
\end{example}
Let us now show that every rule equipped with $\main{k}{}$ is a consistency-maintaining rule at layer $k$.

\begin{theorem}\label{thm_maintaining_ac}
	Given a constraint $c$ in UANF.
	Every rule $\rho = (\ac \wedge \main{k}{\rho'}, \rho')$ where $\rho' = \rle{L}{l}{K}{r}{R}$ and 
	$-1 \leq k < \nlvl(c)$ is odd is a direct consistency maintaining rule w.r.t. $c$ at layer $k$.
	  
\end{theorem} 
\begin{proof}
	Given a graph $G$ with $\kmax = k$ and a transformation $t: G \Longrightarrow_{\rho,m} H$.
	We show that $t$ is a direct consistency maintaining transformation w.r.t. $c$.
	
	 We show that $t$ satisfies the \emph{no new violation by deletion}, \emph{no new violation by insertion}, \emph{no satisfied layer reduction by insertion} and \emph{no satisfied layer reduction by deletion} formulas.
	
	\begin{enumerate}
		\item 
			 Assume that $t$ does not satisfy the \emph{no new violation by deletion} formula.
			 Then $\kmax < \nlvl(c) -1$,
			 $e = \scond{k+2}{c} \neq \false$ and there is a morphism $p: C_{k+2} \inj G$ 
			 such that $p \models \ic{0}{e}{C'}$, $\track_t \circ p$ 
			 is total and $\track_t \circ p \not \models \ic{0}{e}{C'}$ for a 
			 graph $C' \in \ig{C_{k+2}}{C_{k+3}}$. 
			 Therefore, there is an overlap $P$ of $L$ and $C'$ with $i_L^P(L 
			 \setminus K) \cap i_{C'}^P(C' \setminus C_{k+2}) \neq \emptyset$  such that $i_{C'}^P \circ a^r_{k+2} 
			 \models \exists(a^r_{k+2}: C_{k+2} \inj C', \true)$ 
			 and $m \models \exists(i_L^P: L \inj P, \true)$. 
			 Thus, $\wors{k}{\rho'} $ and consequently also $ \main{k}{\rho'}$ 
			 cannot be satisfied. 

		\item 
			Assume that $t$ does not satisfy the \emph{no new violation by insertion} formula. Let

			$$ d := \begin{cases} 
				\ic{0}{\scond{k+2}{c}}{C_{k+3}} & \text{if $\scond{k+2}{c}
				\neq \false$} \\
				\false &\text{otherwise.}
	
				\end{cases}	
			$$

  			Then, there is a morphism $p': C_{k+2} \inj H$ with $p' \not \models d$ 
  			such that no morphism $p :C_{k+2} \inj G$ with $\track_t 
  			\circ p = p'$ exists. Therefore, there is an overlap $P$ of $R$ and $C_{k+2}$ 
  			with $i_R^P(R\setminus K) \cap i_{C_{k+2}}^P(C_{k+2}\setminus C_{k+1}) \neq \emptyset$ 
  			 such that $m \models \shift(\exists(i_R^P: R \inj P, \true), 
  			\rho')$. Hence, $m$ does not satisfy $\ins{k}{\rho'}$. 

		\item 
			Assume that $t$ does not satisfy the \emph{no satisfied layer reduction by insertion} formula. Then, there is a 
			morphism $p: C_j \inj H$ with $0 \leq j <k$ and $C_j$ universally bound such that no morphism $p': C_j \inj G$ with $\track_t 
			\circ p' = p$ exists. Then, there is an overlap $P$ of $C_j$ and $R$ with 
			$i_R^P(R \setminus K) \cap i_{C_j}^P(C_j\setminus C_{j-1}) \neq \emptyset$ 
			such that $m \models \shift(\exists(i_R^P: R \inj P, \true), \rho)$. 
			Hence, $m \not \models  \ins{k}{\rho'}$. 

		\item 
			Assume that $t$ does not satisfy the \emph{no satisfied layer reduction by deletion} formula.  Then, there is a 
			morphism $p :C_j \inj G$ with $j \leq k+1$ and $C_j$  existentially bound and   such that $\track_t \circ p$ is not total. Then, there is an 
			overlap $P$ of $C_j$ and $L$ with $i_L^P(L \setminus K) \cap i_{C_j}
			^P(C_j\setminus C_{j-1}) \neq \emptyset$, such that $m \models \exists(i_L^P:L 
			\inj P, \true)$. Hence, $m \not \models \remain{k}{\rho'}$.
	\end{enumerate}
	It follows that $\rho$ is a direct consistency-maintaining rule at layer $k$
	w.r.t. $c$.
\end{proof}

With the constructions described in Definition \ref{def_main} we are also able to construct direct consistency-maintaining application conditions. 

\begin{theorem}\label{appl-main}
	Given a constraint $c$ in UANF. Every rule $\rho$ equipped with 
	the application condition 
	$$ \Big(\bigwedge_{\substack{-1 \leq i < \nlvl(c) \\ i \textit{ odd}}} \wors{i}{\rho}\Big) \wedge \ \ins{\nlvl(c)-1}{\rho}$$
	is a direct consistency-maintaining rule w.r.t. $c$.
\end{theorem}

\begin{proof}
	Let $\rho = \rle{R}{r}{K}{l}{L}$ be a rule equipped with this application condition. 
	We show that $\rho$ is a consistency-maintaining rule at layer $k$ w.r.t. $c$ 
	for all $-1 \leq k \leq \nlvl(c) -1$.
	Obviously, $\ins{\nlvl(c)-1}{\rho}$ contains $\ins{j}{\rho}$ for all 
	$-1 < j \leq \nlvl(c)-1$.
	The set of intermediate graphs always contains the second graphs on which this set was built, so $\wors{i}{\rho}$ contains the condition 
	$$ \bigwedge_{P' \in 
			\mathbf{P}_{C_{i+3}}} \neg \exists(i_L^{P'}: L \inj P', \true)$$
	which also checks that no occurrence of $C_{i+3 }$ is deleted. Therefore 
	$$\Big(\bigwedge_{\substack{-1 \leq i < \nlvl(c) \\ i \textit{ odd}}} \wors{i}{\rho}\Big)$$ 
	must contain $\remain{\nlvl(c)-1}{\rho}$ and therefore it contains $\remain{j}{\rho}$ for all $-1 \leq j \leq \nlvl(c)-1$.
	So we can rewrite this application condition into the equivalent condition
	$$\Big(\bigwedge_{\substack{-1 \leq i < \nlvl(c) \\ i \textit{ odd}}} \wors{i}{\rho} \wedge \ins{i}{\rho} \wedge \remain{i}{\rho} \Big) =  \Big(\bigwedge_{\substack{-1 \leq i < \nlvl(c) \\ i \textit{ odd}}} \main{i}{\rho} \Big).$$
It follows that $\rho$ is a direct consistency-maintaining rule at layer $k$ for all 
$-1 \leq k < \nlvl(c)$. Since $-1 \leq \kmax < \nlvl(c)$ for each graph $G$, all transformations of $\rho$ are direct consistency-maintaining w.r.t. $c$. Hence, $\rho$ is a consistency-maintaining rule w.r.t. $c$.
\end{proof}
In the following, we will introduce the direct  consistency increasing application conditions at layer. 
For this, we will introduce the notion of \emph{extended overlap}, which will be useful to detect whether a violation is removed by a transformation. 
Intuitively, given an overlap and a morphism $a$, the overlap is extended such that an overlap morphism satisfies $\exists(a, \true)$.

\begin{definition}[\textbf{extended overlaps}]
	Given an overlap $(G, i_{C_0}^G, i_{C_1}^G)$ of $C_0$ and $C_1$ and a 
	morphism $e: C_0 \inj H$. 
	The set of \emph{extended overlaps} of $G$ at $i_{C_0}^G$ with $e$, 
	denoted by $\eol(G, i_{C_0}^G,e)$, is defined as	
	$$\eol(G,i_{C_0}^G, e) := \{P \in \overlay(G,H) \mid
		i_G^P \circ i_{C_0}^G \models \exists(e:C_0 \inj H, \true)
		\}.$$

\end{definition}
In other words, $\eol(G, i_{C_0}^G,e)$ is the set of all overlaps of $G$ and $H$ such that the square in Figure \ref{fig_eol} is commutative, i.e. $i_H^P \circ e = i_G^P \circ i_{C_0}^G$.  
\begin{figure}
\center
	\begin{tikzpicture}
		\node (L) at (0,2) {$C_0$};
		\node (K) at (2,2) {$H$};
		
		\node (G) at (0,0) {$G$};
		
		\node (H) at (2,0) {$P$};
		
		\draw[right hook-stealth] (L) edge node [above] {$e$}  (K);
		\draw[left hook-stealth] (L) edge node [left] {$i_{C_0}^G$}  (G);
		\draw[right hook-stealth] (G) edge node [above] {$i_G^P$}  (H);
		\draw[right hook-stealth] (K) edge node [right] {$i_H^P$}  (H);
		
	\end{tikzpicture}
	\caption{Diagram for the alternative definition of extended overlaps.}\label{fig_eol}
\end{figure}
Using extended overlaps, we will be able to check whether a violation has been 
removed.

For the consistency-increasing application condition at layer, we will use the maintaining application condition at layer.
All that remains is to ensure that at least one violation is removed. 
To do this, we must first check that there is a violation in the match. This means that the match and a violation are overlapping.
Finally, we need to check that this violation is removed.

Again, the increasing application condition has the odd parameter $-1 \leq k < \nlvl(c)-1$, which specifies which constraint graphs are to be considered. Note that $k$ must not be $\nlvl(c)-1$, since all graphs with $\kmax = \nlvl(c)-1$ satisfy the constraint and no consistency-increasing transformations are originating from those graphs.
We also use a second parameter $C'$, which is an intermediate graph of $C_{k+2}$ and $C_{k+3}$ if $c$ contains a graph $C_{k+3}$, i.e. $k < \nlvl(c)-2$, and $C'$ is set to $C_{k+2}$ otherwise. 
A rule equipped with this application condition is a consistency-increasing rule at layer $k$.
It consists of the following parts:
\begin{enumerate}
	\item The maintaining application condition ($\main{k}{}$):
			As already discussed, this application condition ensures that 
			a rule equipped with this application condition is a consistency-maintaining rule at layer $k$.
	\item \emph{Violation exists} ($\nex()$): 
		This condition checks that there is a violation at the match, i.e. there is an occurrence $p$ of $C_{k+2}$ with $m(L) \cap p(C_{k+2}) \neq \emptyset$ not satisfying $\exists(C', \true)$. If $k = \nlvl(c) -2$, then $c$ ends with $\forall(C_{\nlvl(c)}, \false)$ and thus there is no graph $C_{k+3}$ in $c$. In this case, it is sufficient to check only that $m(L) \cap p(C_{k+2}) \neq \emptyset$.
	\item \emph{Violation removed} ($\rep()$):
		This condition checks that the violation is removed.  This can be done in several ways, either by deleting the occurrence $p$ or by inserting elements such that $ p \models \exists(C', \true)$. This leads to case discrimination.  The first case is easy to check, if $m(L \setminus K) \cap p(C_{k+2}\setminus C_{k+1}) \neq \emptyset$, $p$ is removed and this condition can be set to $\true$. Otherwise, we need to check that the violation has been removed by an additional condition that checks whether $p$ satisfies $\exists(C', \true)$ after the transformation. The last case is the special case where the constraint ends with $\forall(C_{\nlvl(c)}, \true)$ and $k = \nlvl(c) -2$. Then there is only one way to remove a violation, by removing the occurrence $p$. So the condition is set to $\true$ if $m(L \setminus K) \cap p(C_{k+2}\setminus C_{k+1}) \neq \emptyset$ and to $\false$ otherwise. 
\end{enumerate}

\begin{definition}[\textbf{consistency increasing application condition at layer}]\label{def_appl_cond}
	Given a rule $\rho = (\ac, \rho')$ with $\rho' = \rle{L}{l}{K}{r}{R}$ and a constraint $c$ in UANF. Let $0 \leq k < \nlvl(c)-1$ be odd and $C' = C_{k+2}$ if $k = \nlvl(c) -2$
	and $C' \in \ig{C_{k+2}}{C_{k+3}}$ otherwise.  
	The \emph{increasing application condition of $c$ for $\rho$ at layer $k$ 
	with $C'$} is defined as
	\begin{equation}
		\incr{k}{C'}{\rho} := \ac \wedge \main{k}{\rho'} \wedge \big(\bigvee_{P
		\in \overlay(L,C_{k+2})} \vFound{}{P}{C'} \wedge \vRepaired{}{P}{C'}\big) 
	\end{equation}
	with
\begin{enumerate}
\item Violation exists: Let $a_{k+2}^r: C_{k+2} \inj C'$ be the restricted morphism of $a_{k+2}$ and $i_L^P$ and $i_P^Q$ be overlap morphisms of $P$ and $Q$, respectively:
	$$\nex(P,C') := 
			\begin{cases}
			\exists (i_L^P: L \inj P, \true) &\textit{if } \scond{k+2}{c} = 
			\false \\
			 \exists(i_L^P: L \inj P, \bigwedge_{Q \in \eol(P,i_{C_{k+2}}^P,a^r_{k+2})} \neg 
			 \exists(i_P^Q: P \inj Q, \true)) &\textit{otherwise}
			\end{cases}			
			$$
			
\item Violation removed: 

\begin{enumerate}
	\item If $i_L^P(L \setminus K) \cap i_{C_{k+2}}^P(C_{k+2}\setminus C_{k+1}) \neq 
		\emptyset$, $\rep(P,C') := \true $.
	\item If $\scond{k+2}{c} = \false$, i.e. $k = \nlvl(c)-2$,
		$$\rep(P,C') :=
		\begin{cases} 
			\true &\text{if $i_L^P(L \setminus K) \cap i_{C_{k+2}}^P(C_{k+2}\setminus C_{k+1}) \neq \emptyset$}\\
			\false &\text{otherwise}.
		\end{cases}$$
	\item Otherwise, let $P'$ be the overlap of $R$ and $C_{k+2}$ such that there is a transformation $P' \Longrightarrow_{\rho^{-1}, i_R^{P'}} P$. 
If this overlap or transformation does not exist, we set $\rep(P,C') := \false$ and otherwise 
$$\rep(P,C') := \shift(\forall (i_R^{Q}: R \inj P', 
				\bigvee_{Q \in \eol(P',i_{C_{k+2}}^{P'}, a^r_{k+2})}  \exists(i_{P'}^Q:P' \inj Q,\true)), \rho).$$
\end{enumerate}
\end{enumerate}

\end{definition}

\begin{figure}
\centering
\includegraphics[scale=0.8]{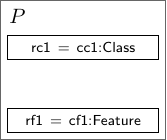}
\par
\vspace{1cm}
\includegraphics[scale=0.75]{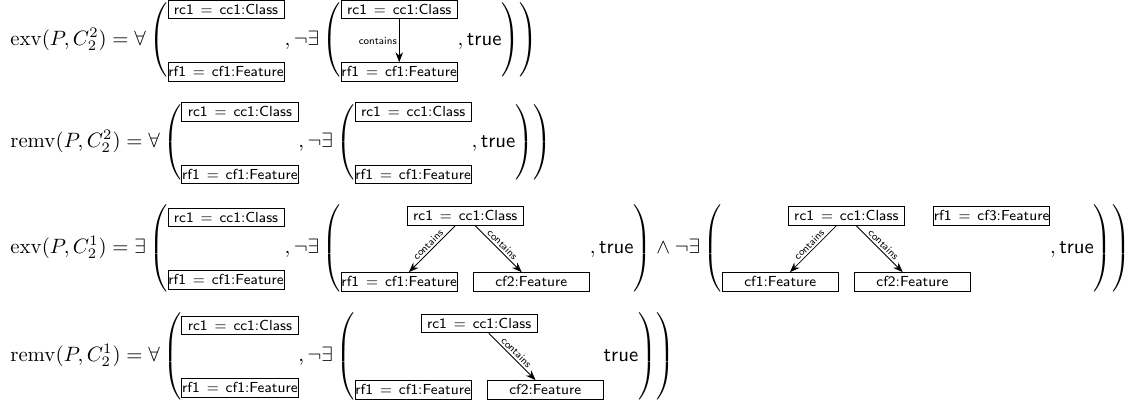}
\vspace{0.1cm}

\caption{Examples for $\nex(P,C_2^2), \nex(P,C_2^1), \rep(P,C_2^2)$ and  $\rep(P,C_2^1)$ using the rule \texttt{assignFeature} and constraint $c_1$ given in Figure \ref{fig:constraints}.}\label{fig_increasing_1}
\end{figure}

\begin{figure}
\centering
\includegraphics[scale=0.7]{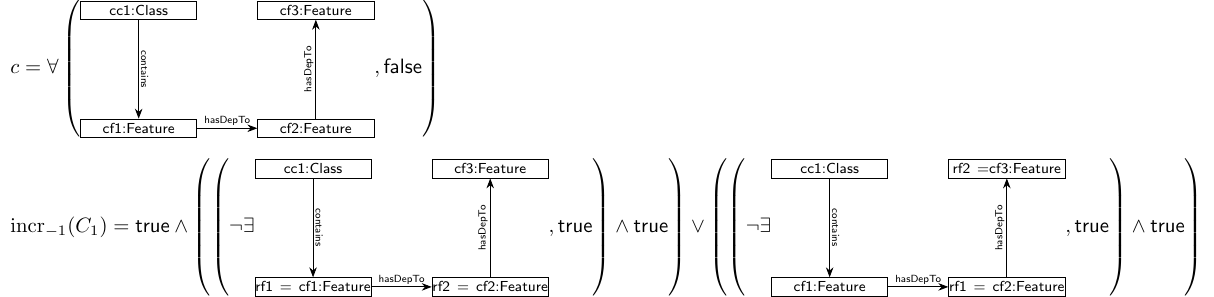}

\caption{Example of $\incr{-1}{C_1}{\rho}$ using this constraint and the rule \texttt{removeDependency}.}\label{fig_increasing_2}
\end{figure}

\begin{example}
\begin{enumerate}
	\item 
	Consider constraint $c_1$ given in Figure \ref{fig:constraints} and the rule 
\emph{\texttt{as\-sign\-Fea\-ture}}.
There is only one overlap $P$ of $L$ and $C_1^1$ which is shown in Figure \ref{fig_increasing_1}. 
The $\nex(P,C')$ and $\nex(P, C')$ parts of $\incr{-1}{C'}{\emph{\texttt{assignFeature}}}$ with $C' = C_2^2$ and $C' = C_2^1$, respectively, are also given in Figure \ref{fig_increasing_1}.

	\item
	Consider the constraint and rule given in Figure \ref{fig_increasing_2}.
	The consistency-increasing application condition of this rule and at layer $-1$ with $C_1$ of the constraint is also given in this Figure.
\end{enumerate}
\end{example}


Let us now show that a rule equipped with the consistency-increasing transformation condition at layer $k$ is indeed a consistency-increasing rule at layer $k$.

\begin{theorem}
	Given a constraint $c$ in UANF. 
	Every rule $\rho = (\ac \wedge \incr{k}{C'}{\rho'},\rho')$ with $-1 \leq k < \nlvl(c)-1$ odd and $C' \in \ig{C_{k+2}}{C_{k+3}}$ if $k < \nlvl(c)-2$ and 
	$C' = C_{k+2}$ otherwise is a direct consistency-increasing rule w.r.t. $c$ at layer $k$.
	
\end{theorem}
\begin{proof}
	Let a transformation $t: G \Longrightarrow_{\rho} H$ with $\maxk{c}{G} = k$
	be given. 
	Since  $\incr{k}{C'}{\rho}$ contains $\main{k}{\rho}$, $t$ is a direct consistency maintaining transformation at layer $k$ according to Theorem \ref{thm_maintaining_ac}. It remains to show that $t$ satisfies 
	the general or special increasing formula respectively.

	\begin{enumerate}
		\item	
			If $k = \nlvl(c) -2$, i.e. $c$ ends with a condition of the form $\forall(C_{\nlvl(c)},\false)$, assume that $t$ does not satisfy the special increasing formula. Then there is no morphism $p: C_{k+2} \inj G$ such that $\track_t \circ p$ is not total.
			So there is no overlap $P$ of $L$ and $C_{k+2}$ with $i_L^P(L\setminus K) \cap i_{C_{k+2}}^P(C_{k+2}\setminus C_{k+1} ) \neq \emptyset$ such that $m \models \exists(i_L^P: L \inj P, \true)$.
			Since, $k = \nlvl(c) -2$ it follows that either $\rep(P',C') = \false$ or $m \not \models \nex(P',C')$ for all $P \in \overlay(L,P)$. Therefore $m \not \models \incr{k}	{C}{\rho}$, this is a contradiction.
		
		\item 
			Otherwise let $P \in \overlay(L, C_{k+2})$. We show that 
			$m \models \vFound{}{P}{C'} \wedge \vRepaired{}{P}{C'}$ implies  that $t$ satisfies the general increasing formula.
			If $m \models \vFound{}{P}{C'}$, there is a morphism
			$p: P \inj G$ with $m = p \circ i_L^P$ and $p \models \neg \exists(
			i_P^Q:P \inj Q, \true)$ for all $Q \in \eol(P,i_{C_{k+2}}^P, a^r_{k+2})$.
			Therefore, $q:= p \circ 
			i_{C_{k+2}}^P\not \models 
			\exists(a_{k+2}^r:C_{k+2} \inj C', \true)$. 
			
			For $\vRepaired{}{P}{C'}$ there are two cases. Either $\vRepaired{}{P}{C'} = \true$ or $\vRepaired{}{P}{C'}$ is the condition described in Definition \ref{def_appl_cond} and $m \models \vRepaired{}{P}{C'}$. 
			In the first case, if $\vRepaired{}{P}{C'} = \true$ it follows that 
			$i_L^P(L \setminus K) \cap i_{C_{k+2}}^P(C_{k+2} \setminus C_{k+1}) \neq \emptyset$. 
			Therefore $\track_t \circ q$ is not total and the general increasing formula is satisfied. 
			
			Otherwise,  if $\vRepaired{}{P}{C'}$ is the condition described in Definition \ref{def_appl_cond} and $m \models \vRepaired{}{P}{C'}$, $\track_t \circ q$ is total and there is 
			a morphism $p': P' \inj H$ with 
			$\track_t\circ q = p' \circ i_{C_{k+2}}^{P'}$.
			Since $m \models \vRepaired{}{P}{C'}$,  all morphisms $p'':P' \inj H$ with $n = p'' \circ i_R^Q$ satisfy $\bigvee_{Q \in \eol(P',i_{C_{k+2}}^{P'}, a^r_{k+2})}  \exists(i_{P'}^Q:P' \inj Q,\true)$. 
			It follows that $p'' \circ i_{C_{k+2}}^P \models \exists(C', \true)$ and in particular that $\track_t\circ q = p' \circ i_{C_{k+2}}^{P'} \models \exists(C', \true)$. Therefore, the general increasing formula is satisfied.
	\end{enumerate}	 
	In summary, $\rho$ is a direct consistency-increasing rule at layer $k$.
\end{proof}

Note that $\incr{k}{C'}{\rho}$ is only evaluated to $\true$ if 
an occurrence $p: C_{k+2} \inj G$ that does not satisfy $\exists(C', \true)$ is either removed or satisfies $\exists(C', \true)$ in the derived graph.
For all smaller improvements, i.e. a similar improvement for a subgraph $C'' \in \ig{C_{k+2}}{C'}$ of $C'$, $\incr{k}{C'}{\rho}$ would be evaluated to $\false$.
For any larger improvements, i.e. the same improvement for a supergraph $C'' \in \ig{C'}{C_{k+3}}$ of $C'$, $\incr{k}{C'}{\rho}$ will also be evaluated to $\false$ if the repaired occurrence of $C_{k+2}$ satisfies $\exists(C', \true)$.
In both cases, the application condition would prohibit the transformation, even though it would be direct consistency-increasing.
To solve this problem, several application conditions could be combined by
$$\bigvee_{C' \in \ig{C_{k+2}}{C_{k+3}}} \incr{k}{C'}{\rho}.$$ 
This application condition will be evaluated to $\true$ if the cases described above occur, with the drawback that this results in a huge condition, even if duplicate conditions are removed.
At least all duplicates of $\main{}{}$ can be removed, since they are identical for each $\incr{k}{C'}{\rho}$ and only need to be constructed once.

In general, these application conditions are a compromise between condition size and restrictiveness. 
They are very restrictive because they do not allow deletions of occurrences of existentially bound graphs and insertions of universally bound graphs.
For example, any of these application conditions for the rule \texttt{moveFeature} and the constraint $c_1$ will be equivalent to $\false$; the maintaining part of the condition will always be evaluated to $\false$, since \texttt{moveFeature} always removes occurrences of the existentially bound graph $C_2^1$.
Changing the conditions constructed by $\maintaining()$ to check whether two nodes of type \texttt{Feature} are connected to a node of type \texttt{Class} will give application conditions that can be satisfied with \texttt{moveFeature}. However, for a similar rule moving two nodes of type \texttt{Feature}, this newly constructed $\maintaining()$ would still be evaluated to $\false$.
So this would only lead to a slight decrease in restrictiveness. 

The conditions constructed by $\rem()$ and $\nin()$ could be modified in a similar way.
For $\rem()$ and an occurrence $p$ of the universally bound graph $C_{j}$, by checking whether there are two occurrences $p_1,p_2$ of $C_{j}$ with $p = p_1 \circ a_j = p_2 \circ a_j$, and for $\nin()$, by checking whether an introduced occurrence $p$ of $C_j$ satisfies $\exists(C_{j+1}, \true)$.
As above, this only leads to a small decrease of restrictiveness. 
Also, the consistency-increasing application condition becomes more and more restrictive as $k$ increases, since the number of conditions and in particular the number of negative application conditions also increases.

\subsection{Basic Consistency-Maintaining and Consistency-Increasing Rules}
 
The construction of the application conditions introduced in the previous section, as well as the constructed application conditions themselves, is very complex.
For a certain set of rules, which we will call \emph{basic consistency-increasing rules}, we are able to construct application conditions with the same property, namely that a rule equipped with this application condition is consistency-increasing at a layer $k$, in a less complex way.  
The main idea is that these rules (a) are not able to delete occurrences of existentially bound graphs or insert occurrences of universally bound graphs and (b) are able to increase consistency at a certain layer.
That is, given a basic increasing rule $\rho$, there exists a transformation $t: G \Longrightarrow_{\rho} H$ such that $t$ is a consistency increasing transformation with respect to a constraint $c$.

To ensure that (a) is satisfied, we first introduce \emph{basic consistency-maintaining rules at layer}, which means that, given a constraint, a plain rule is not able to delete existentially bound and insert universally bound graphs up to a certain layer.  For the definition, we use the notion of direct consistency maintaining rules at layer.  The set of basic consistency-maintaining rules up to layer is a subset of the set of direct consistency-maintaining rules at layer, since these rules must be plain rules, whereas direct consistency-maintaining rules at layer are allowed to have application conditions, e.g. $\maintaining(\cdot,\cdot)$.

\begin{definition}[\textbf{basic consistency maintaining rule at layer}]
	Given a plain rule $\rho$ and a constraint $c$ in UANF.
	Then, $\rho$ is called  \emph{basic consistency maintaining at layer
	$-1 \leq k < \nlvl(c)$ w.r.t. $c$} if it is a direct consistency-maintaining rule at layer $k$ w.r.t. $c$.
	The rule $\rho$ is called \emph{basic consistency-maintaining w.r.t. $c$ } if it is a direct consistency-maintaining rule w.r.t. $c$.
\end{definition}

\begin{example}
	Consider the rules \emph{\texttt{moveFeature}},
	\emph{\texttt{assignFeature}} and \emph{\texttt{addDependency}}  given in Figure \ref{fig:rules} and constraints
	$c_1$ and $c_2$ given in Figure \ref{fig:constraints}.
	The rule \emph{\texttt{assignFeature}} is a basic consistency maintaining 
	rule at layer $1$ w.r.t. $c_1$, whereas \emph{\texttt{moveFeature}} is 
	not a basic consistency maintaining rule w.r.t. $c_1$.
	The rule \emph{\texttt{addDependency}} is a basic consistency-maintaining rule at layer $-1$ w.r.t. $c_2$, but is not a basic consistency-maintaining rule at layer $1$ w.r.t. $c_2$ since it can insert occurrences of $C_3^2$.
\end{example}

Since there are infinitely many transformations via a plain rule $\rho$, it is impossible to check whether $\rho$ is a basic consistency-maintaining rule at a layer based on the definition above. Therefore, we present a characterisation of basic consistency-maintaining rules that relies only on $\rho$ itself. 

First, let us assume that $\rho$ is able to create occurrences of a universally bound graph $C_j$. This is possible if (a) $\rho$ inserts an edge of $C_j \setminus C_{j-1}$ connecting pre-existing nodes of $C_j$, since it is unclear whether this would create a new occurrence of $C_j$, or (b) if $\rho$ inserts a node $v$ of $C_j$, so that all edges $e \in E_{C_j}$ with $\src(e) = v$ or $\tar(e) = v$ are also inserted.
If at least one of these edges is not inserted, it is guaranteed that this insertion will not create an occurrence of $C_j$, since $v$ is only connected to edges that have also been inserted by $\rho$.

Second, suppose $\rho$ is able to delete occurrences of an existentially bound graph $C_j$.
This is possible if (a) $\rho$ deletes an edge of $C_j \setminus C_{j-1}$ or (b) $\rho$ deletes a node $v$ of $C_j \setminus C_{j-1}$ such that all edges $e \in E_{C_j}$ with $\src(e) = v$ or $\tar(e) = v$ are also deleted.
If $\rho$ deletes a node $c$ of $C_{j} \setminus C_{j-1}$ without all its connected edges in $C_j$, there is no transformation via $\rho$ such that an occurrence of $C_j$ is deleted by deleting that node, since the dangling edge condition would not be satisfied. 
A rule that satisfies these properties does not reduce the largest satisfied layer.

We also need to ensure that the number of violations is not increased. To do this, we have to check that $\rho$ is not able to insert occurrences of the corresponding universally bound graph, as described above, and that $\rho$ is not able to remove occurrences of any intermediate graph. This is only ensured if $\rho$ does not remove any elements of $C_{k+1}\setminus C_k$ when the set of intermediate graphs is given by $\ig{C_k}{C_{k+1}}$.

To check that a plain rule satisfies these properties, we use the dangling edge condition, or in other words, we check that the rule is not applicable at certain overlaps of $L$ and an existentially bound graph, or that the inverse rule is not applicable at certain overlaps of $R$ and a universally bound graph.

\begin{lemma}\label{def:non-decreasing}
	Given a plain rule $\rho = \rle{L}{l}{K}{r}{R}$ and a constraint $c$ in UANF.
	Let $-1 \leq k < \nlvl(c)$ be odd, then $\rho$ is a basic consistency-maintaining rule up to layer $k$ w.r.t. $c$ if \ref{category:non-decreasing_1} and \ref{category:non-decreasing_2} hold for all $k$, and \ref{category:non-decreasing_3} holds if $k < \nlvl(c) -2$.  
	\begin{enumerate}
		\item \label{category:non-decreasing_1}
			For each existentially bound graph $C_j$ with $2 \leq j \leq k+1$ and each overlap $P \in \overlay(L,C_j)$ 
			with $i_L^P(L \setminus K) \cap i_{C_j}^P(C_j \setminus C_{j-1} ) \neq \emptyset$, 
			the rule $\rho$ is not applicable at match $i_L^P$.
		\item \label{category:non-decreasing_2}
			
			For each universally bound graph $C_j$ with $1 \leq j \leq k+2$ and each overlap $P \in \overlay(R,C_j)$ 
			with $i_R^P(R \setminus K) \cap i_{C_j}^P(C_j) \neq \emptyset$, the rule $\rho^{-1}$ is not applicable at match $i_r^P$.
		\item \label{category:non-decreasing_3}
			For all graphs $P \in \overlay(L, C_{k+3})$ it holds that
			$$i_L^P(L \setminus K) \cap i_{C_{k+3}}^P(C_{k+3} \setminus C_{k+2})
			= \emptyset
			.$$
\end{enumerate}

\end{lemma}
\begin{proof}
	Let $\rho = L \xhookleftarrow{l} K \xhookrightarrow{r} R$ be a rule 
	that satisfies the characterisations listed in Lemma 
	\ref{def:non-decreasing} with $-1 \leq k < \nlvl(c)$ odd. Suppose $\rho$ is not a direct consistency-maintaining rule up to layer $k$ w.r.t. $c$.
	Then, there is a transformation $t: G \Longrightarrow_{\rho,m} H$ with $\maxk{c}{G} = k$ such that $t$ is not direct consistency-maintaining w.r.t. $c$.
	Therefore, either the no new violation by deletion, the no new violation by insertion, the no satisfied layer reduction by insertion or the no satisfied layer reduction by deletion formula is not satisfied.
	\begin{enumerate}
		\item 
			If the no new violation by deletion formula is not satisfied, then $k < \nlvl(c) 
			-2$. There is an occurrence $p:C_{k+2} \inj G$ such that 
			$p \models \ic{0}{\scond{k+2}{c}}{C'}$ and $\track_t \circ p \not \models 
			\ic{0}{\scond{\kmax+2}{c}}{C'}$ with $C' \in \ig{C_{k+2}}{C_{k+3}}$.
			So an overlap $P \in \overlay(L, C_{k+3})$ with $i_L^P(L \setminus K) \cap i_{C_{k+3}}^P(C_{k+3} \setminus 
			C_{k+2})= \emptyset$ must exist. This is a contradiction.
			
		\item 
			If  the no new violation by insertion formula is not satisfied, there is
			an occurrence $p:C_{k+2} \inj H$ such that no morphism 
			$q: C_{k+2} \inj G$ with $p = \track_t \circ q$ exists 
			and $p \not \models \false$ if $\scond{{k+2}}{c} = \false$ and 
			$p \not \models \ic{0}{\scond{{k+2}}{c}}{C_{j+3}}$ otherwise.
			So there is an overlap $P \in \overlay(R, C_{k+2})$ with 
			$i_R^P(R\setminus K)\cap i_{C_{k+2}}^P(C_{k+2}) \neq \emptyset$
			such that $\rho^{-1}$ is applicable at match $i_R^P$.
			This is a contradiction.

		\item 
			If the no satisfied layer reduction by insertion formula is not satisfied, there is
			an occurrence $p : C_j \inj H$ of an universally bound graph $C_j$ 
			with $1 \leq j \leq k+2$ such that no morphism $q:C_j \inj G$ with 
			$\track_t \circ q = p$ exists.
			So there is an overlap $P \in \overlay(R, C_j)$ with 
			$i_R^P(R\setminus K)\cap i_{C_j}^P(C_j) \neq \emptyset$ 
			such that $\rho^{-1}$ is applicable at match $i_R^P$.
			This is a contradiction.
		\item
			If the no satisfied layer reduction by deletion formula is not satisfied, there is
			an occurrence $p : C_j \inj H$ of an existentially bound graph $C_j$ 
			with $2 \leq j \leq k+1$ such that $\track_t \circ p$ is not total.
			So there is an overlap $P \in \overlay(L, C_j)$ with 
			$i_L^P(L \setminus K) \cap i_{C_j}^P(C_j\setminus C_{j-1}) \neq \emptyset$ such that the rule $\rho$ is applicable at match $i_L^P$.
			This is a contradiction.
	\end{enumerate}	    
	In summary, $\rho$ is a basic consistency-maintaining rule up to layer $k$.
\end{proof}

Now we are ready to introduce \emph{basic increasing rules at layer $k$}, where $k$ is odd. The set of basic increasing rules is a subset of the set of consistency-maintaining rules at layer $k$, which ensures that the largest satisfied layer as well as the number of violations will not increase.
In addition, the left-hand side of this rule contains an occurrence $p$ of the universally bound graph $C_{k+2}$, such that either this occurrence is removed, i.e. elements of $C_{k+2}\setminus C_{k+1}$ are deleted, or an intermediate graph $C \in \ig{C_{k+2}}{C_{k+3}}$ is inserted.
Of course, this second case only occurs if $k < \nlvl(c)-2$, where $c$ is the corresponding constraint.
This property has the advantage that the application conditions for basic increasing rules are less complex and smaller, since it can be determined exactly how this rule removes a violation, and therefore no overlaps need to be considered.

This, at first sight, seems to be a restriction of the set of basic increasing rules, but the context of any rule $\rho$ that satisfies all the properties of a basic increasing rule except that $C_{k+2}$ is a subgraph of the left-hand side can be extended so that this new rule $\rho'$ is a basic increasing rule and the semantic of $\rho'$ is a subset of the semantic of $\rho$. 
A characterisation for these derived rules will be presented later. In particular, derived rules are all amalgamated rules \cite{biermann2010parallel} of 
$\rle{L'}{l'}{K}{r'}{R}$, $\rle{C_{k+2}}{\id}{C_{k+2}}{id}{C_{k+2}}$ and $\rho= \rle{L}{l}{K}{r}{R}$ such that injective morphisms 
$l_1 : L' \inj C_{k+2}$, $l_2 : L' \inj L$, $k_1 : K' \inj C_{k+2}$, $k_2 : K' \inj K$, $r_1 : R' \inj C_{k+2}$ and $r_1 : R' \inj R$ exists. 

Basic increasing rules at layer $k$ are called \emph{deleting basic increasing rules} when $p$ is removed and \emph{inserting basic increasing rules} when an intermediate graph is inserted.

\begin{definition}[\textbf{basic increasing rule}]\label{def_basicIncreasing}
	Given a constraint $c$ in UANF and a  direct consistency-maintaining rule $\rho = (\ac,  L \xhookleftarrow{l} K \xhookrightarrow{r} R)$ at layer $-1 \leq k \leq \nlvl(c) -2$, where $k$ is odd. Then, $\rho$ is a  
	\emph{basic increasing rule w.r.t $c$ at layer 
	$k$} if a morphism $p: C_{k+2} 
	\inj L$, called the \emph{increasing morphism}, exists such that either 
	\ref{incr_1} or \ref{incr_2} holds.
	
	\begin{enumerate}
		\item \label{incr_1}
			\emph{Universally deleting}:
			$r \circ l^{-1} \circ p$ is not total. Then, $\rho$ is called
			a \emph{deleting basic increasing rule}.
		\item \label{incr_2}	
			\emph{Intermediate inserting}:
			If $k < \nlvl(c)-2$, there is an intermediate graph
			$C' \in \ig{C_{k+2}}{C_{k+3}}$ such that $p \not \models \exists(
			a_{k+2}^r:C_{k+2} \inj C', \true)$, $r \circ l^{-1} \circ p$ is 
			total and 
			$r \circ l^{-1} \circ p \models \exists(a_{k+2}^r:C_{k+2} \inj C',
			\true)$. Then, $\rho$ is called a \emph{inserting basic increasing 
			rule with $C$}. 
	\end{enumerate} 
\end{definition}

\begin{example}
	Consider the rule \emph{\texttt{assignFeature}} given in Figure \ref{fig:rules} and constraint $c_1$ given in Figure \ref{fig:constraints}. Then, \emph{\texttt{assignFeature}} is an inserting basic rule with $C_2^2 \in \ig{C_1^1} {C_2^1}$ w.r.t. $c_1$ but is not an inserting basic rule with respect to the constraint $\forall(C_2^2, \exists( C_2^1, \true))$ since the left-hand side of  \emph{\texttt{assignFeature}} does not contain an occurrence of $C_2^2$.
\end{example}

As mentioned above, given a direct consistency-maintaining rule $\rho$, we can derive basic increasing rules that are applicable when $\rho$ is applicable by extending the context of that rule so that it contains an occurrence of the graph $C_{k+2}$.

\begin{figure}
\center
	\begin{tikzpicture}
		\node (L) at (0,2) {$L$};
		\node (K) at (2,2) {$K$};
		\node (R) at (4,2) {$R$};
		\node (G) at (0,0) {$L'$};
		\node (D) at (2,0) {$K'$};
		\node (H) at (4,0) {$R'$};
		\node (1) at (1,1) {(1)};
		\node (2) at (3,1) {(2)};
		
		\draw[left hook-stealth] (K) edge node [above] {$l$}  (L); 
		\draw[right hook-stealth] (K) edge node [above] {$r$}  (R); 
		\draw[left hook-stealth] (D)   edge node[above]{$l'$}(G); 
		\draw[right hook-stealth] (D)  edge node [above]{$r'$}(H);
		\draw[left hook-stealth] (K) edge node[fill = white] {$k$} (D);  
		\draw[left hook-stealth] (L) -- node[left]{$i_L^{L'}$} (G);
		\draw[left hook-stealth] (R) edge node [right] {$i_R^{R'}$} (H);    
	\end{tikzpicture}
	\caption{Pushout diagram for Lemma \ref{derived_appl}.}\label{fig_dpo_construction}
\end{figure}
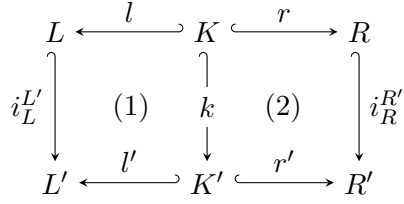

\begin{definition}[\textbf{derived rules}]
	Given a constraint $c$ in UANF and a rule $\rho = (\ac, \rle{L}{l}{K}{r}{R})$. The set of \emph{derived rules from $\rho$} at layer $-1 \leq k \leq \nlvl(c)-2$, where $k$ is odd, contains rules characterised as follows:
	Let 
	$$\mathbf{G} := \begin{cases}
						 \{C_{k+2}\} & \text{if $k = \nlvl(c)-2$ is 
						existentially bound} \\
						\ig{C_{k+2}}{C_{k+3}} &\text{otherwise.}
					\end{cases}
	$$
	For $P \in \mathbf{G}$ and $(L', i_L^{L'}, i_P^{L'}) \in \overlay(L,P)$: If the diagram shown 
	in Figure 
	\ref{fig_dpo_construction} is a transformation, i.e. (1) and (2) are 
	pushouts. 
	The rule $$\rho' = (\shiftm(\ac, i_L^{L'}), \rle{L'}{l'}{K'}{r'}{R'})$$
	is a derived rule of $\rho$ at layer $k$.
\end{definition}

\begin{figure}
	\centering
	\includegraphics[scale=0.8]{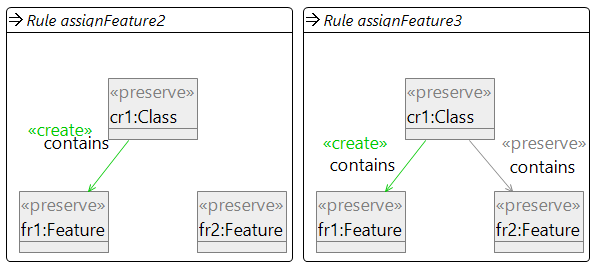}

	\caption{Derived rules of \texttt{assignFeature} and $c_1$.}\label{fig:derived}
\end{figure}

\begin{example}
	Consider the rule \emph{\texttt{assignFeature}} given in Figure 
	\ref{fig:rules} and constraint $c_1$ given in Figure \ref{fig:constraints}.
	The set of derived rules from $\rho$ at layer $1$ is given in Figure 
	\ref{fig:derived}.
\end{example}

The following lemma shows that a derived rule $\rho'$ of $\rho$ is only applicable at a graph $G$ if $\rho$ is also applicable at $G$ and the derived graphs of these transformations are identical. 

\begin{lemma}\label{derived_appl}
	Given a graph $G$ and rules $\rho = \rle{L}{l}{K}{r}{R}$ and $\rho' = \rle{L'}{l'}{K'}{r'}{R'}$ such that $\rho'$ is a derived rule of $\rho$. 
	Then, $\rho'$ is only applicable at match $m': L' \inj G$ if $\rho$ is applicable at match $m' \circ i_L^{L'}$ where $i_L^{L'}$ is the morphism shown in Figure \ref{fig_dpo_construction}.
\end{lemma}
\begin{proof}
	Consider the transformation composed of the pushouts $(3)$ and $(4)$ given in Figure \ref{fig_derived_rules_applcation}.
	Since $\rho'$ is a derived rule of $\rho$, the squares $(1)$ and $(2)$ are pushouts. 
	Therefore, the squares $(1)+(3)$ and $(2)+(4)$ are also pushouts \cite{hartmut2006fundamentals}. It follows that $\rho$ is applicable at match $m' \circ i_L^{L'}$.
\end{proof}
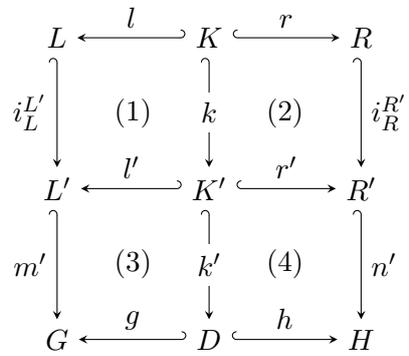
\begin{figure}
\center

	\begin{tikzpicture}
		\node (L) at (0,2) {$L$};
		\node (K) at (2,2) {$K$};
		\node (R) at (4,2) {$R$};
		\node (G) at (0,0) {$L'$};
		\node (D) at (2,0) {$K'$};
		\node (H) at (4,0) {$R'$};
		\node (1) at (1,1) {(1)};
		\node (2) at (3,1) {(2)};
		\node (3) at (1,-1) {(3)};		
		\node (4) at (3,-1) {(4)};
		\node (G') at (0, -2) {$G$};
		\node (D') at (2, -2) {$D$};		
		\node (H') at (4, -2) {$H$};		
		
		\draw[left hook-stealth] (K) edge node [above] {$l$}  (L); 
		\draw[right hook-stealth] (K) edge node [above] {$r$}  (R); 
		\draw[left hook-stealth] (D)   edge node[above]{$l'$}(G); 
		\draw[right hook-stealth] (D)  edge node [above]{$r'$}(H);
		\draw[left hook-stealth] (K) edge node[fill = white] {$k$} (D);  
		\draw[left hook-stealth] (L) -- node[left]{$i_L^{L'}$} (G);
		\draw[left hook-stealth] (R) edge node [right] {$i_R^{R'}$} (H);    
		\draw[left hook-stealth] (D') edge node [above] {$g$}  (G');
		\draw[right hook-stealth] (D') edge node [above] {$h$}  (H');
		\draw[left hook-stealth] (G) -- node[left]{$m'$} (G');
		\draw[left hook-stealth] (D) edge node[fill = white]{$k'$} (D');
		\draw[left hook-stealth] (H) -- node[right]{$n'$} (H');
	\end{tikzpicture}
	
	\caption{Pushout diagram for the construction of basic increasing rules.}\label{fig_derived_rules_applcation}
\end{figure}
Therefore extending a rule set $\mathcal{R}$ by the set of all derived rules for each rule of $\mathcal{R}$ does not extend the semantic of $\mathcal{R}$.
The main idea of the concept of derived rules is to extend a given set of rules by as many basic increasing rules as possible without extending its semantics, and to use the less complex and less restrictive basic application conditions instead of the general ones.

%

In transformations via a rule $\rho$ such that the match intersects an occurrence of a universally bound graph $C_{k+2}$, $\rho$ can be replaced by a derived rule of $\rho$ at layer $k$. 

\begin{lemma}
Given a constraint $c$ in UANF and a rule $\rho = (\ac, \rle{L}{l}{K}{r}{R})$.
	Then, for each transformation $$t: G \Longrightarrow_{\rho,m}H$$ such that an occurrence $p:C_{k+2} \inj G$ of a universally bound graph $C_{k+2}$ with $p(C_{k+2}) \cap m(L) \neq \emptyset$ exists, there is a transformation $$t': G \Longrightarrow_{\rho', m'}H$$ where $\rho'$ is a derived rule of $\rho$ at layer $k$.
\end{lemma}

\begin{proof}
	Since $p(C_{k+2}) \cap m(L) \neq \emptyset$ there is an overlap $P \in \overlay(C_{k+2},L)$ such that there exists a morphism $q: P \inj G$ with $m = q \circ i_L^P$ and $p = q \circ i_{C_{k+2}}^P$. 
	Since $t$ exists $\rho$ and $m$ satisfy the dangling edge condition. One sees easily that $GP_{\rho,m} = q(GP_{\rho, i_L^P})$ and $DP_{\rho,m} \subseteq q(DP_{\rho,i_L^P})$. It follows that $DP_{\rho, i_L^P} \subseteq GP_{\rho,i_L^P}$ and therefore, there is a transformation $t:P \Longrightarrow_{\rho, i_L^P} R'$. In particular, there is a derived rule $\rho' = (\ac', 
	\rle{L'}{l'}{K'}{r'}{R})$ of $\rho$ at layer $k$, where $L' = P$.
	We set $m' = q$, since $m = m' \circ i_L^{L'} \models \ac$, it follows that $m' = \ac'$, and since $\rho'$ removes and inserts the same elements as $\rho$, there is the transformation $t': G \Longrightarrow_{\rho',m'} H$. 
\end{proof}

This allows us to replace consistency-increasing transformations via a direct con\-sis\-ten\-cy-maintaining rule $\rho$ at layer $k$ by a rule derived from $\rho$ at layer $k$, i.e. a basic increasing rule at layer $k$.

\subsection{Application Conditions for Basic Rules}
Let us now introduce the application conditions for basic increasing rules.
Since basic rules are direct consistency-maintaining at a certain layer $k$ it is sufficient to check whether $m \circ i \not\models \exists(C_{k+3}, \true)$ if $\rho$ is a deleting rule, and whether $m \circ i \not\models \exists(C', \true)$ if $\rho$ is an inserting rule, where $m$ is the match of the transformation and $i$ is the increasing morphism of $\rho$.

\begin{definition}[\textbf{application conditions for basic increasing rules}]\label{def_basic_ap}
Given a constraint $c$ in UANF and a basic increasing rule  $\rho = (\ac,L \xhookleftarrow{l} K \xhookrightarrow{r} R)$ w.r.t. $c$ at layer $-1 \leq k\leq \nlvl(c)-2$, where $k$ is odd.
The \emph{basic application condition of $\rho$ w.r.t. $c$ at layer $-1 \leq j \leq \nlvl(c)-2$} is given by 
$$\ac' = \ac \wedge \apb{j}{\rho}$$
with 
$$\apb{j}{\rho} := \begin{cases}
						\bigwedge_{P \in \eol(L,a,i)} \neg \exists(i_L^P: L 								\inj P,\true) &\text{if j = k and $k < \nlvl(c) -2$} \\
						\true &\text{if $k = \nlvl(c)-2$} \\
						\false &\text{otherwise}
				   \end{cases}
$$
where $a= a_{k+2}$, if $\rho$ is a deleting rule, $a = a_{k+2}^r: C_{k+2} \inj C'$ if $\rho$ is an inserting rule with $C'$ and $i$ is the increasing morphism of $\rho$.
\end{definition}

These application conditions are much easier to construct and smaller than those constructed by Definition \ref{def_appl_cond}. 
Note that in the case of an inserting rule $\rho$ which inserts an intermediate graph $C$, the application condition only checks whether the increasing morphism does not satisfy $\exists(C, \true)$. But an application of this rule could also lead to a consistency increasing transformation w.r.t. $c$ if the increasing morphism satisfies $\exists(C, \true)$ and another intermediate graph $C'$ is inserted. To check this, conditions similar to those constructed via Definition \ref{def_appl_cond} must be constructed.
At first sight, this seems like a restriction, but via the notion of derived rules, we are able to dissolve this restriction, since the set of derived rules of $\rho$ will contain an inserting basic increasing rule with $C'$, so that this rule, equipped with the corresponding basic application condition, can be used to perform this consistency-increasing transformation. For example, consider the rule \texttt{assignFeature} and constraint $c_1$ given in Figure \ref{fig:constraints}, there is a consistency increasing transformation $t: C_2^2 \Longrightarrow_{\texttt{assignFeature},m} C_2^1$ such that $m \not \models \apb{-1}{\texttt{assignFeature}}$, but there is also a transformation $t: C_2^2 \Longrightarrow_{\texttt{assignFeature3},m'} C_2^1$ with 
$m' \models \apb{-1}{\texttt{assignFeature3}}$.

Let us now show that basic increasing rules equipped with the application condition constructed by Definition \ref{def_basic_ap} are indeed direct consistency increasing rules at layer.
\begin{theorem}
	Given a constraint $c$ in UANF and a basic increasing rule $\rho = (\ac, 
	\rle{L}{l}{K}{r}{R})$ w.r.t $c$ at layer $-1 \leq k \leq \nlvl(c)-2$, where $k$ is odd. 
	
	Then, $\rho' = (\ac \wedge \apb{k}{\rho}, \rle{L}{l}{K}{r}{R})$ is a direct 
	consistency increasing rule at layer $k$.
\end{theorem}

\begin{proof}
	Given a graph $G$ with $\kmax = k$. We show that each transformation 
	$t: G \Longrightarrow_{\rho', m}H$ is direct consistency increasing w.r.t. $c$.
	Since, $\rho'$ is a basic increasing rule at layer $k$, $\rho'$ is also a consistency maintaining transformation at layer $k$ and $t$ satisfies the no new violation by deletion, no new violation by insertion, no satisfied layer reduction by insertion and no satisfied layer reduction by deletion formulas.
Therefore, we only need to show that $t$ satisfies the special increasing or general increasing formula respectively.
	\begin{enumerate}
		\item 
			If $\rho'$ is a deleting rule, $r \circ l^{-1} \circ i$ is not 
			total, where $i$ is the increasing morphism of $\rho'$. 
			If $k = \nlvl(c)-2$, the transformation satisfies
			the special increasing formula, since one occurrence of $C_{k+2}$ is removed.
			If $k < \nlvl(c)-2$, since $m \models \apb{k}{\rho}$, the morphism $m \circ i$ does not satisfy $\exists (C_{k+3}, \true)$. Since this occurrence is destroyed, $t$ satisfies the general increasing formula.
		\item
			If $\rho'$ is an inserting rule with $C' \in \ig{C_{k+2}}{C_{k+3}}$,
			then $k \leq \nlvl(c)-2$.
			The morphism $m \circ i$ does not satisfy $\exists(C', \true)$, since 
			$m \models \apb{k}{\rho}$. Since $\rho'$ is a inserting rule it holds that $\track_t \circ m 
			\circ i = n \circ r \circ l^{-1} \circ i \models \exists(C', \true)$ and therefore $t$ satisfies 
			the general increasing formula.  
	\end{enumerate}
	In summary, $\rho'$ is a basic direct consistency increasing rule at layer $k$ w.r.t. $c$.
\end{proof}

\begin{figure}
	\centering
	\includegraphics[scale=0.8]{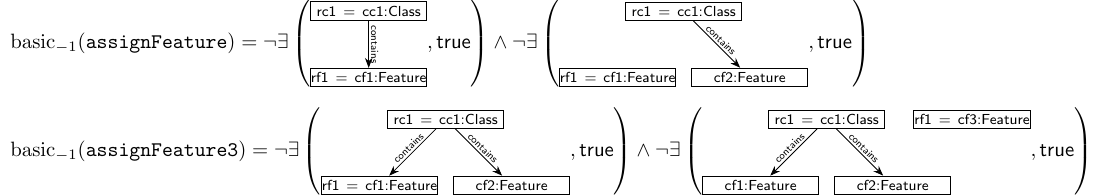}

	\caption{Basic application condition for \texttt{assignFeature} and \texttt{assignFeature3} with $c_1$ at layer $1$.}\label{fig:example_basic_ap}
\end{figure}

\begin{example}
	Again, consider the rule \emph{\texttt{assignFeature}}, its derived rule 
	\emph{\texttt{assignFeatur3}} and $c_1$.
	The basic application condition for these rules at layer $-1$ w.r.t. $c_1$ 
	is given in Figure  \ref{fig:example_basic_ap}.
\end{example}

\section{Rule-based Graph Repair}\label{repair}

In the following, we present our rule-based graph repair approach.
First, we propose a graph repair process for a constraint in UANF, and second, a repair process for a set of constraints in UANF, both based on a given set of rules $\mathcal{R}$. In addition, we need to make further assumptions for these constraints and sets of constraints, namely that they are \emph{circular conflict free}, in order to guarantee that our approach terminates. 
Intuitively, a constraint is \emph{circular conflict free} if, during a repair of an occurrence of a universally bound graph $C_k$, no new occurrences of $C_k$ that do not satisfy $\exists(C_{k+1}, \true)$ are inserted, and any occurrences of $C_k$ that satisfied $\exists(C_{k+1}, \true)$ also satisfy $\exists(C_{k+1}, \true)$ after the repair. 
A set of constraints $\mathcal{C}$ is \emph{circular conflict free} if there exists a sequence $c_1, \ldots, c_n$ with $c_i \in \mathcal{C}$ such that repairing a constraint $c_i$ does not destroy the satisfaction of $c_j$ for all $j < i$.

\subsection{Conflicts within Conditions}
During a repair process, inserting elements of an existentially bound constraint $C_j$ could also insert new occurrences of universally bound graphs $C_i$.
This insertion is unproblematic if $i > \kmax+2$, but if $i \leq \kmax+2$ it could lead either to the insertion of new violations or a reduction of the largest satisfied layer.
In addition, removing elements of a universally bound graph $C_j$ may destroy occurrences of an existentially bound graph $C_i$.
Again, this can lead to the insertion of new violations or a reduction of the largest satisfied layer.

We will now introduce the notion of \emph{conflicts within conditions}, which states that $C_j$ has a conflict with $C_i$ if and only if one of the cases described above can occur. 
Note that conflicts can only occur between existentially and universally bound graphs, and vice versa. There cannot be a conflict between two existentially bound or two universally bound graphs, since the insertion of elements cannot destroy occurrences of existentially bound graphs, and the removal of elements cannot insert new occurrences of universally bound graphs.

\begin{definition}[\textbf{conflicts within conditions}]\label{def_conflicts}
	Given a condition $c$ in UANF.
	An existentially bound graph $C_k$ \emph{causes a conflict for} a universally bound 
	graph $C_j$ if there is a transformation $t: G \Longrightarrow_{\rho} H$ with 
	$\rho = \rle{C_{k-1}}{\id}{C_{k-1}}{a_{k-1}}{C_k}$ such that
	$$\exists p:C_j \inj H(\neg \exists 	q:C_j \inj G(\track_t \circ q = p)).$$
	A universally bound graph $C_j$ \emph{causes a conflict for} an existentially bound 
	graph $C_k$ if there is a transformation $t: G \Longrightarrow_{\rho} H$ with 
	$\rho = \rle{C_j}{a_{j-1}^r}{C}{\id}{C}$ for any $C \in \ig{C_{j-1}}{C_j}$ 
	such that
	$$\exists p: C_k \inj G(\track_t \circ p \text{ is not total}).$$

\end{definition}

In addition, we introduce \emph{conflict graphs}, which represent the conflicts within a constraint via a graph.
With these, we are able to define \emph{transitive conflicts}, \emph{circular conflicts} and their absence, which will be a necessary property for the termination of our repair process.
Intuitively, as the name suggests, a condition $c$ contains a circular conflict if a graph $C_k$ has a conflict with itself or if there exists a sequence $C_k= C_{j_1}, \ldots, C_{j_n}= C_k$ of graphs such that $C_{j_i}$ has a conflict with $C_{j_{i+1}}$. We can check this property by checking whether the 
conflict graph contains cycles.
Note that conflict graphs contain additional edges that do not correspond to the conflicts within the constraint. 
These edges ensure that during repair it is possible to choose whether a violation is removed by deletion or insertion. Otherwise, it must be done alternately. That is after a violation is removed by deletion, all violations introduced by that deletion must be removed by insertion, and vice versa. The absence of these additional edges would also lead to a more restrictive definition of repairing sets. 
Intuitively, if a graph $C_k$ causes a conflict for $C_j$, the conflict graph of that constraint contains an edge from a node labelled $k'$ to a node labelled $j'$ if $C_k$ is the domain or co-domain of the morphism $a_{k'}$, $C_j$ is the domain or co-domain of the morphism $a_{j'}$ and $k' \neq j'$.

\begin{figure}
\centering
\includegraphics[scale=0.85]{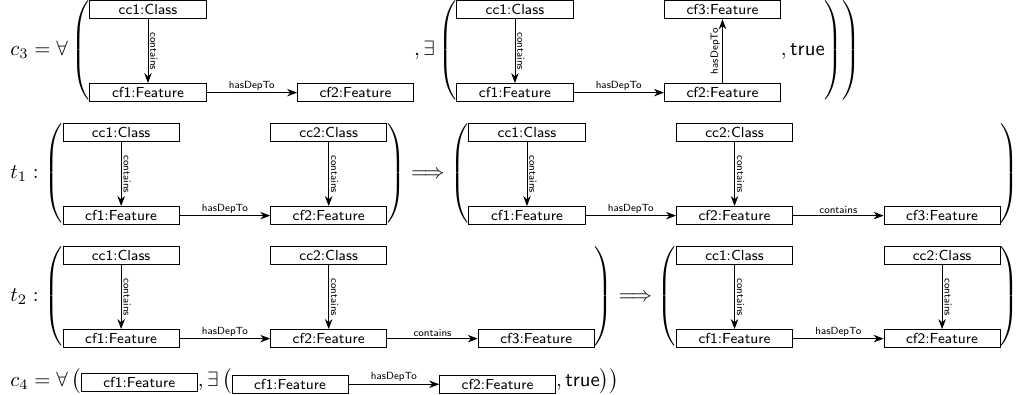}

\caption{Constraint $c_3$ and the transformation that show the existence of 
conflicts between $C_1$ and $C_2$ and $C_2$ and $C_1$.}\label{fig:conflict_example}
\end{figure}

\begin{figure}
\centering
\includegraphics[scale=1]{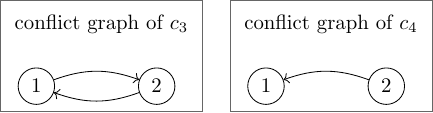}

\caption{Conflict graphs of $c_3$ and $c_4$.}\label{fig:conflict_graph}
\end{figure}

\begin{definition}[\textbf{conflict graph, circular conflicts}]\label{conflicts_within}
	Let a condition $c$ in UANF be given. The \emph{conflict graph of $c$} is 
	constructed in the following way. 
	For every $0 \leq k < \nlvl(c)$ there is a node labelled $k$.	If  $C_k$ causes a conflict for $C_j$, there is an edge $e$ with $\src(e) = k'$ and $\tar(e) = j'$ if either $k = k'$ or $k = k'+1$, either $j = j'$ or $j = j'+1$ and $j' \neq k'$.
	
	A graph $C_k$ causes a \emph{transitive conflict} with $C_j$ if there exists a path from $k$ to $j$ in the conflict graph of $c$.
	A graph $C_k$ has a \emph{circular conflict} if $C_k$ has a transitive conflict with itself.
A condition $c$ is called \emph{circular conflict free} if $c$ does not contain a circular conflict.
\end{definition}
In other words, a condition $c$ is \emph{circular conflict free} if its conflict graph is acyclic.
\begin{example}
	Consider constraint $c_3$ and the transformations $t_1$ and $t_2$ shown in 
	Figure \ref{fig:conflict_example}.
	Transformation $t_1$ shows that $C_1$ has a conflict with $C_2$ because the
	rule $\rho = \rle{C_1}{\id}{C_1}{a_1}{C_2}$ has been applied and there is a newly inserted occurrence of $C_1$.
	Transformation $t_2$ shows that $C_2$ has a conflict with $C_1$, since the 
	rule $\rle{C_2}{a_1}{C_1}{\id}{C_1}$ has been applied and one occurrence 
	of $C_1$ has been destroyed. 
	So $c_3$ contains a circular conflict, the conflict graph 
	of $c_3$ is shown in Figure \ref{fig:conflict_graph}.
	
	In general, the statement \enquote{ $C_j$ causes a conflict for $C_k$} 
	does not imply that \enquote{ $C_k$ causes a conflict for $C_j$} as shown by 
	constraint $c_4$ given in Figure \ref{fig:conflict_example}.
	The conflict graph of $c_4$ is also shown in Figure \ref{fig:conflict_graph}. 
	It can be seen that $c_4$ is a circular conflict-free constraint.
	
\end{example}

 
We will now present two characterisations of conflicts. The first one is based on the applicability of certain rules and the second one is based on the notion of basic maintaining rules.
 
For $C_k$, which is existentially bound, and $C_j$, which is universally bound, the first characterisation checks whether for each overlap of $C_k$ and $C_j$, such that the overlap morphisms restricted to $C_k \setminus C_{k+1}$ and $C_j$ overlap, the rule that only deletes $C_k\setminus C_{k-1}$ is applicable.
If this is not possible, there is no transformation as described in Definition \ref{conflicts_within}.
If $C_j$ is universally bound and $C_k$ is existentially bound, the characterisation checks whether for each overlap of $C_k$ and $C_j$ such that 
the elements of $C_j \setminus C_{j-1}$ and $C_k$ overlap, a rule is applicable which only removes elements of $C_j \setminus C_{j-1}$. 
Again, if this is not possible, there is no transformation as described in Definition \ref{conflicts_within}.

\begin{lemma}\label{charact_conflict}
Given a constraint $c$ in UANF.
\begin{enumerate}
	\item 
		Let $C_k$ be an existentially bound and $C_j$ a universally bound graph of 
		$c$. Then, $C_k$ causes a conflict for $C_j$, if and only if there is an overlap $P 
		\in \overlay(C_k,C_j)$ with 
		$$ i_{C_k}^P(C_k\setminus C_{k-1}) 
		\cap i_{C_j}^P(C_j) \neq \emptyset$$ 
		and the rule $\rho = \rle{C_k}{a_{k-1}}{C_{k-1}}{\id}{C_{k-1}}$ is 
		applicable at match $i_{C_k}^P$.

	\item 
		Let $C_j$ be a universally bound and $C_k$ an existentially bound graph of 
		$c$. Then, $C_j$ causes a conflict with $C_k$ if an only if there is an overlap $P \in \ig{C_j}{C_k}$ with 
		$$i_{C_j}^P(C_j \setminus C_{j-1}) \cap i_{C_k}^P(C_k) 
		\neq \emptyset$$ 
		and a rule $\rho = \rle{C_j}{a_{j-1}^r}{C}{\id}{C}$ with $C \in 
		\ig{C_{j-1}}{C_j}$ and $i_{C_j}^P(C_k \setminus C) \cap i_{C_k}^P(C_k 
		)\neq \emptyset $ is applicable at match $i_{C_j}^P$.

\end{enumerate}
\end{lemma}

\begin{proof}
Given a condition $c$ in UANF.
\begin{enumerate}
\item 
\enquote{$\Longrightarrow$}: Let $C_k$ be an existentially bound graph that causes a conflict for a universally bound graph $C_j$. 
Then, there is a transformation $t: G \Longrightarrow_{\rho,m} H$ with $\rho = \rle{C_{k-1}}{\id}{C_{k-1}}{a_{k-1}}{C_k}$ such that a new occurrence $p$ of $C_j$ is inserted. 
Since only elements of $C_k \setminus C_{k-1}$ are inserted, it holds that $p(C_j) \cap n(C_k\setminus C_{k-1}) \neq \emptyset$, where $n$ is the co-match of $t$. 
The overlap $(P, p,i_{C_k}^P)$ with $P = p(C_j) \cup n(C_k)$ and $i_{C_k}^P = n$ is the one we are looking for. Since $\rho^{-1}$ is applicable at match $n$, it holds that $DP_{\rho^{-1},n} \subseteq GP_{\rho^{-1},n}$. It also holds that $GP_{\rho^{-1},n} = GP_{\rho^{-1},i_{C_k}^P}$ and $DP_{\rho^{-1}, i_{C_k}^P} \subseteq  DP_{\rho^{-1}, n}$. It follows that $DP_{\rho^{-1}, i_{C_k}^P} \subseteq GP_{\rho^{-1},n} = GP_{\rho^{-1},i_{C_k}^P}$ and therefore, the rule $\rho^{-1}$ is applicable at the match $i_{C_k} ^P$. 
\\
\enquote{$\Longleftarrow$}:
Let $C_k$ be an existentially and $C_j$ a universally bound graph such that there exists an overlap $P \in \overlay(C_k,C_j)$ with $i_{C_k}^P(C_k\setminus C_{k-1}) \cap i_{C_j }^P(C_j\setminus C_{j-1}) \neq \emptyset$ so that the rule $\rho = \rle{C_{k}}{a_{k-1}}{C_{k-1}}{\id}{C_{k-1}}$ is applicable at match $i_{C_k}^P$.
Then the inverse transformation of $t: P \Longrightarrow_{\rho,i_{C_k}^P} H$ is the transformation we are looking for and $C_k$ causes a conflict for $C_j$. 

\item
\enquote{$\Longrightarrow$}: Let $C_j$ be a universally bound graph that causes a conflict for an existentially bound graph $C_k$. 
Then, there is a transformation $t: G \Longrightarrow_{\rho,m} H$ with $\rho = \rle{C_j}{a_{j-1}^r}{C}{\id}{C}$ and $C \in \ig{C_{j-1}}{C_j}$ such that $\track_t \circ p$ is no total for an occurrence $p : C_k \inj G$. 
The overlap $(P, i_{C_k}^P, i_{C_j}^P)$ with $P = p(C_k) \cup m(C_j)$, $i_{C_k}^P = p$ and $i_{C_j}^P =m$  is the one we are looking for and  $i_{C_j}^P(C_j \setminus C) \cap i_{C_k}^P(C_k) \neq \emptyset$ must hold since $\rho$ only deletes elements of $C_j\setminus C$.
Since $\rho$ is applicable at match $m$, it holds that $DP_{\rho,m} \subseteq GP_{\rho,m}$. It also holds that $GP_{\rho,m} = GP_{\rho,i_{C_j}^P}$ and $DP_{\rho,i_{C_j}^P} \subseteq DP_{\rho,m}$. It follows that $DP_{\rho,i_{C_j}^P} \subseteq GP_{\rho,i_{C_j}^P}$ and therefore, $\rho$ is applicable at match $i_{C_j}^P$.
\\
\enquote{$\Longleftarrow$}:
Let $C_j$ be universally and $C_k$ existentially bound such that there is an  overlap $P \in \overlay(C_j,C_k)$ with $i_{C_j}^P(C_j \setminus C_{j-1}) \cap i_{C_k}^P(C_k) \neq \emptyset$  such that a rule $\rho = 	\rle{C_j}{a_{j-1}^r}{C}{\id}{C}$ with $C \in \ig{C_{j-1}}{C_j}$ is applicable at match $i_{C_j}^P$.
Then, the transformation of $t: P \Longrightarrow_{\rho, i_{C_j}^P} H$ is the transformation we are looking for and $C_j$ causes a conflict for $C_k$.
\end{enumerate}
\end{proof}

Note that this characterisation can also be expressed via the notions of \emph{conflicts between rules } and  \emph{parallel independency} \cite{lambers2006conflict}.
Using these notions, the first part of Lemma \ref{charact_conflict} can be expressed as: An existentially bound graph $C_k$ causes a conflict for a universally bound graph $C_j$ if and only if the rules $\rho = \rle{C_k}{a_{k-1}}{C_{k-1}}{\id}{C_{k-1}}$ and $\rle{C_j}{\id}{C_{j}}{\id}{C_{j}}$ are parallel independent. 
The second part can be expressed as: A universally bound graph $C_j$ causes a conflict for an existentially bound graph $C_j$ if and only if the rules $\rle{C_j}{a_{j-1}^r}{C}{\id}{C}$ and $\rle{C_k}{\id}{C_{k}}{\id}{C_{k}}$ are parallel independent for all $C \in \ig{C_{j-1}}{C_j}$.

Our second characterisation of conflicts is based on the notion of basic maintaining rules. 
\begin{lemma}\label{basic_conflict}
	Let a condition $c$ in UANF be given. 
	\begin{enumerate}
		\item 
			Let $C_k$ be an existentially and $C_j$ be a universally bound graph 
			of $c$. Then, $C_k$ causes a conflict for $C_j$ if and only if  
			the rule $\rho = \rle{C_{k-1}}{\id}{C_{k-1}}{a_{k-1}}{C_k}$
			is not a basic consistency maintaining rule w.r.t. 
			$\forall(a_{j-1} \circ \ldots 
			\circ a_0: C_0 \inj C_j, \false)$.
		\item
			Let $C_j$ be a universally and $C_k$ be an existentially bound graph 
			of $c$. Then, $C_j$ causes a conflict for $C_k$ if and only if a
			rule
			$\rho = \rle{C_j}{a_{j-1}^r}{C}{\id}{C}$ with $C \in 
			\ig{C_{j-1}}{C_j}$ is not a basic consistency  maintaining rule w.r.t. 
			$\exists(a_{k-1}\circ \ldots \circ a_0: C_0 \inj C_k, \true)$.
	\end{enumerate}
\end{lemma}
\begin{proof}
	\begin{enumerate}
		\item 
			Let $C_k$ be an existentially and $C_j$ a universally bound graph of 
			$c$.
			
			\enquote{$\Longrightarrow$}: Assume that $C_k$ causes a conflict for
			$C_j$. Therefore, there is a transformation $t: G 
		\Longrightarrow_{\rho} H$ with $\rho = \rle{C_{k-1}}{\id}{C_{k-1}}
			{a_{k-1}}{C_k}$ such that a new occurrence $p: C_j \inj H$ has been 
			inserted. 
			Then, $t$ does not satisfy the no satisfied layer reduction by insertion formula and
			$\rho$ is not a basic maintaining rule w.r.t.$\forall(a_{j-1} \circ \ldots 
			\circ a_0: C_0 \inj C_j, \false)$ .
			
			\enquote{$\Longleftarrow$}: Assume that $\rho = \rle{C_{k-1}}{\id}
			{C_{k-1}}{a_{k-1}}{C_k}$ is not a basic maintaining rule w.r.t. $\forall(a_{j-1} \circ \ldots \circ a_0: C_0 \inj C_j, 
			\false)$. Since this constraint only contains universally 
			bound graphs, there must exist a transformation $t: G 
			\Longrightarrow_{\rho} H$ that does not satisfy 
			the no satisfied layer reduction by insertion formula. Therefore, a new occurrence of $C_j$ 
			has been inserted by $t$ and with Definition \ref{def_conflicts} 
			follows that $C_k$ causes a conflict for $C_j$.
			
		\item 
			Let $C_j$ be a universally and $C_k$ be an existentially bound graph 
			of $c$ and $c' = \exists(a_{k-1}\circ \ldots \circ a_0: C_0 \inj C_k, 
		\true)$.
			
			\enquote{$\Longrightarrow$}: Assume that $C_j$ causes a conflict for 
			$C_k$. Therefore, there is a transformation $t: G 
			\Longrightarrow_{\rho} H$ with $\rho = \rle{C_j}{a_{j-1}^r}{C}{\id} 
			{C}$, for a $C \in \ig{C_{j-1}}{C_j}$ such that an occurrence of 
			$C_k$ has been destroyed. Then, $t$ does not satisfy 
			the no satisfied layer reduction by insertion formula.
			Therefore, $\rho$ is not a basic consistency-maintaining rule 
			w.r.t. $c'$. 
			
			\enquote{$\Longleftarrow$}: Assume that $\rho = \rle{C_j}{a_{j-1}^r}
			{C}{\id} {C}$ is a not a basic increasing rule w.r.t. $c'$.  
			Therefore, there is transformation $t: G \Longrightarrow_{
			\rho} H$ that does not satisfy the no satisfied layer reduction by insertion formula and an occurrence of $C_k$ has been removed by $t$.
			It follows that  $C_j$ causes a conflict for $C_k$.

	\end{enumerate}
\end{proof}

\subsection{Repairing rule Sets}

Given a set of rules and a constraint, it is unclear whether or not it is possible to repair a graph using the rules of that set. 
Therefore, we introduce the notion of \emph{repairing rule sets}, which is a characterisation of rule sets that are able to repair a graph w.r.t. a circular conflict free constraint.
First, we introduce the notion of \emph{repairing sequences}. 
A repairing sequence is a sequence of rule applications that either destroys an occurrence of a universal or inserts an occurrence of an existentially bound graph, and is applicable to each occurrence of these respective graphs. 
To ensure that these sequences are applicable to each occurrence, it is necessary to ensure that no nodes of these occurrences are removed and that the left-hand side of the first rule of the repairing sequence is contained in that occurrence. 
In other words, every repairing sequence of $C_k$ starts with a transformation
originating in $C_k$ if $C_k$ is universally bound and $C_{k-1}$ if 
$C_k$ is existentially bound.

\begin{definition}[\textbf{repairing sequence}]\label{rep_sequence}
	Let a constraint $c$ in UANF and a set of rules $\mathcal{R}$ be given. 
	\begin{enumerate}
		\item 
			If $C_k$ is existentially bound, a sequence of transformations 
			$$C_{k-1} = G_0 \overset{t_1}{\Longrightarrow}_{\rho_1,m_1} G_1 
			\overset{t_2}{\Longrightarrow}_{\rho_2,m_2} \ldots 
			\overset{t_n}{\Longrightarrow}_{\rho_n, m_n} 
			G_n$$
			with plain rules $\rho_i \in \mathcal{R}$ is called a \emph{repairing sequence 
			for $C_k$} if $G_n \models_{k}c$, $\track_{t_i} \circ \ldots \circ 
			\track_{t_1} \circ \id_{C_{k-1}}$ is total for all $1 \leq i \leq n$. And for each universally bound graph $C_j$ such that $C_k$ does not cause a conflict for $C_j$,  the concurrent rule 
			of this sequence is a basic consistency-maintaining rule w.r.t.
			$\forall(C_j, \false)$. 
		\item	
			If $C_k$ is universally bound, a sequence of transformations  
			$$C_k = G_0 \overset{t_1}{\Longrightarrow}_{\rho_1,m_1} G_1 
			\overset{t_2}{\Longrightarrow}_{\rho_2,m_2} \ldots 
			\overset{t_n}{\Longrightarrow}_{\rho_n, m_n} 
			G_n$$
			with plain rules $\rho_i \in \mathcal{R}$ is called a \emph{repairing sequence 
			for $C_k$} if $G_n \models_k c$, for each node $v \in V_{G_0}$ there 
			is a node $v' \in V_{G_n}$ with $v' = \track_{t_n}(\ldots 
			\track_{t_1}(v))$ and the concurrent rule 
			of this sequence  
			is a basic consistency-maintaining 
			rule w.r.t. $\forall(C_j, \false)$ for all universally bound graphs 
			$C_j$.
	\end{enumerate}
\end{definition}
Note that this definition prohibits the deletion of nodes of $V_G$. If nodes are deleted, there is no guarantee that the repairing sequence is applicable at every occurrence of the corresponding graph since the dangling edge condition may be violated.
Also note that a repairing sequence for a universally bound graph $C_j$ can only insert occurrences of an existentially bound graph $C_k$ if $C_j$ causes a conflict for $C_j$.
 
\begin{figure}
\centering
\includegraphics[scale=0.7]{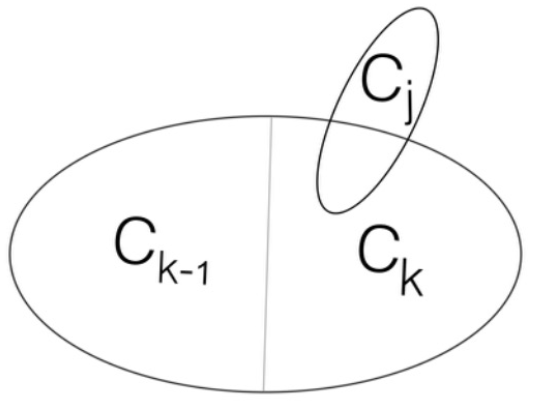}

\caption{Sketch of an overlap of an existentially bound graph $C_k$ and a universally bound graph $C_j$ that could lead to an insertion of $C_j$ via repairing sequences.}\label{fig:example_univ}
\end{figure}

In both cases of Definition \ref{rep_sequence} the insertion of additional elements, i.e. $G_n \neq C_{k+1}$ if $C_k$ is existentially bound and $G_n \neq C$ for all $C \in \ig{C_{k-1}}{C_k}$ if $C_k$ is universally bound, could lead to the insertion of universally bound graphs. 
For an existentially bound graph, this can happen if there is an overlap with a universally bound graph in a similar way as shown in Figure \ref{fig:example_univ}. 
To ensure that this does not happen, we need the additional condition that the concurrent rule is a basic consistency maintaining rule with respect to certain constraints.
If $G_n = C_{k+1}$ if $C_k$ is existentially bound or $G_n = C$ with $C \in \ig{C_{k-1}}{C_k}$ if $C_k$ is universally bound, this condition is not needed as  the following theorem shows.

\begin{theorem}\label{special_rep_sequence}
	Let a constraint $c$ in UANF and a set of rules $\mathcal{R}$ be given. 
	\begin{enumerate}
		\item 
			If $C_k$ is existentially bound and there is sequence 
			$$C_{k-1} \Longrightarrow_{\rho_1,m_1} \ldots 
			\Longrightarrow_{\rho_n,
				m_n} C_k$$ 
			with plain rules $\rho_i \in \mathcal{R}$ such that $\track_{t_i} \circ 
			\ldots \track_{t_1} \circ 
			 \id_{C_{k-1}}$ is 
			total for all $1 \leq i \leq n$ and $C_k \models_{k}c$. Then, this is a repairing sequence 
			for $C_k$. 
		\item
			If $C_k$ is universally bound and there is  a sequence 
			$$C_k \Longrightarrow_{\rho_1,m_1} \ldots \Longrightarrow_{\rho_n, 	
			m_n} C$$ 
			with plain rules $\rho_i \in \mathcal{R}$, $C \in \ig{C_{k-1}}{C_k}$ such that $C\models_k c$ and for each 
			node $v \in V_{G_0}$ there is a node $v' \in V_{G_n}$ with 
			$v' = \track_{t_n}(\ldots \track_{t_1}(v))$.
			Then, this is a repairing sequence for $C_k$.
	\end{enumerate}
\end{theorem} 
\begin{proof}
	\begin{enumerate}
		\item 
			If $C_k$ is existentially bound, we need to show that the concurrent
			$\rho = \rle{C_{k-1}}{\id}{C_{k-1}}{a_k}{C_k}$ is a basic consistency maintaining rule w.r.t. $\forall(C_j, \false)$ for all universally bound $C_j$ such that $C_k$ does not cause a conflict with $C_j$.
			Assume that $C_j$ is a universally bound graph such that $C_k$ does not cause a 
			conflict for $C_j$ and  $\rho$ is not a 
			basic consistency maintaining rule w.r.t. $\forall(C_j, \false)$.
			With Lemma \ref{basic_conflict} follows immediately that 
			$C_k$ causes a conflict for $C_j$, this is a contradiction.
		\item
			If $C_k$ is universally bound, the concurrent rule is given  by 
			$\rho = \rle{C_k}{a_{k-1}^r}{C}{\id}{C}$. 
			Then, $\rho$ is a basic consistency maintaining rule w.r.t. 
			$\forall(C_j, \false)$ for all universally bound graphs since 
			$\rho$ does not insert any elements. 
	\end{enumerate}
\end{proof}
 
Note that this characterisation is similar to the construction of repairing rules in \cite{sandmann2019rule}. 
Let us now show that a repairing sequence for a graph $C_k$ of a constraint $c$ is applicable at each occurrences of $C_k$. 

\begin{theorem}
	Given a graph $G$, a constraint $c$ in UANF and and a graph $C_k$ of $c$ with $0 \leq k \leq \nlvl(c)$. 
	\begin{enumerate}
		\item If $C_k$ is existentially bound. Then, each repairing sequence 
		$$C_{k-1} = G_0 \overset{t_1}{\Longrightarrow}_{\rho_1,m_1} G_1 
			\overset{t_2}{\Longrightarrow}_{\rho_2,m_2} \ldots 
			\overset{t_n}{\Longrightarrow}_{\rho_n, m_n} 
			G_n$$ for $C_k$ is applicable at every occurrence $p: C_{k-1} \inj G$ of $C_{k-1}$. 
		\item If $C_k$ is universally bound. Then each repairing sequence 
		$$C_k = G_0 \overset{t_1}{\Longrightarrow}_{\rho_1,m_1} G_1 
			\overset{t_2}{\Longrightarrow}_{\rho_2,m_2} \ldots 
			\overset{t_n}{\Longrightarrow}_{\rho_n, m_n} 
			G_n$$
			for $C_k$ is applicable at every occurrence $p: C_k \inj G$ of $C_k$. 
	\end{enumerate}		
\end{theorem}
\begin{proof}
We show that the concurrent rule of the repairing sequence is applicable. Then, the applicability of the repairing sequence follows immediately.
	\begin{enumerate}
		\item If $C_k$ is existentially bound, let a repairing sequence $$C_{k-1} = G_0 \overset{t_1}{\Longrightarrow}_{\rho_1,m_1} G_1 
			\overset{t_2}{\Longrightarrow}_{\rho_2,m_2} \ldots 
			\overset{t_n}{\Longrightarrow}_{\rho_n, m_n} 
			G_n$$ for $C_k$ be given. 
			Since $\track_{t_i} \circ \ldots \circ \id_{C_{k-1}}$ is total for all $0 \leq i \leq n$, the concurrent rule is given by 
			$\rho = \rle{C_{k-1}}{\id} {C_{k-1}}{r}{G_n}$ and $\rho$ does not delete any elements. Therefore, the dangling edge condition cannot be violated since $GP_{\rho,p} = V_{C_{k-1}}$. It follows that $\rho$ and in particular the repairing sequence is applicable at match $p$.
		\item If $C_k$ is universally bound, let $$C_k = G_0 \overset{t_1}{\Longrightarrow}_{\rho_1,m_1} G_1 
			\overset{t_2}{\Longrightarrow}_{\rho_2,m_2} \ldots 
			\overset{t_n}{\Longrightarrow}_{\rho_n, m_n} 
			G_n$$ be a repairing sequence for $C_j$. 
			Since for each node $v \in V_{G_0}$ there 
			is a node $v' \in V_{G_n}$ with $v' = \track_{t_n}(\ldots 
			\track_{t_1}(v))$, the concurrent rule of the sequence is given by $\rho = \rle{C_k}{a^r_{k-1}}{C}{r}{G_n}$
			where $C \in \ig{C_{k-1}}{C_k}$ and $E_C = E_{C_k}$. It follows that $\rho$ does not delete any nodes and therefore, the dangling edge condition cannot be violated since $GP_{\rho,p} = V_{C_k}$. It follows that  $\rho$ and in particular the repairing sequence is applicable at match $p$. 
	\end{enumerate}
\end{proof}
We will now define under which circumstances our repair approach is able to repair a constraint given a rule set $\mathcal{R}$.
\begin{definition}[\textbf{repairing rule set}]
	Let a set of rules $\mathcal{R}$ and a circular conflict free constraint $c$ 
	in UANF be given. 
	Then, $\mathcal{R}$ is called a \emph{repairing rule set for 
	$c$} if there is a repairing sequence for each existentially bound 
	graph of $c$ and, if $\nlvl(c)$ is odd, i.e. $c$ ends with a condition 
	of the form $\forall(C_{\nlvl(c)}, \false)$, $\mathcal{R}$ contains 
	a repairing sequence for $C_{\nlvl(c)}$.
\end{definition}

Note that there cannot exist a repairing sequence for a universally bound graphs $C_k$ such that $C_k \setminus C_{k-1}$ does not contain any edges.
Therefore, there is no repairing set for all constraints of the form $\forall(C_1, \false)$ such that $E_{C_1} = \emptyset$. 

\begin{theorem}
	Let a circular conflict free constraint $c$ in UANF and a repairing set 
	$\mathcal{R}$ of $c$ be given.
	Then, for each graph $G$ with $G \not \models c$, there is a 
	sequence of transformations 
	$$G = G_0 \Longrightarrow_{\rho_1, m_1} \ldots \Longrightarrow_{\rho_n, m_n} 
	G_n$$  
	with $\rho_i \in \mathcal{R}$ such that $G_n \models c$.
\end{theorem}
We will postpone the proof of this Theorem, as it follows immediately from the termination of our repair process.

\begin{figure}
\centering
\includegraphics[scale=0.8]{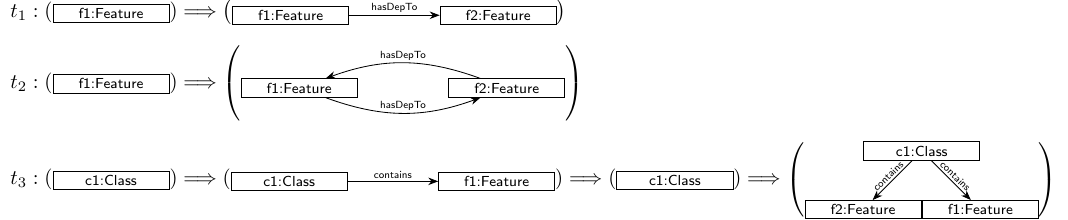}

\caption{Repairing sequences for $c_1$ and $c_4$.}\label{fig:rep_set}
\end{figure}

\begin{example}
	Consider the constraints $c_1$ (Figure \ref{fig:constraints}), $c_4$ (Figure \ref{fig:conflict_example}) and the sequences shown in Figure \ref{fig:rep_set}. The first sequence is not a repairing sequence for the existentially bound graph of $c_4$, since $G_1 \not \models_1 c_4$ and therefore a rule set containing only this rule is not a repairing set for $c_4$.
	The second sequence is a repairing sequence for the existentially bound 
	graph of $c_4$, since the last graph satisfies $c_4$. The existentially bound graph causes a conflict for the universally bound graph. Therefore, the condition for the concurrent rule is also satisfied, and 
	a rule set containing this rule is a repairing set for $c_4$.

	The third sequence is a repairing sequence for $c_1$ since the last graph 
	satisfies $c_1$ and the sequence satisfies the criteria given in Theorem 
	\ref{special_rep_sequence}. Note that this sequence consists of two applications of the same rule. A set of rules containing this rule is a repairing set for $c_1$.
\end{example}

\subsection{Rule-based Graph Repair for one Constraint}

In the following, we present our graph repair process for a circular conflict-free constraint in UANF.
We start with an algorithm that computes $\kmax$ given a graph $G$ and a constraint $c$ in UANF, as shown in Algorithm \ref{Algo_kmax}.  
\SetKw{KwBy}{by}
\begin{figure}
\centering
\begin{algorithm}[H]
	\KwData{A graph $G$, a constraint $c$ in UANF.}
	\KwResult{$\kmax$.}
	
	\If{$G \models c$}{
		\KwRet{$\nlvl(c) -1$};	
	}

	\For{$i\gets -1$ \KwTo $\nlvl(c)-1$ \KwBy $2$ }{
		\If{$G \not \models \cut{i}{c}$}{
			\KwRet{$i-2$}\;
		}
	}

\caption{Determine $\kmax$.}\label{Algo_kmax}

\end{algorithm}
\end{figure}

\begin{theorem}
	Given a graph $G$ and a constraint $c$ in UANF, Algorithm \ref{Algo_kmax} returns $\maxk{c}{G}$.  
\end{theorem}
\begin{proof}
	If $G \models c$, then $\kmax = \nlvl(c)-1$ which will be returned by the algorithm. 
	Otherwise, if $G \not \models c$, Corollary \ref{corol:satisfaction} implies that $G \not \models_k c$ for all even $-1 \leq k < \nlvl(c)$. Therefore, $\kmax$ must be odd. 
	If the algorithm returns $k$, which is odd,  it holds that $G \not \models_{k+2} c$ and Lemma \ref{lem_ex_lower} implies that $k$ must be equal to $\kmax$. 

\end{proof}
The repair process is shown in Algorithm \ref{Algo_conflict_free} and proceeds as follows.
The algorithm starts by finding all potentially increasing occurrences  $p$ of $C_{\kmax +2}$ at layer $\kmax$ w.r.t. $c$. Recall, these occurrences satisfy the following: 
\begin{enumerate}
	\item $p \not \models d$.
	\item $p = a_{\kmax+1} \circ \ldots \circ a_0 \circ q$ where  
	$a_i \circ \ldots \circ q \models \scond{i+1}{\cut{\kmax}{c}}$ for all  $0 \leq i \leq \kmax$ and $q: \emptyset \inj G$ is the empty morphism.
\end{enumerate}
The condition $d$ is equal to $\false$ if 
$\kmax +2 = \nlvl(c)-2$ and equal to $\exists(C_{\kmax +2}, \true)$ otherwise.
All these occurrences are contained in the set $P$ (line 3).  
If $P$ is empty,  Lemma \ref{num_violations} implies $G \models_{\kmax+2} c$, and so we only will apply repairing sequences at occurrences contained in this set.
It may be sufficient to repair only some of these occurrences. 
Since we do not know which of these are likely to increase the 
consistency, we choose one at random (line 4). 
For example, for an existential constraint $c$, i.e. their equivalent constraint in UANF is $\forall(\emptyset,c)$, there may exist occurrences of $C_{\kmax +2}$ whose repair will never lead to an increase of the largest satisfied layer. 

There are two ways to repair the selected occurrence, either by destroying it or by inserting elements such that the occurrence satisfies $\cut{0}{\scond{\kmax+2}{c}}$. The algorithm chooses one of these options (line 5) and applies the appropriate repair sequence 
(lines 6--12). Note that there may be no repair sequence for $C_{\kmax+2}$ since this graph is universally bound. If this is the case, we use the repairing sequence for $C_{\kmax+3}$. This must exist because $C_{\kmax+3}$ is existentially bound and $\mathcal{R}$ is a repairing set for $c$.

If the repairing sequence for  $C_{\kmax+2}$ was applied, occurrences of existentially bound graphs may have been destroyed.
Note that these can only be occurrences of graphs $C_i$ such that $C_{\kmax+2}$ has a conflict with $C_i$.
This could lead to a reduction of the largest satisfied layer. Therefore the algorithm finds all these destroyed occurrences, in particular, it finds all occurrences $p$  of universally bound graphs $C_i$ such that an occurrence $q$ of $C_{i+1}$ with $p = q \circ a_j$ has been removed (line 8).
If the repairing sequence for $C_{\kmax+3}$ has been applied, occurrences of universally bound graphs may have been inserted. 
Again, these can only be occurrences of graphs $C_i$ such that 
$C_{\kmax+2}$ has a conflict with $C_i$. This could lead to a reduction of the largest satisfied layer and the algorithm finds all new occurrences of these universally bound graphs (line 11). In both cases, the algorithm only finds potentially increasing occurrences at the respective layer. 
If the largest satisfied layer has not been reduced, the algorithm chooses the next occurrence in $P$.

Otherwise, the largest satisfied layer must be restored. To do this, all newly inserted potentially increasing occurrences of universally bound graphs and potentially increasing occurrences $p$ of   universally bound graphs $C_j$ such that $p \models \exists(C_{j+1}, \true)$ and  $\track \circ p \not\models \exists(C_{j+1}, \true)$ where $\track$ is the morphism of the repairing sequence application are collected in the set $M$ and must be repaired. Again, it might be sufficient to only repair some of these occurrences in order to restore the largest satisfied layer. 
Since it is unclear for which  of these occurrences a deletion or extension leads to an increase of the largest satisfied layer, we select one at random. 
Repairing these occurrences may again result in the insertion of universally bound graphs or the removal of existentially bound graphs. 
These occurrences are added to $M$, and this process is repeated until 
the largest satisfied layer is restored, i.e. $H \models_{\kmax}c$ (line 13 -- 26).
The whole process is repeated until a graph satisfying $c$ is derived. 

This shows why $c$ must be free of circular conflicts. 
For a constraint with circular conflicts, a new occurrence of $C_{\kmax +2}$ can be inserted and an occurrence of $C_{\kmax +3}$ can be removed during the recovery phase.
In certain cases, this could lead to an infinite loop, so there is no guarantee that this algorithm will terminate.
For example, consider the constraint $c_3$ given in Figure \ref{fig:conflict_example}.
The set of rules used for the transformations $t_1$ and $t_2$ in figure \ref{fig:conflict_example} forms a repairing set. 
During a repair process using Algorithm \ref{Algo_conflict_free}, where the starting graph is the first graph of $t_1$, Algorithm \ref{Algo_conflict_free} can enter an infinite loop by alternately applying $t_1$ and $t_2$.

Optimisation of the repair algorithm in terms of the number of elements inserted or deleted can be achieved by using partial repairing sequences where possible.
For example, consider the repairing sequence
$$ C_k \Longrightarrow C' \Longrightarrow \ldots \Longrightarrow C_{k+1}$$
with $C' \in \ig{C_k}{C_{k+1}}$. 
For an occurrence $p$ of $C_k$, which already satisfies the condition $\exists(C_1, \true)$, it may be sufficient to apply only the sequence
$$C_1 \Longrightarrow \ldots \Longrightarrow C_{k+1}$$ at $p$. 
Then, we need to check that no occurrences of existentially bound graphs have been destroyed, and that no occurrences of universally bound graphs $C_i$ such that $C_k$ does not cause a conflict for $C_i$ have been inserted. This can be achieved by replacing the application of the repairing sequence by an application of its concurrent rule, and by equipping this concurrent rule with its (basic) consistency-increasing application condition of $c$ at layer $k$ with $C_{k+1}$. 
If the application condition is not satisfied, another (partial) repairing sequence must be used.
Although this would lead to an optimisation in terms of the number of elements inserted and deleted, it would lead to an increase in runtime due to the construction of application conditions.

For any circular conflict-free constraint, Algorithm \ref{Algo_conflict_free} is correct and will always terminate according to the following Theorem.

\begin{figure}
\centering
\begin{algorithm}[H]
	\KwData{A graph $G$, a circular conflict free constraint $c$ in UANF and 
	a repairing set $\mathcal{R}$ for $c$.}
	\KwResult{A graph $H$ with $H \models c$.}
	\While{$G \not \models c$}{
		Determine $\kmax$ using Algorithm \ref{Algo_kmax} \;
		$P \gets \{p: C_{\kmax + 2} \inj H \mid p \not \models \cut{0}{\scond{\kmax 
		+2}{c}} $ and $p$ is a potentially increasing occurrence at layer $\kmax$ w.r.t. $c \}$\;
		Choose $p \in P$ uniformly at random \;
		Choose $r \in \{0,1\}$ uniformly at random\;

		\uIf{$r = 0$ and $\mathcal{R}$ contains a repairing sequence for 
		$C_{\kmax +2}$}{
			Apply the repairing sequence for $C_{\kmax+2}$ at match $p$ and let 
			$H$ be the derived graph \;
			$M \gets \{q :C_j \inj H \mid j < \kmax+2 \text{ odd and } \neg \exists q': C_j: 
			\inj G(\track\circ q' = q) \text{ and $q$ is a potentially increasing occurrence at layer $j-2$ w.r.t. $c$} \}$\;
		}\Else{
			Apply the repairing sequence for $C_{\kmax+3}$ at match $p$ and let 
			$H$ be the derived graph and $\track$ the track morphism \;
			$M \gets \{q: C_j \inj H \mid j< \kmax +2 \text{ odd and } \exists q':C_{j} 
			\inj G(q = \track \circ q' \text{ is total, } q' \models\exists(C_{j+1}, \true) \text{ and }  q \not \models\exists(C_{j+1}, \true))\text{ and $q$ is a potentially increasing occurrence at layer $j-2$ w.r.t. $c$}\}$\;
		}
		\While{$H \not \models_{\maxk{c}{G}}c$}{
			Choose $p:C_j \inj H \in M$ uniformly at random \;
			Choose $r \in \{0,1\}$ uniformly at random \;
			
			\uIf{$r = 0$ and $\mathcal{R}$ contains a repairing sequence for 
			$C_j$}{
				Apply the repairing sequence for $C_{j}$ at match $p$ and 
				let $H'$ be the derived graph \;
				$M' \gets \{q: C_{j'} \inj H' \mid {j'} \text{ odd and } \neg \exists 
				q':C_{j'} \inj H(\track \circ q' = q) \text{ and $q$ is a potentially increasing occurrence at layer $j'-2$ w.r.t. $c$} \} $ \;
				 
			}\Else{
				Apply the repairing sequence for $C_{j+1}$ at match $p$ and 
				let $H'$ be the derived graph and $\track$ the track morphism\;
				
				$M' \gets \{q: C_{j'} \inj H' \mid j< \kmax +2 \text{ odd and } \exists q':C_{j'} 
			\inj H(q = \track \circ q' \text{ is total, } q' \models\exists(C_{j'+1}, \true) \text{ and }  q \not \models\exists(C_{j'+1}, \true)) \text{ and $q$ is a potentially increasing occurrence at layer $j'-2$ w.r.t. $c$}\}$\;
			}

			$M \gets M' \cup \{\track \circ q \mid q \in M\setminus\{p\} \text{ and } \track \circ q \text{ is total}\}$ \;
			$H \gets H'$\;
		}	
		$G \gets H$\;	
	}
	\KwRet{G}\;

\caption{Repair for one circular conflict free constraint}\label{Algo_conflict_free}

\end{algorithm}
\end{figure}

\begin{theorem}
Given a graph $G$,  a circular conflict free constraint $c$ in UANF and 
	a repairing set $\mathcal{R}$ of $c$. 
	Then, Algorithm \ref{Algo_conflict_free} with input $G$, $c$ and $\mathcal{R}$ 
	terminates and returns a graph $H$ with $H \models c$.
\end{theorem}
\begin{proof}
	If Algorithm \ref{Algo_conflict_free} terminates, it returns a graph 
	that satisfies $c$. Therefore, it is sufficient to show that Algorithm 
	\ref{Algo_conflict_free} terminates.
	Since $G$ is finite, the set $P$ must also be finite. If a repairing sequence has been applied, the set $M$ contains only occurrences of graphs $C_j$ such that $C_{\kmax +2}$ causes a (transitive) conflict with $C_j$, since the repairing sequence is not able to destroy or insert occurrences of $C_i$ such that $C_{\kmax +2}$ does not cause a conflict with $C_i$. 
	Note that this is also the case for repairing sequence for universally bound graphs. Since $G$ is finite, $|M|$ must also be finite.
	
	If the derived graph does not satisfy $\cut{\maxk{c}{G}}{c}$, we need to restore the largest satisfied layer.
	Since the largest satisfied layer only decreases if an occurrence of an existentially bound graph $C_j$ is destroyed such that an occurrence $p$ of $C_{j-1}$ satisfies $\exists(C_j, \true)$  and $\track \circ p$ does not satisfy $\exists(C_j, \true)$ or an occurrence of universally bound graphs is inserted, and $M$ contains all these occurrences that can also improve the satisfaction at layer (see Lemma \ref{num_violations}). So we only need to consider the occurrences contained in $M$.
	Applying repairing sequences to occurrences $p :C_j \inj H \in M$ could again lead to the insertion of universally bound graphs or the removal of existentially bound graphs. The set $M'$ contains all these occurrences, and again these are only occurrences of $C_i$ such that $C_j$ causes a (transitive) conflict for $C_i$.
	Since $c$ is free of circular conflict, $M'$ cannot contain any occurrences of $C_{\kmax +2}$, otherwise, $C_j$ would have caused a (transitive) conflict for $C_{\kmax +2}$ and therefore $C_{\kmax +2}$ has a circular conflict. Therefore no occurrences of $C_{\kmax +3}$ are destroyed and no occurrences of $C_{\kmax+2}$ are inserted.
	In addition, $C_{\kmax +2}$ causes a (transitive) conflict for $C_i$, and repairing any $p \in M'$ will not lead to the insertion of an occurrence of $C_{\kmax +2}$ or removal of an occurrence of $C_{\kmax +3}$.
	
	Since $c$ is circular conflict free, there must exist graphs $C_i$, such that $C_i$ does not cause a conflict with any other graph $C_{i'}$ and $C_{\kmax+2}$ causes a (transitive) conflict for $C_i$. 
Therefore, the application of repairing sequences at occurrences of these graphs will not lead to the insertion or removal of any universally or existentially bound graph, respectively.
	Since $c$ is finite, the number of graphs $C_i$ 
	such that $C_{\kmax+2}$ causes (transitive) conflict with $C_i$ is finite.
	Since $|M'|$ is also finite, after a finite number of applications of  repairing sequences, $M'$ contains only occurrences of graphs that do not cause any conflicts. After a repairing sequence has been applied to all these occurrences, $M'$ is empty and $H \models_{\maxk{c}{G}} c$, since all occurrences $p$ of $C_j$ which have either been inserted or an occurrence $q$ of $C_{j+1}$ with $p = a_j \circ q$ has been removed satisfy $\exists (C_{j+1},\true)$ or have been removed. 
	
	Therefore, after a finite number of iterations, the set $P$ is empty and Lemma \ref{num_violations} implies that the largest satisfied layer has been increased by at least $1$.
It follows that after a finite number of iterations $G \models c$. 
Then Algorithm \ref{Algo_conflict_free} terminates and returns $G$.
\end{proof}

\begin{figure}
	\centering
	\includegraphics[scale=1]{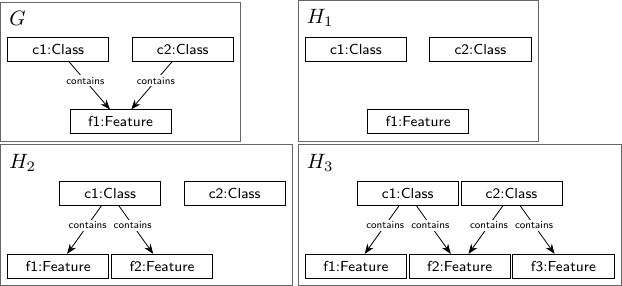}

	\caption{Possible outputs of the repairing process for $G$ and $\forall(C_2^2, \exists(C_2^1, \true))$ using the rule set $\{\texttt{removeFeature},\texttt{createFeature}\}$.}\label{fig:rep}
\end{figure}

\begin{example}
	Consider constraint $c =\forall(C_2^2, \exists(C_2^1, \true))$ which is composed of
	the graphs shown in Figure \ref{fig:constraints}.
	This constraint is circular conflict-free and a repairing set for $c$ is 
	given in Figure \ref{fig:rep}.
	There is a repairing sequence for $C_2^2$ via the rule 		
	\emph{\texttt{remove}} and a repairing sequence for $C_2^1$ via the rule 
	\emph{\texttt{insert}}.
	Using the rule set $\{\emph{\texttt{remove}},\emph{\texttt{insert}}\}$,
	Algorithm \ref{Algo_conflict_free} could return one of the graphs 
	$G_1, G_2$ or $G_3$ given in Figure \ref{fig:rep}, depending on the repairing sequences used.
\end{example}


\subsection{Rule-based Graph Repair for multiple Constraints}

We will now present our rule-based repair approach for a set of constraints in UANF.

\begin{definition}[\textbf{satisfaction of constraint sets}]
	Let $\mathcal{C}$ be a set of constraints. A graph $G$ satisfies $\mathcal{C}$, denoted by $G \models \mathcal{C}$, if $G \models \bigwedge_{c \in \mathcal{C}} c$. The set $\mathcal{C}$ is called \emph{satisfiable } if there exists a graph $G$ with $G \models \mathcal{C}$.
\end{definition}

To guarantee that a set of constraints can be repaired by a set of rules, we need to extend the notion of repairing sets such that a set of rules is called a \emph{repairing set} for a set of constraints if it is a repairing set for every constraint in the constraint set.
\begin{definition}[\textbf{repairing set for a set of constraints}]
	Given a set $\mathcal{C}$ of constraints in UANF  and a set of rules $\mathcal{R}$. 
Then $\mathcal{R}$ is called a \emph{repairing set for $\mathcal{C}$} if $\mathcal{R}$ is a repairing set for all constraints $c \in \mathcal{C}$.

\end{definition}

We also extend the notion of conflicts to \emph{conflicts between constraints}. 
Intuitively, a constraint $c$ causes a conflict for another constraint $c'$ if one of its graphs causes a conflict for a graph of $c'$.

\begin{definition}[\textbf{conflict between constraints}]
	Let the constraints $c$, $c'$ in UANF and a set of rules $\mathcal{R}$ be given. 
Then $c$ \emph{causes a conflict} for $c'$ w.r.t. $\mathcal{R}$ if a repairing sequence 
$$C_k = G_0 \Longrightarrow_{\rho_1,m_1} \dots \Longrightarrow_{\rho_n,m_n}
	G_n$$
	for a graph, $C_k$ of $c$ exists such that the concurrent rule of that 
	sequence is not a basic consistency maintaining rule w.r.t. $\forall(C_j, 
	\false)$ or $\exists(C_j,\true)$ for any universally or existentially bounded 
	graph $C_j$ of $c'$.

\end{definition}

The following lemma is a useful statement for proving the correctness of our repair approach. 
It states that applying a repairing sequence to a constraint $c$ cannot destroy the satisfaction of $c'$ if $c$ causes no conflict for $c'$.

\begin{lemma}\label{lemma_preserving}
	Given a set of rules $\mathcal{R}$ and constraints $c$ and $c'$ in UANF such that $c$ causes no conflict for $c'$ w.r.t. $\mathcal{R}$.
Then, the concurrent rule $\rho$ of any graph of $c$ is a $c'$-preserving rule.
\end{lemma}
\begin{proof}
	Suppose $\rho$ is not a $c'$-preserving rule.  Then there exists a transformation $t : G \Longrightarrow_{\rho,m} H$ such that $G \models c'$ and $H \not \models c'$. 
	Therefore, either a universally bound graph of $c'$ has been inserted or an existentially bound graph of $c'$ has been removed.
	It follows that $\rho$ is not a basic maintaining rule w.r.t. $\forall(C_j, \false)$ for all universally bound graphs $C_j$ of $c'$ or $\rho$ is not a basic maintaining rule w.r.t. $\exists(C_j, \true)$ for all existentially bound graphs $C_j$ of $c'$, which is a contradiction.
\end{proof}

\begin{figure}
\centering
\begin{algorithm}[H]
	\KwData{A graph $G$, circular constraint-conflict free set of constraints $
	\mathcal{C}$ and a repairing set $\mathcal{R}$ for $\mathcal{C}$.}
	\KwResult{A graph $H$ with $H \models \bigwedge_{c \in \mathcal{C}}c$.}
	$(c_1, \ldots ,c_n) \gets $ topological ordering of $\mathcal{C} w.r.t. \mathcal{R}$ \;
	\For{$i \gets 1$ \KwTo $n$}{
		
		Repair $c_i$ in $G$ with Algorithm \ref{Algo_conflict_free}, let $H$ be the 		returned graph \;
		$G \gets H$ \;
	
	}
	\KwRet{G}\;

\caption{Repair for a circular constraint-conflict free set of constraints }\label{Algo_non-conflict_free}

\end{algorithm}
\end{figure}

The \emph{conflict graph} for a set of constraints and \emph{circular conflicts of a set of constraints} are defined in a similar way to the conflict graph and circular conflicts for one constraint. 
A set of constraints is called \emph{circular conflict free} if each of its constraints is circular conflict-free and there is no sequence $c = c_0, \ldots, c_n = c$ such that $c_i$ has a conflict with $c_{i+1}$ for all $0\leq i < n$. In other words, the conflict graph of this set is acyclic.

\begin{definition}[\textbf{conflict graphs, circular conflicts}]
	Given a set of rules $\mathcal{R}$ and a set $\mathcal{C}$ of constraints  in UANF. The \emph{conflict graph} of $\mathcal{C}$  w.r.t. $\mathcal{R}$ is constructed in the following way. For each constraint $c \in \mathcal{C}$ there is a node. If $c$ causes a conflict for $c'$ w.r.t. $\mathcal{R}$, there is an edge $e$ with $\src(e) = c$ and $\tar(e) = c'$.
	
	A constraint $c$ causes a \emph{transitive conflict } for $c'$ w.r.t. $\mathcal{R}$ if the 
conflict graph of $\mathcal{C}$ w.r.t. $\mathcal{R}$ contains a path from $c$ to $c'$. 
A constraint $c$ has a \emph{circular conflict w.r.t. $\mathcal{R}$} if $c$ has a transitive conflict with itself. 
A set of constraints $\mathcal{C}$ is called \emph{circular conflict free w.r.t. $\mathcal{R}$} if every constraint in $\mathcal{C}$ is circular conflict-free and $\mathcal{C}$ contains no circular conflicts w.r.t. $\mathcal{R}$.
\end{definition}

\begin{figure}
	\centering
	\includegraphics[scale=1]{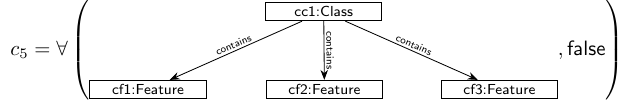}
	\par
\vspace{1cm}

	\includegraphics[scale=1]{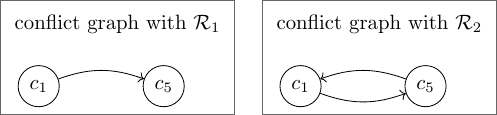}

	\caption{Constraints $c_5$ and conflicts graphs of the constraint set 
	$\{c_1,c_5\}$ with the rule sets $\mathcal{C}_1=\{\texttt{resolve},\texttt{createFeatures}\}$ and $mathcal{C}_2 =\{\texttt{resolve2},\texttt{createFeatures}\}$. }\label{fig:constraint_conflict}
\end{figure}

\begin{example}
	Consider the rules \emph{\texttt{resolve, resolve2, createFeatures}} and constraints $c_1$ and $c_5$ given in Figures \ref{fig:constraint_conflict} and \ref{fig:constraints}.
The constraint set $\mathcal{C} = \{c_1, c_5\}$ is a multiplicity which expresses that \enquote{Each node of type \emph{\texttt{Class}} is connected to exactly two nodes of type \emph{\texttt{Feature}}}.
	With the rule set $\mathcal{R}_1 = \{\emph{\texttt{resolve}}, \emph{\texttt{createFeatures}}\}$, there is only one conflict in $\mathcal{C}$; $c_1$ causes a conflict for $c_5$, since applying \emph{\texttt{createFeatures}} could lead to inserting the universally bound graph of $c_5$. With the rule set $\mathcal{R}_2 = \{\emph{\texttt{resolve2}}, \emph{\texttt{createFeatures}}\}$ there are two conflicts. Again, $c_1$ causes a conflict for $c_5$ and  $c_5$ causes a conflict for $c_1$, since applying \emph{\texttt{resolve}} can destroy an occurrence of the existentially bound graph of $c_1$.
	
	Therefore, our approach will terminate with the rule set $\mathcal{R}_1$ 
	but not with  $\mathcal{R}_2$ because $\mathcal{C}$ is 
	not circular conflict free w.r.t. $\mathcal{R}_2$.
	
\end{example}

Our repair process exploits the fact that the conflict graph of a circular conflict-free set of constraints in UANF is acyclic. 
In particular, our approach uses the \emph{topological ordering} of this 
conflict graph.

\begin{definition}[\textbf{topological ordering of a graph}]
	Given is a graph $G$. A sequence $(v_1, \ldots, v_n)$ of nodes of $G$ is called a \emph{topological ordering} of $G$ if no edge $e \in E_G$ exists with $\src(e) = v_i$, $\tar(e) = v_j$ and $i \geq j$. The topological ordering of a circular conflict-free set of constraints $\mathcal{C}$ w.r.t. a rule set $\mathcal{R}$ is the topological order of its conflict graph w.r.t. $\mathcal{R}$.
\end{definition}
It is well known that every directed acyclic graph has a topological ordering that can be computed in $\Theta(|V| + |E|)$ where $V$ and $E$ are the set of nodes and edges of the respective graph \cite{cormen2022introduction}. Therefore every conflict graph of a circular conflict-free set of constraints also has a topological ordering.

The repair process is given in Algorithm \ref{Algo_non-conflict_free} and proceeds as follows. 
First, the topological ordering of the constraint set is determined (line 1). Then Algorithm \ref{Algo_conflict_free} is used to repair each constraint of $\mathcal{C}$ in the order of the topological ordering (lines 2 -- 4). 
This ensures that the satisfaction of a constraint that has already been repaired is not destroyed by the repair of another constraint. 
\begin{theorem}
	Given rule set $\mathcal{R}$, a graph $G$ and a satisfiable circular conflict free set of constraints $\mathcal{C}$, such that $\mathcal{R}$ is a repairing set for $\mathcal{C}$.
Then Algorithm \ref{Algo_non-conflict_free} terminates and returns a graph $H$ with $H \models \mathcal{C}$.
\end{theorem}
\begin{proof}
	Since $\mathcal{C}$ is finite and every $c \in \mathcal{C}$ is circular conflict-free,  Algorithm \ref{Algo_conflict_free} terminates for every $c \in \mathcal{C}$. Therefore,  Algorithm \ref{Algo_non-conflict_free} will also terminate.
	It remains to show that the returned graph satisfies $\mathcal{C}$.
	Let $(c_1, \ldots, c_n)$ be a topological ordering of $\mathcal{C}$ w.r.t. $\mathcal{R}$. Then, no constraint $c_j$ with $j \neq 1$ causes a conflict for $c_1$, and  Lemma \ref{lemma_preserving} implies that the concurrent rule of every repairing sequence for every graph of $c_i$ with $2 \leq i \leq n$ is a $c_1$-preserving rule.
	In general, the concurrent rule of each repairing sequence for graphs of $c_j$ is a
	$c_i$-preserving rule if $i < j$.
	After one iteration it holds that $G \models c_1$. Suppose that after $m$ iterations it holds that $G \models c_i$ for all $1 \leq i \leq m$.
	In iteration $m+1$, $c_{m+1}$ is repaired by Algorithm 
 \ref{Algo_conflict_free}. Since each concurrent rule of each repairing sequence for graphs of  $c_{m+1}$ is a $c_i$-preserving rule for all $1 \leq i \leq m$ and the application of repairing sequence can be replaced by an application of its concurrent rule, it follows that  $ H \models c_i$ for all $1 \leq i \leq m+1$.  
	Therefore, after $n$ iterations, $H \models c_i$ for all $1 \leq i \leq n$ 
	and the returned graph satisfies 
	$\mathcal{C}$.
\end{proof}

\section{Related Work}\label{rel_work}

\begin{figure}
\centering
\begin{tikzpicture}
\draw[fill=lightgray](0,-0.5) ellipse (2cm and 3.5cm);
\draw{(0,0) circle(4cm)};
\node at (0,3.5) {ANF};
\draw{(0,-1) circle(3cm)};
\node at (0,0.5) {proper};
\draw{(0,-2) circle(2cm)};
\node at (0,-1.5) {w-f};
\draw{(0,-3) circle(1cm)};
\node at (0,-3) {mult};

\end{tikzpicture}
\caption{Relationships between the sets of constraints where \enquote{w-f} is an abbreviation for well-formedness and \enquote{mult} is an abbreviation for multiplicity.
The set of circular conflict-free constraints is highlighted in grey.}\label{venn}
\end{figure}
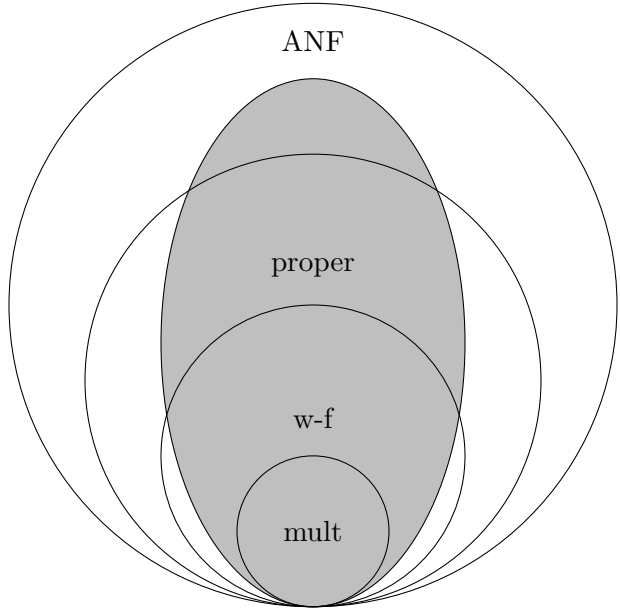

In this section, we summarise other approaches for rule-based graph repair and compare them to our approach.
An overview of the relations of all constraint types mentioned below is shown in Figure \ref{venn}. \\ \\
\textbf{Iterative Development of Consistency-Preserving Rule-Based Refactorings}: Becker et al. \cite{becker2011iterative} introduced an interactive approach to construct con\-sis\-tency-\-pre\-ser\-ving transformations based on their invariant checker introduced in \cite{becker2006symbolic} for so-called \emph{well-formedness constraints}. 

These are constraints of the form $\neg \exists (c_1, \true)$ or $\forall (C_1,  \exists(C_2, \true))$.
Given a consistent graph, a well-formedness constraint and a refactoring specification, which is a set of rules in the single-pushout approach \cite{hartmut2006fundamentals}, the invariant checker constructs all minimal counterexamples that lead to a non-consistency preserving transformation via rules of the refactoring specification. If there are no such counterexamples, then any transformation with a rule of the refactoring specification is consistency-preserving.
This approach is designed to be fully interactive, requiring the user to revise the refactoring specification until no counterexamples are returned.

Our approach allows the automatic construction of consistency-preserving rule sets w.r.t. a set of constraints in ANF. The results of section \ref{comp_general} show that every direct consistency-preserving rule is also a consistency-preserving rule, and therefore all rules of a rule set $\mathcal{R}$ can be equipped with the direct consistency-maintaining application condition introduced in section \ref{appl_conds}. This newly created rule set contains only consistency-preserving rules.
\\ \\
\textbf{Ensuring Consistency of Conditional Graph Grammars:} Heckel and Wagner \cite{heckel1995ensuring} have presented an approach to construct consistency-preserving application conditions for rules in the single-pushout approach and  constraints of the form $\forall(C_1, \exists (C'_1, \true)) \wedge \ldots \wedge \forall(C_n, \exists (C'_n, \true))$.
Although the constructed application conditions are not presented as nested conditions, they can be transformed into nested conditions of the form $\forall(C_1, \exists(C_1^1, \true) \vee \ldots \vee \exists(C_1^{k_1}, \true)) \wedge \ldots \wedge \forall(C_n, \exists(C_n^1, \true) \vee \ldots \vee \exists(C_1^{k_n}, \true))$.

Our approach allows the construction of consistency-preserving application conditions for such constraints. A constraint $c$ of the form described above is a conjunction of constraints in UANF. Therefore, we can construct the direct consistency-maintaining application conditions for all these constraints. The conjunction of these application conditions is a direct consistency-maintaining and therefore a consistency-preserving application condition for $c$.
\\ \\
\textbf{Sustaining and Improving graduated Graph Consistency}:
Kosiol et al. \cite{kosiol2022sustaining} have introduced the notions of (direct) consistency-sustaining and (direct) consistency-improving transformations as already introduced in section \ref{sec_consistency}.
This approach is designed for rules in the double-pushout approach and nested conditions in ANF. 
They have introduced a method for constructing consistency-sustaining application conditions, a sufficient criterion for consistency-sustaining  transformations and a necessary criterion for consistency-improving transformations. Both criteria have been implemented and evaluated.

As already discussed, our notions of (direct) consistency-maintaining and (direct) consistency-increasing application conditions are more fine grained and in generally not related to those described in \cite{kosiol2022sustaining}. 
If the nesting level of a constraint $c$ is $1$, these notions are identical and any consistency-increasing application condition at layer $-1$ for $c$ is also a consistency-increasing application condition for $c$. 
Therefore, a consistency-increasing application condition at layer $-1$ for $c$ is also consistency-improving application condition for $c$.

Any consistency-maintaining application condition constructed by Theorem \ref{appl-main} contains a consistency-sustaining application condition constructed by the approach of \cite{kosiol2022sustaining}. Therefore, these consistency-maintaining application conditions are also consistency-sustaining ones. However, because these consistency-maintaining application conditions also contain additional conditions, a con\-sis\-ten\-cy-maintaining application condition is more restrictive and complex than a application condition constructed using the method introduced in \cite{kosiol2022sustaining}. 
\\ \\
\textbf{Constructing optimized constraint-preserving application conditions for mo\-del transformation rules}:
Nassar et al. \cite{nassar2020constructing} have introduced a method for constructing consistency-sustaining and consistency-preserving application conditions in the framework of $\mathcal{M}-$adhesive categories.
Due to some optimisations, these application conditions are less restrictive and less complex than those described in \cite{habel2009correctness} and \cite{kosiol2022sustaining}. They have introduced the notion of \emph{weakest application conditions}.
As the name suggests, a weakest application condition is implied by any other application condition with the same property. 
For example, a weakest consistency-preserving application condition is implied by every other consistency-preserving application condition. The construction of the application conditions has been implemented as an eclipse plug-in called \emph{OCL2AC}, which is able to construct consistency-guaranteeing, weakest consistency-preserving or consistency-sustaining application conditions.

Some of the optimisations presented could also be used to optimise the application conditions we have introduced. 
In particular, these optimisations could be used to reduce the complexity of the direct consistency-maintaining and consistency-increasing application conditions at layer for general rules. 
We have already indicated that the application conditions introduced in Section \ref{appl_conds} are not weakest conditions, 
and that constructing such weakest direct consistency-maintaining or direct consistency-increasing application conditions would probably lead to huge application conditions.
\\ \\
\textbf{Rule-based Graph Repair}:
Sandmann and Habel \cite{sandmann2019rule} have introduced a repair process for so-called \emph{proper constraints} based on so-called \emph{repair programs}. 
A constraint in ANF is called \emph{proper} if it ends with $\exists(C, \true)$ or is of the form $\exists(C_1, \forall(C_2, \false))$ or $\forall(C_1, \false)$.
The authors describe a method for inductively constructing a repair program consisting of rules in the double-pushout approach that, when applied to a non-consistent graph, returns a consistent graph. 
They have also introduced a graph repair approach given a set of rules $\mathcal{R}$. 
The approach described above can be used to repair a graph with rules from $\mathcal{R}$ if there is a repair program such that for every rule in the repair program there is an equivalent rule in $\mathcal{R}$.

Our approach can repair circular conflict-free constraints. The set of circular conflict-free constraints and the set of proper constraints intersect, but the set of circular conflict-free constraints is not contained in the set of proper constraints. Note that a constraint $c$ in ANF is not proper if it ends with $\forall(C_{\nlvl(c)}, \false)$ and $\nlvl(c) > 2$. Therefore, a non-proper circular conflict-free constraint can be easily constructed. 
Furthermore, our approach can repair a set $\mathcal{C}$ of circular conflict-free constraints, if $\mathcal{C}$ is a circular conflict-free set of constraints w.r.t. to a rule set $\mathcal{R}$ and $\mathcal{R}$ is a repairing set for $\mathcal{C}$. 
\\ \\
\textbf{Rule-based Repair of EMF Models}:
Nassar et al. \cite{nassar2017rule, nassar2017rule1} have introduced a repair approach for models of the \emph{eclipse modeling framework} (EMF) \cite{steinberg2008emf}. In particular, this approach is able to repair multiplicities of a given EMF metamodel. 
Multiplicities can be described as nested conditions of the form $\forall(C_1, \exists(C_2, \true))$ and $\forall(C_1, \false)$. 
The approach was implemented using two Eclipse plugins in Henshin \cite{arendt2010henshin}. One plugin derives rules for the repair process from a metamodel. The other plugin is an implementation of the repairing process.

Every upper bound of a multiplicity can be described as a constraint of the form $\forall(C_1, \true)$. Such a constraint is always circular conflict-free. In addition, a lower bound of a multiplicity can be described as a constraint of the form $\forall(C_1, \exists(C_2, \true))$ where $C_1$ contains exactly one node and no edges. Therefore, such a constraint is also always circular conflict-free. 
This implies, that our approach can repair a set of multiplicities, if an appropriate set of rules $\mathcal{R}$ is used. That is, the set of multiplicities is a circular conflict-free set of constraints w.r.t. $\mathcal{R}$, and $\mathcal{R}$ is a repairing set for $\mathcal{C}$.

\section{Conclusion}\label{conclusion}

In summary, we have introduced a rule-based graph repair approach for circular conflict-free constraints and, in particular, for a circular conflict-free set of constraints, and have shown its correctness and termination. 
Our approach can also be used for constraints and sets of constraints that are not circular conflict-free. However, there is no guarantee that the algorithm will terminate and thus return a consistent graph. 

We have introduced new notions of consistency, called (direct) consistency-main\-tain\-ing and (direct) consistency-increasing transformations and rules, which are finer-grained than previously known concepts. To do this, we have introduced the notion of \emph{satisfaction up to layer}, which is a notion of partial consistency, and showed what can be deduced from it. For example, if $k$ is even and a graph satisfies a constraint up to layer $k$, then the graph also satisfies $c$.

We have used the notions of direct consistency-maintaining and direct consistency-increasing transformations to characterise the circumstances under which a given rule set $\mathcal{R}$ is able to repair a constraint or set of constraints. 
To do this, we first introduced the notions of \emph{conflicts within conditions} and \emph{conflicts between conditions}. 
Intuitively, given a constraint $c$, a universally bound graph $C_k$ causes a conflict with an existentially bound graph $C_j$ if an occurrence of $C_j$ can be destroyed by destroying an occurrence of $C_k$. An existentially bound graph $C_j$ causes a conflict for a universally bound graph $C_k$ if an occurrence of $C_k$ can be inserted by inserting elements of $C_j \setminus C_{j-1}$. 
A set of rules $\mathcal{R}$ is capable of repairing a circular conflict-free constraint if there exists a sequence of transformations that repairs an occurrence of a universally bound graph $C_k$ so that it satisfies $\exists(C_{k+1}, \true)$ or destroys its occurrence. Using the notions of direct consistency-maintaining rules and transformations, we ensure that these sequences cannot create or delete occurrences of graphs such that either $C_k$ or $C_{k+1}$ does not cause a conflict for them.

We have compared the notions of (direct) consistency-maintaining and (direct) con\-sis\-ten\-cy-increasing transformations and rules with the notions of 
consistency-gua\-ran\-tee\-ing, con\-sis\-ten\-cy-preserving, (direct) consistency-improving and (direct) consistency sustaining transformations and rules \cite{kosiol2022sustaining, habel2009correctness}. In general, the notions of (direct) consistency-sustaining and  (direct) consistency-improving transformations  are not at all related to our newly introduced notions.
Only in the special case $\nlvl(c) = 1$, our notions are identical to those of (direct) consistency-sustaining and (direct) consistency-improving transformations. 
Our notions are related to those of consistency-preserving and consistency-guaranteeing  transformations. 
If a constraint $c$ is not satisfied, a $c-$guaranteeing transformation is consistency-increasing w.r.t. $c$.
In general, a consistency-guaranteeing transformation is also consistency-maintaining and a consistency-maintaining transformation is also consistency-pre\-ser\-ving.

Furthermore, we have introduced the weaker notions of \emph{(direct) consistency-main\-tain\-ing rules at layer} and \emph{(direct) consistency-increasing rules at layer}, which allow the construction of less complex application conditions. A rule $\rho$ is direct consistency-maintaining or direct consistency-increasing at layer $k$ if all of its applications at graphs $G$ with $\kmax = k$ are (direct) consistency-maintaining or (direct) consistency-increasing. Using these notions, we have introduced four constructions for application conditions and shown that rules equipped with these conditions are direct consistency-maintaining at layer, direct consistency maintaining and direct consistency-increasing at layer. In particular, we have introduced two types of consistency-increasing application conditions at layer.
First, application conditions for general rules, and second, application conditions for a special set of rules, called basic rules.
Since basic rules, by definition, are not able to introduce new violations or decrease the satisfaction up to layer, it is sufficient to check that at least one violation is removed. Since the left-hand side of a basic increasing rule $\rle{L}{l}{K}{r}{R}$  at layer $k$ must contain an occurrence $p$ of $C_k$ such that either $r \circ l^{-1} \circ p \models \exists(C', \true)$ and $p \not \models \exists(C, \true)$ for an intermediate graph $C' \in \ig{C_k}{C_{k+1}}$ or $r \circ l^{-1} \circ p$ is not total, it is sufficient to check whether $m \circ p$ does not satisfy $\exists(C', \true)$. This leads to less complex application conditions compared to the application conditions of other approaches. Compared to the general ones, these application conditions are less complex and less restrictive.

We have introduced \emph{derived rules} to ensure that the condition that $L$ must contain an occurrence of $C_k$ is not a restriction on the notion of basic rules. The left-hand sides of derived rules contain an occurrence of $C_k$, and we have shown that these rules are only applicable if the rule from which this rule is derived is applicable at a smaller match such that both transformations construct the same graph.

Future work is to extend the notions of consistency-maintaining and consistency-increasing transformations for all types of nested constraints, and a rule-based repair process for all satisfiable nested constraints, i.e. constraints that also use Boolean operators. 
Although we have presented characterisations for circular conflict-free constraints, it remains unclear for which practical applications our approach is suitable. This may require implementation and further evaluation of the repair process, characterisations and construction of application conditions.
The notion of conflict between constraints is very strict, and we are confident that there exists a repair process that uses the repair process for one constraint to repair a set of constraints in parallel. To do this, the conflict graph for all graphs of all constraints must be acyclic, i.e. there is no circular conflict in the set of all graphs of all constraints. Then, we can repair all constraints using Algorithm \ref{Algo_conflict_free}. The challenge of this approach is to decide which occurrences of which graphs need to be repaired in order to increase consistency.

We also suggest the following more technical optimisations for the general application conditions.
First, not all overlaps in the set $\mathbf{P}_{C'}$ need to be considered when constructing the no violation inserted part of the maintaining application condition since some of them are implied by others. 
Second, the application condition constructed from the violation removed part of the consistency increasing application condition is far too restrictive, since all occurrences of $P'$ that extend $R$ must satisfy $\exists(C', \true)$ after the transformation.
We are confident that, with some additional theory, the $\nex(P,C' )$ and $\rep(P,C' )$ parts of the application condition can be combined to construct less restrictive and less complex application conditions.

\newpage

\bibliographystyle{abbrv}
\bibliography{Literatur.bib}

\begin{thebibliography}{10}

\bibitem{arendt2010henshin}
T.~Arendt, E.~Biermann, S.~Jurack, C.~Krause, and G.~Taentzer.
\newblock
  \href{https://d1wqtxts1xzle7.cloudfront.net/78132946/Henshin_Advanced_Concepts_and_Tools_for_20220105-29458-lznu2g-with-cover-page-v2.pdf?Expires=1653240209&Signature=b-TMhXY1k8tHf6XHIuHPz3tbCBJ1q0vtXwCYkbM7TM1pLKXORlbwG5feVaVgcMKBYHEJ75irQ35lN6m9NBeCf-CvurclLnSa-ZL-cmxTuYtdiiPc3PRI89EARqKb8gBSNqwrkHb4zH6nguJnPTZzF7hb97diSIeN8sfkv4EeNmdJnon857ynWjcw2PcNWl2LPmTzXrZMFCDVjFCKBcgHeVZrhh0S1YEozS~~K3bl4yk4Ili7TgiibF40zyw25r3fjYjpXFmOZOh2FncobwANsj6qmqVizMOWrugKZlbUu2WSfsq2vx2c-zks4~58tdmY9LhE6ToKZmqfvSIxsgjr2w__&Key-Pair-Id=APKAJLOHF5GGSLRBV4ZA}{Henshin:
  advanced concepts and tools for in-place EMF model transformations}.
\newblock In {\em International Conference on Model Driven Engineering
  Languages and Systems}, pages 121--135. Springer, 2010.

\bibitem{becker2006symbolic}
B.~Becker, D.~Beyer, H.~Giese, F.~Klein, and D.~Schilling.
\newblock \href{https://dl.acm.org/doi/abs/10.1145/1134285.1134297}{Symbolic
  invariant verification for systems with dynamic structural adaptation}.
\newblock In {\em Proceedings of the 28th international conference on Software
  engineering}, pages 72--81, 2006.

\bibitem{becker2011iterative}
B.~Becker, L.~Lambers, J.~Dyck, S.~Birth, and H.~Giese.
\newblock
  \href{https://link.springer.com/chapter/10.1007/978-3-642-21732-6_9}{Iterative
  development of consistency-preserving rule-based refactorings}.
\newblock In {\em International Conference on Theory and Practice of Model
  Transformations}, pages 123--137. Springer, 2011.

\bibitem{behr2020efficient}
N.~Behr, R.~Heckel, and M.~G. Saadat.
\newblock \href{https://arxiv.org/abs/2003.11010}{Efficient Computation of
  Graph Overlaps for Rule Composition: Theory and Z3 Prototyping}.
\newblock {\em arXiv preprint arXiv:2003.11010}, 2020.

\bibitem{biermann2010parallel}
E.~Biermann, H.~Ehrig, C.~Ermel, U.~Golas, and G.~Taentzer.
\newblock
  \href{https://link.springer.com/chapter/10.1007/978-3-642-17322-6_7}{Parallel
  independence of amalgamated graph transformations applied to model
  transformation}.
\newblock {\em Graph Transformations and Model-Driven Engineering: Essays
  Dedicated to Manfred Nagl on the Occasion of his 65th Birthday}, pages
  121--140, 2010.

\bibitem{cormen2022introduction}
T.~H. Cormen, C.~E. Leiserson, R.~L. Rivest, and C.~Stein.
\newblock {\em
  \href{https://books.google.de/books?hl=de&lr=&id=RSMuEAAAQBAJ&oi=fnd&pg=PR13&dq=Introduction+to+algorithms&ots=a2n7YT2HUL&sig=OwTm-BWxfMT6wZMEDtUQ9QohHA0&redir_esc=y#v=onepage&q=Introduction%20to%20algorithms&f=false}{Introduction
  to algorithms}}.
\newblock MIT press, 2022.

\bibitem{hartmut2006fundamentals}
H.~Ehrig, K.~Ehrig, U.~Prange, and G.~Taentzer.
\newblock
  \href{https://scholar.google.com/scholar?hl=de\&as_sdt=0\%2C5\&q=Fundamentals+of+algebraic+graph+transformation\&btnG=}{Fundamentals
  of algebraic graph transformation}.
\newblock {\em Monographs in theoretical computer science. An EATCS series.
  Springer}, 2006.

\bibitem{ehrig2010parallelism}
H.~Ehrig, A.~Habel, and L.~Lambers.
\newblock
  \href{https://journal.ub.tu-berlin.de/eceasst/article/view/363}{Parallelism
  and concurrency theorems for rules with nested application conditions}.
\newblock {\em Electronic Communications of the EASST}, 26, 2010.

\bibitem{habel2005nested}
A.~Habel and K.-H. Pennemann.
\newblock
  \href{https://link.springer.com/chapter/10.1007/978-3-540-31847-7_17}{Nested
  constraints and application conditions for high-level structures}.
\newblock In {\em Formal methods in software and systems modeling}, pages
  293--308. Springer, 2005.

\bibitem{habel2009correctness}
A.~Habel and K.-H. Pennemann.
\newblock
  \href{https://www.cambridge.org/core/journals/mathematical-structures-in-computer-science/article/abs/correctness-of-highlevel-transformation-systems-relative-to-nested-conditions/680F641F9356C8C67079778660C31291}{Correctness
  of high-level transformation systems relative to nested conditions}.
\newblock {\em Mathematical Structures in Computer Science}, 19(2):245--296,
  2009.

\bibitem{heckel1995ensuring}
R.~Heckel and A.~Wagner.
\newblock
  \href{https://www.sciencedirect.com/science/article/pii/S1571066105801884}{Ensuring
  consistency of conditional graph grammars-a constructive approach}.
\newblock {\em Electronic Notes in Theoretical Computer Science}, 2:118--126,
  1995.

\bibitem{kosiol2022sustaining}
J.~Kosiol, D.~Str{\"u}ber, G.~Taentzer, and S.~Zschaler.
\newblock
  \href{https://www.sciencedirect.com/science/article/abs/pii/S0167642321001222}{Sustaining
  and improving graduated graph consistency: A static analysis of graph
  transformations}.
\newblock {\em Science of Computer Programming}, 214:102729, 2022.

\bibitem{lambers2006conflict}
L.~Lambers, H.~Ehrig, and F.~Orejas.
\newblock \href{https://link.springer.com/chapter/10.1007/11841883_6}{Conflict
  detection for graph transformation with negative application conditions}.
\newblock In {\em Graph Transformations: Third International Conference, ICGT
  2006 Natal, Rio Grande do Norte, Brazil, September 17-23, 2006 Proceedings
  3}, pages 61--76. Springer, 2006.

\bibitem{nassar2020constructing}
N.~Nassar, J.~Kosiol, T.~Arendt, and G.~Taentzer.
\newblock
  \href{https://www.sciencedirect.com/science/article/abs/pii/S2352220820300493}{Constructing
  optimized constraint-preserving application conditions for model
  transformation rules}.
\newblock {\em Journal of Logical and Algebraic Methods in Programming},
  114:100564, 2020.

\bibitem{nassar2017rule1}
N.~Nassar, J.~Kosiol, and H.~Radke.
\newblock
  \href{https://www.researchgate.net/profile/Nebras-Nassar/publication/319068717_Rule-based_Repair_of_EMF_Models_Formalization_and_Correctness_Proof/links/5efdcab5299bf18816fa5e62/Rule-based-Repair-of-EMF-Models-Formalization-and-Correctness-Proof.pdf}{Rule-based
  repair of EMF models: formalization and correctness proof}.
\newblock In {\em Electronic Pre-Proc. Intl. Workshop on Graph Computation
  Models}, 2017.

\bibitem{nassar2017rule}
N.~Nassar, H.~Radke, and T.~Arendt.
\newblock
  \href{https://www.researchgate.net/profile/Nebras-Nassar/publication/318172242_Rule-Based_Repair_of_EMF_Models_An_Automated_Interactive_Approach/links/5efdcaa6a6fdcc4ca444ab8f/Rule-Based-Repair-of-EMF-Models-An-Automated-Interactive-Approach.pdf}{Rule-based
  repair of EMF models: An automated interactive approach}.
\newblock In {\em International Conference on Theory and Practice of Model
  Transformations}, pages 171--181. Springer, 2017.

\bibitem{plump2005confluence}
D.~Plump.
\newblock
  \href{https://link.springer.com/chapter/10.1007/11601548_16}{Confluence of
  graph transformation revisited}.
\newblock In {\em Processes, Terms and Cycles: Steps on the Road to Infinity},
  pages 280--308. Springer, 2005.

\bibitem{sandmann2019rule}
C.~Sandmann and A.~Habel.
\newblock \href{https://arxiv.org/abs/1912.09610}{Rule-based graph repair}.
\newblock {\em arXiv preprint arXiv:1912.09610}, 2019.

\bibitem{steinberg2008emf}
D.~Steinberg, F.~Budinsky, E.~Merks, and M.~Paternostro.
\newblock {\em \href{https://sisis.rz.htw-berlin.de/inh2009/12368099.pdf}{EMF:
  eclipse modeling framework}}.
\newblock Pearson Education, 2008.

\end{thebibliography}

\end{document}